\documentclass[%
  reprint,
  superscriptaddress,
  twocolumn,
  amsmath,
  amssymb,
  10pt,
  aip,
  jcp,
  citeautoscript,
  notitlepage,
  longbibliography
]{revtex4-2}

\usepackage[utf8]{inputenc}
\usepackage[T1]{fontenc}
\usepackage{xcolor}
\definecolor{blue}{RGB}{0,100,200}
\usepackage[%
  colorlinks=true,
  allcolors=blue
]{hyperref}
\usepackage{url}
\usepackage{enumitem}
\usepackage{amsfonts}
\usepackage{amssymb}
\usepackage{amsmath}
\usepackage{bm}
\usepackage{mathtools}
\usepackage[ruled,vlined]{algorithm2e}
\usepackage{lmodern}
\usepackage[normalem]{ulem}

\newcommand{\nocontentsline}[3]{}
\newcommand{\tocless}[2]{\bgroup\let\addcontentsline=\nocontentsline#1{#2}\egroup}

\DeclarePairedDelimiter\ppar{(}{)}              
\DeclarePairedDelimiter\pnrm{\lVert}{\rVert}    
\DeclarePairedDelimiter\pbkt{[}{]}              
\DeclarePairedDelimiter\pset{\{}{\}}            

\newcommand{\rfig}[1]{Fig.~\ref{#1}}

\newcommand{\rsct}[1]{Sec.~\ref{#1}}

\newcommand{\req}[1]{Eq.~\ref{#1}}

\newcommand{\dd}[1]{\operatorname{d#1}}
\newcommand{\dit}[1]{\mathrm{d}#1}

\newcommand{\bz}{\mathbf{z}}
\newcommand{\bx}{\mathbf{x}}

\newcommand{\dz}{\dd{\mathbf{z}}}

\newcommand{\dx}{\dd{\mathbf{x}}}

\newcommand{\e}{\operatorname{e}}
\newcommand{\kT}{k_{\mathrm{B}}T}

\newcommand{\cond}{\left.\right\vert}

\begin{document}
\title{%
  Machine Learning of Slow Collective Variables and Enhanced Sampling via Spatial Techniques}

\author{Tuğçe Gökdemir}
\affiliation{%
  Institute of Physics,
  Faculty of Physics, Astronomy and Informatics,
  Nicolaus Copernicus University,
  Grudziadzka 5, 87-100 Toru\'n, Poland
}

\author{Jakub Rydzewski}
\email[Email: ]{jr@fizyka.umk.pl}
\affiliation{%
  Institute of Physics,
  Faculty of Physics, Astronomy and Informatics,
  Nicolaus Copernicus University,
  Grudziadzka 5, 87-100 Toru\'n, Poland
}


\begin{abstract}
Understanding the long-time dynamics of complex physical processes depends on our ability to recognize patterns. To simplify the description of these processes, we often introduce a set of reaction coordinates, customarily referred to as collective variables (CVs). The quality of these CVs heavily impacts our comprehension of the dynamics, often influencing the estimates of thermodynamics and kinetics from atomistic simulations. Consequently, identifying CVs poses a fundamental challenge in chemical physics. Recently, significant progress was made by leveraging the predictive ability of unsupervised machine learning techniques to determine CVs. Many of these techniques require temporal information to learn slow CVs that correspond to the long timescale behavior of the studied process. Here, however, we specifically focus on techniques that can identify CVs corresponding to the slowest transitions between states without needing temporal trajectories as input, instead using the spatial characteristics of the data. We discuss the latest developments in this category of techniques and briefly discuss potential directions for thermodynamics-informed spatial learning of slow CVs.
\end{abstract}

\maketitle

\tableofcontents

\section{Introduction}
Complex systems in chemical physics often exhibit dynamics with multiple temporal scales, characterized by infrequent transitions between long-lived metastable states that occur on timescales orders of magnitude slower than fast molecular motions~\cite{chandler1987introduction,stoltz2010free,dfrenkel:mc}. This significant disparity in timescales is known as timescale separation. Understanding such physical processes depends on our ability to recognize patterns in molecular dynamics (MD) simulations. We typically simplify the dynamics by introducing a set of reaction coordinates, customarily referred to as order parameters or collective variables (CVs)~\cite{rogal2021reaction}, which are meant to describe it on the macroscopic level. Then, we can estimate a free-energy landscape in CV space, which, to a large extent, is responsible for the thermodynamics and kinetics of physical processes~\cite{bolhuis2002transition,abrams2014enhanced,pietrucci_strategies_2017,valsson2016enhancing,yang2019enhanced,bussi2020using,henin2022enhanced,chen2022enhancing,ray2023kinetics}. 

However, the determination of CVs has proved challenging even for simpler systems~\cite{geissler1999kinetic,bolhuis2000reaction,ma2005automatic}. The most interesting properties of complex processes are often hidden in slow dynamics to which fast variables are adiabatically constrained. Therefore, CVs should describe transitions between states that occur when crossing free-energy barriers significantly higher than thermal energy ($\gg\kT$). This picture is based on the transition state theory and Kramers' theory for reaction dynamics, where the reactant and product states are separated by the energy barrier locating the transition state~\cite{hanggi1990reaction}. Processes such as protein folding~\cite{shaw2010atomic,lindorff2011fast}, crystallization~\cite{neha2022collective}, nucleation~\cite{beyerle2023recent}, glass transitions~\cite{berthier2011theoretical,hohenberg2015introduction}, aqueous systems~\cite{10.1063/5.0207567}, catalysis~\cite{piccini2022ab}, or molecular recognition~\cite{baron2013molecular,chipot2014frontiers,rydzewski2017ligand,bernetti2019kinetics} are only a few examples where these characteristics are present and that have frequently profited from such a reduced description.

Due to the rapid development of machine learning (ML) libraries~\cite{scikit-learn,paszke2019pytorch}, using neural networks has become relatively straightforward and readily available for applications in chemical physics and MD. Interestingly, a fundamental challenge in ML is to develop simple and interpretable representations for complex data~\cite{hastie2009elements,fukunaga2013introduction,bengio2013representation,xie2020representation,karniadakis2021physics,meilua2024manifold}, which closely resembles the task of developing CVs for dynamical systems. Consequently, ML methods have been employed to extract meaningful information from simulations due to their ability to recognize statistical patterns~\cite{glielmo2021unsupervised,brunton2021modern}. These techniques can be harnessed to devise algorithms for learning CVs hidden in data to explain the dynamics on the macroscopic scale. A variety of such data-driven methods has been developed at the intersection of statistical physics, MD, and ML. Many of them were recently reviewed~\cite{rohrdanz2013discovering,li2014recent,peters2016reaction,noe2017collective,ceriotti2019unsupervised,wang2020machine,wu2020variational,klus2018data,sidky2020machine,rogal2021reaction,chen2021collective,chen2023chasing,rydzewski2023manifold,mehdi2024enhanced}.

Nonetheless, learning slow CVs remains a challenging task, presenting several difficulties. One key issue is that the quality of CVs is often significantly hampered by the inability to effectively capture longer timescales during standard simulations within a reasonable computing time. This is commonly known as the sampling problem in MD. As such, the construction of training datasets for ML techniques can be problematic as it cannot be known if every state is sufficiently sampled. Additionally, the scarcity of observations between states makes the representation of transition states in reduced space problematic. Enhanced sampling methods can partly alleviate the problem of poor statistics. However, they require correct data reweighting to obtain equilibrium characteristics, often tailored to a particular class of ML algorithms. These problems cause a circular dependency between sampling and learning that poses a major obstacle in developing these techniques. 

In this brief review, we will focus on a specific type of ML methods for building slow CVs. Unlike other reviews that cover techniques using trajectories and their time-delayed versions as input to calculate kinetic quantities, such as correlation functions directly, our priority will be on unsupervised techniques that do not rely on temporal characteristics; instead, they estimate kinetics indirectly by analyzing the thermodynamic properties of MD data. These methods aim to learn the reduced space of CVs by capturing spatial characteristics of simulation data encoded in configuration or reduced space, such as the proximity between samples, density estimates, and weights derived from enhanced sampling simulations. We will explore various techniques, including spectral methods such as diffusion maps and their extensions and recently developed algorithms that leverage deep neural networks to learn slow CVs. Lastly, we will discuss potential avenues for future advancements in this field.

\section{Background}

\subsection{Collective Variables}
\label{sec:cvs}
In statistical mechanics, we consider a system described by the microscopic coordinates $\bx = (x_1, \dots, x_n)$ whose dynamics at temperature $T$ evolves according to a potential energy function $U(\bx)$. This dynamics can be described by the following overdamped Langevin equation:
\begin{equation}
  \dd{\bx} = -\nabla U(\bx)\, \dit{t}+\sqrt{2 \beta^{-1}}\dd{\mathbf{w}},
\label{eq:ger_co}
\end{equation}
where $\beta=1/k_{\mathrm{B}}T$ is the inverse temperature, $k_{\mathrm{B}}$ is the Boltzmann constant, and $\dd{\mathbf{w}}$ is the Brownian motion. The time-evolution of the system results in a canonical equilibrium distribution given by the Boltzmann density:
\begin{equation}
  p(\bx) = \frac{1}{Z}\e^{-\beta U(\bx)},  
\end{equation}
where $Z=\int\dx\e^{-\beta U(\bx)}$ is the partition function of the system~\cite{zwanzig2001nonequilibrium}. We reduce the representation of the system by mapping it into reduced space defined by a set of $d$ functions of the microscopic coordinates, commonly referred to as CVs:
\begin{equation}
  \label{eq:target-mapping}
  \bz = f(\bx) \equiv \pset[\big]{f_k(\bx)}_{k=1}^d,
\end{equation}
where $d \ll n$. The dynamics of the system in reduced space samples a marginal equilibrium density:
\begin{equation}
  p(\bz) = \int\dx p(\bx)\delta\ppar[\big]{\bz - f(\bx)}
\end{equation}
that is defined by weighting each slice through configuration space $\bx$, denoted by the delta function $\delta(\cdot)$, with the Boltzmann factors $p(\bx)\propto\e^{-\beta U(\bx)}$. The marginal probability $p(\bz)$ contains information about the free-energy landscape:
\begin{equation}
  \label{eq:fel}
  F(\bz) = -\frac{1}{\beta}\log p(\bz)
\end{equation}
Even for simple systems, the free-energy landscape contains many stable states that are separated by barriers much larger than thermal energy, leading to significant timescale disparities in the dynamics.

As summarized in a review by Peters~\cite{peters2013reaction}, a general requirement for optimal CVs is to preserve dynamical self-consistency: The dynamics projected onto the free-energy landscape should remain consistent with trajectories sampling configuration space. Taking this apart, we can list more specific characteristics that define optimal CVs:
\begin{enumerate}
  \item[(a)] CVs must accurately recognize metastability; that is, distinguish between long-lived metastable states and identify transition states~\cite{hummer2004transition}. Accurate metastability recognition is often difficult to achieve due to the sampling problem~\cite{bolhuis2002transition,bussi2020using}.

  \item[(b)] CVs need to model reduced dynamics as primarily corresponding to transitions on longer timescales, with the dynamics of fast variables being negligible~\cite{coifman2008diffusion}. Slow and fast variables should be unmixed in such a way that they induce a significant separation of timescales. Moreover, CVs should not be degenerate; each should describe a different slow mode.
  
  \item[(c)] CVs need to, preferably, be Markovian for the ability to describe slow dynamics as evolution in the free-energy landscape with configuration-dependent diffusion coefficients~\cite{berezhkovskii2005one,berezhkovskii2011time,berezhkovskii2018single}.

  \item[(d)] CVs must be applicable in CV-based enhanced sampling methods (i.e., smooth and differentiable), such as umbrella sampling~\cite{torrie1977nonphysical,mezei1987adaptive,Maragakis-JPCB-2009,Kastner2011umbreallsampling}, metadynamics~\cite{laio2002escaping,barducci2008well,Invernizzi2020opus,invernizzi2020unified}, or variationally enhanced sampling~\cite{valsson2014variational,yang2018refining,bonati2019neural}, to improve sampling in MD and drive it toward long-timescale processes.

\end{enumerate}

\subsection{Timescale Separation}
\label{sec:ts}
To illustrate the problem of timescale separation, let us focus on the spectral theory of dynamical systems and reversible Markov processes~\cite{roux2022transition}. Consider the forward Fokker--Planck equation for time-propagation of a probability distribution $p(\bx, t)$: $\partial p/\partial t=-L p$, where $L$ is the generator of a Markov process. Through a change of variables, the Fokker--Planck equation can be solved~\cite{shuler1959relaxation}. The solution to this equation can be written in closed form as an eigenfunction expansion~\cite{risken1996fokker}:
\begin{equation}
  \label{eq:prob}
  p(\bx, t) = \varphi_0(\bx) + \sum_{k=1}^{\infty} a_k \e^{-\mu_k t} \varphi_k(\bx),
\end{equation}
where $a_k$ are coefficients and $t$ is a time variable. For time $t\to\infty$, the solution of the forward equation converges to the equilibrium Boltzmann distribution $p(\bx)$. 

Under general conditions, the generator of the diffusion process $L$ has a discrete eigenspectrum of eigenvalues ${\mu_k}$, and the corresponding eigenfunctions $\varphi_k(\bx)$. The zeroth eigenfunction is the equilibrium density $\varphi_0(\bx) \propto \e^{-\beta U(\bx)}$ with the eigenvalue $\mu_0=0$. The eigenvalues are non-negative and sorted in increasing order:
\begin{equation}
  \mu_0 = 0 < \mu_1 < \mu_2 < \dots < \mu_{\infty}.
\end{equation}

\begin{figure}
  \includegraphics{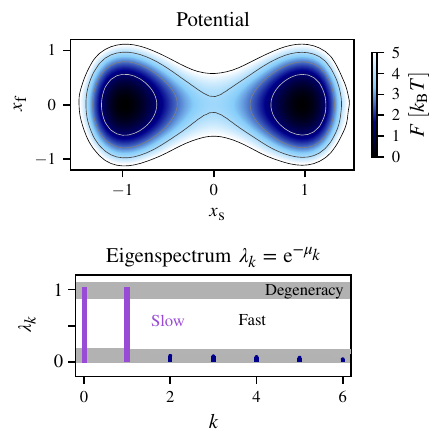}
  \caption{Model potential with two metastable states whose long-time behavior can effectively be described by the slow variable $x_{\mathrm{s}}$, with the fast variable $x_{\mathrm{f}}$ responsible only for fluctuations within the states. The corresponding eigenspectrum of the diffusion generator $\lambda_k=\e^{-\mu_k}$ shows timescale separation, which is indicated by the spectral gap $\lambda_{k-1} - \lambda_k$, where $k=2$ is the number of states.}
  \label{fig:model}
\end{figure}

The dominant eigenvalues of the Markov generator decay exponentially and are linked to the slowest relaxation timescales in the system. Each eigenvalue can be matched with an effective timescale $t_k=1/{\mu_k}$. In systems with timescale separation, only a few slow processes related to rare transitions between metastable states remain. As a result, the eigenspectrum of $L$ has a spectral gap, i.e., the largest difference between eigenvalues $\mu_{k+1}$ and $\mu_{k}$. This implies that the eigenvalues much lower than $\mu_{k+1}$ can be neglected as they correspond with rapid fluctuations within states and decay much faster, leading effectively to $k$ slow processes (\rfig{fig:model}).

The spectral properties of reversible Markov processes can be related to the concept of metastability~\cite{bovier2002metastability}. Although this relation can be understood intuitively, Gaveau and Schulman~\cite{gaveau1996master,gaveau1998theory}, drawing on the extensive work of Davies~\cite{davies1983spectral,davies1982metastable-1,davies1982metastable-2}, developed a spectral definition of metastability. They formally showed that dominant and nearly degenerate eigenvalues are related to metastable timescales. This concept relies on the presence of the spectral gap. If an eigenvalue is nearly degenerate, the equilibrium distribution separates into metastable states with infrequent transitions between them. Conversely, eigenvalue degeneracy exists if the equilibrium density breaks into metastable states separated by a free-energy barrier much larger than thermal energy. The eigenfunctions related to the dominant eigenvalues are linked to distributions that remain stable longer than transient processes. Furthermore, sign changes in these eigenfunctions indicate transitions between metastable states. The theory is summarized in a monograph by Bovier and Den Hollander~\cite{bovier2016metastability}.

\subsection{Enhanced Sampling}
\label{sec:es}
Acquiring an informative training dataset from unbiased MD trajectories is a crucial challenge. These trajectories need to spontaneously and repeatedly cross over all significant free-energy barriers in the system. However, the metastability leads to kinetic entrapment in a single state, making transitions between metastable states rare. To alleviate this issue, enhanced sampling methods can be used to improve sampling efficiency~\cite{abrams2014enhanced,pietrucci_strategies_2017,valsson2016enhancing,yang2019enhanced,bussi2020using,henin2022enhanced}.

Enhanced sampling methods that require CVs to improve sampling are based on employing a nonphysical bias potential. To such methods, we can include umbrella sampling introduced by Torrie and Valeau~\cite{torrie1977nonphysical}, adiabatic biasing force~\cite{Darve-JCP-2001}, adiabatic free-energy dynamics~\cite{rosso2002use}, metadynamics proposed by Laio and Parrinello~\cite{laio2002escaping} and improved to the well-tempered variant by Barducci et al.~\cite{barducci2008well}, mean-force dynamics~\cite{morishita2012free}, or variationally enhanced sampling~\cite{valsson2014variational}. Biasing the system can cause the probability distribution of collective variables (CVs) to significantly deviate from equilibrium, resulting in sampling according to a biased distribution:
\begin{equation}
  \label{eq:pbias}
  p_V(\bz, t) \propto \e^{-\beta \pbkt*{F(\bz) + V(\bz,t)}},
\end{equation}
where $V(\bz,t)$ is a time-dependent bias potential. To calculate equilibrium properties, such as free-energy landscapes, the bias must be reverted during postprocessing. This is customarily done by reweighting, where each sample is associated with a statistical weight to counter the effect of biasing. In general, the weights are given by the likelihood ratio between the equilibrium and the biased probability distributions (\req{eq:pbias}):
\begin{equation}
  \label{eq:w}
    w(\bz,t) = \frac{p(\bz)}{p_V(\bz,t)}.
\end{equation}
For methods using a quasi-stationary bias potential~\cite{torrie1977nonphysical,valsson2014variational,giberti2020iterative} (e.g., umbrella sampling), or when the simulation is converged and the bias does not change significantly, the weights are given as: 
\begin{equation}
  w(\bz) \propto \frac{\e^{-\beta F(\bz)}}{\e^{-\beta \pbkt*{F(\bz) + V(\bz)}}} = \e^{\beta V(\bz)}.
\end{equation}
In contrast, in metadynamics~\cite{barducci2008well}, the bias potential changes over time and requires accounting for a time-dependent offset~\cite{tiwary2015time}. Thus, the functional form of weights may vary depending on an enhanced sampling method and a reweighting algorithm~\cite{bonomi2009reconstructing,Sch_fer_2020,giberti2020iterative,linker2020connecting}. A summary of such methods was recently published by Kamenik et al~\cite{kamenik2021enhanced}.

To efficiently sample and drive complex physical processes, high-quality CVs are required for biasing. However, learning CVs demands using exhaustively sampled data. This problem creates a challenging circular dependency, which is referred to as the ``chicken-and-egg'' problem~\cite{wang2020machine}. Advances in the determination of CVs help address this problem and contribute to the development and implementation of enhanced sampling methods.

\section{Spatial Learning}
Due to recent extensive advancements in data-driven temporal methods~\cite{hernandez2013identification,wehmeyer2018time,mccarty2017variational,yang2018refining,bonati2021deep,mardt2018vampnets,chen2019nonlinear}, there are numerous reviews summarizing this topic~\cite{wang2020machine,wu2020variational,chen2021collective,chen2023chasing,rydzewski2023manifold,mehdi2024enhanced}. In this work, however, we consider techniques that are ``spatial,'' i.e., algorithms for learning slow CVs that do not need to exploit temporal information in MD simulations. We can describe spatial techniques as those that rely on pairwise relations between samples in the dataset (usually through a distance metric) instead of counting transitions within a specified lag time. The development of such techniques can be traced back to the work of Shi and Malik~\cite{shi2000normalized} on image segmentation and the classic Laplacian eigenmaps introduced by Belkin and Niyogi~\cite{belkin2001laplacian,belkin2003laplacian,belkin2004semi,belkin2008towards}; and is closely related to graph spectral theory~\cite{chung1997spectral} based on graphs, kernels, and random walks~\cite{scholkopf1998nonlinear,szummer2001partially,10.5555/645531.655996}. 

The primary difference between spatial and temporal techniques lies in how kinetics is estimated. Spatial techniques estimate kinetics indirectly by analyzing the thermodynamic characteristics of MD data, such as equilibrium probabilities, in contrast to temporal techniques. Additionally, in spatial techniques, we assume that MD data closely approximates overdamped Langevin dynamics (see \rsct{sec:cvs}). For these reasons, we can refer to these methods as thermodynamics-informed learning.

\subsection{Anisotropic Kernels}
The core of most spatial learning methods involves establishing similarity between samples, typically through a distance metric and a kernel~\cite{izenman2012introduction}. For example, Laplacian eigenmaps construct a Gaussian kernel to model relations between $N$ samples in a dataset $X=\pset{\bx_k}_{k=1}^N$~\cite{belkin2001laplacian,belkin2003laplacian,belkin2004semi,belkin2008towards}:
\begin{equation}
  \label{eq:gaussian-kernel}
  G_\varepsilon(\bx_k,\bx_l) = \exp\ppar*{-\pnrm{\bx_k - \bx_l}^2/\varepsilon^2},
\end{equation}
where $\varepsilon>0$ is a scale parameter. This kernel is then used to define a Laplacian matrix and parametrize reduced space using its eigenvectors. However, methods that use a Gaussian kernel, such as Laplacian eigenmaps, cannot be used to compute slow CVs as their construction implicitly assumes that data is distributed uniformly. As the equilibrium density is often far from uniform, Laplacian eigenmaps have not seen many applications for analyzing trajectories. However, they are often used as a baseline for developing more advanced techniques.

Based on Laplacian eigenmaps, Coifman et al.~\cite{coifman2005geometric} developed the diffusion map algorithm that is especially suited for learning the reduced space of slow CVs. Diffusion maps use a density-preserving kernel for data sampled from any underlying probability distribution. For this, an anisotropic kernel is constructed on the dataset $X$~\cite{nadler2006diffusion}:
\begin{equation}
  \label{eq:kernel}
  K(\bx_k,\bx_l) = \frac{G_{\varepsilon}(\bx_k,\bx_l)}{{\rho^\alpha(\bx_k)\rho^\alpha(\bx_l)}},
\end{equation}
where $\varepsilon$ is a scale parameter, $\rho(\bx_k)=\sum_l G_\varepsilon(\bx_k,\bx_l)$ is a density estimate that allows us to include information about non-uniformly sampled data into the kernel, and $\alpha\in[0, 1]$ is the anisotropic diffusion constant. Next, a Markov transition matrix is constructed by row-normalizing $K$:
\begin{equation}
\label{eq:markov}
  M(\bx_k,\bx_l) = \frac{K(\bx_k,\bx_l)}{\sum_i K(\bx_k,\bx_i)}
\end{equation}
to build a discrete Markov chain on the data:
\begin{equation}
  m_{kl} = \mathrm{Pr}\ppar{\bx_{i+1}=\bx_l \cond \bx_i=\bx_k}
\end{equation}
that expresses a transition probability between $\bx_k$ and $\bx_l$. Note that this construction does not depend on the physical time. The local scale parameter $\varepsilon$ plays an important role in determining the quality of slow CVs, as it defines the scale within which the relation between two samples contributes to the Markov transition matrix. 

Depending on the anisotropic diffusion constant $\alpha$, several kernel normalizations are available, which can change the long-time convergence of the Markov chain to a particular operator. This group of constructions is known as anisotropic diffusion maps~\cite{coifman2005geometric,NIPS2005_2a0f97f8,nadler2006fundamental,nadler2006diffusion,coifman2006diffusion}. For example, with $\alpha=1/2$, the Markov chain approaches the time asymptotics of the system by describing the dynamics by the Fokker--Planck anisotropic diffusion with the potential $U(\bx)$. As such, this normalization is commonly used to extract information from MD trajectories. Two other frequently considered values are $\alpha=0$ and 1. The former results in the classical normalized graph Laplacian, while the latter yields the Laplace-Beltrami operator with a uniform probability density~\cite{coifman2005geometric,NIPS2005_2a0f97f8,nadler2006fundamental,nadler2006diffusion,coifman2006diffusion}.

The advancements of the diffusion map algorithm and anisotropic Markovian kernels often involve using a kernel that captures more aspects of the data. For instance, self-tuning local kernels were introduced by Zelnik-Manor and Perona~\cite{zelnik2004self}. Following this works by Rohrdanz et al.~\cite{rohrdanz2011determination} and Zhang et al.~\cite{zheng2013rapid,zheng2013molecular} demonstrated that estimating the scale parameter as configuration-dependent $\varepsilon(\bx_k) \varepsilon(\bx_l)$, where each term can be calculated as the distance between $\bx$ and its $n$-th nearest neighbor, improves the overall quality of slow CVs~\cite{rohrdanz2013discovering}. A more general method for computing the local scale parameters was later proposed by Berry et al.~\cite{berry2015nonparametric,berry2016variable}

In works by Dsilva et al.~\cite{dsilva2013nonlinear,dsilva2015data} and Singer et al. \cite{singer2009detecting}, it was proposed to use a heterogeneous Gaussian kernel to improve properties of the resulting CVs. Instead of using the Euclidean distance, this kernel introduces a Mahalanobis-like distance, which incorporates a covariance matrix. The implication of this is that the Mahalanobis kernel, by including the correlations in the dataset, can be used to remove the effect of observing the underlying space through a complex nonlinear function~\cite{singer2009detecting,dsilva2013nonlinear,dsilva2015data}:
\begin{equation}
  G_{\Sigma}(\bx_k,\bx_l) = \exp\ppar*{-d^2_{{\Sigma}}(\bx_k,\bx_l)/\varepsilon^2},
\end{equation}
where the squared Mahalanobis distance is:
\begin{equation}
  d^2_{{\Sigma}}(\bx_k,\bx_l)=\ppar*{\bx_k - \bx_l}^\top \ppar*{{\Sigma}_k + {\Sigma}_l}^\dagger \ppar*{\bx_k - \bx_l}.
\end{equation}
The local covariance matrix ${\Sigma}_k$ can be estimated as a sample covariance matrix at configuration $\bx_k$ in its immediate neighborhood~\cite{singer2008non,singer2009detecting,dsilva2013nonlinear} and $\dagger$ denotes a pseudo-inverse (as $\Sigma$ can be rank-deficient). 

Subsequently, Berry and Sauer~\cite{berry2016local} developed a generalization of diffusion maps to local kernels by introducing diffusion and drift terms in the distance metric, which should be additionally computed from the data~\cite{mugnai2015extracting,domingues2024estimating}. It was shown by Berry et al.~\cite{berry2013time} that it is possible to improve anisotropic kernels by including Taken's delay coordinates in datasets, especially when observations are scarce. Diffusion map was also embedded in a framework for coarse-graining and clustering~\cite{lafon2006diffusion}.

\subsection{Reweighted Transitions}
\label{sec:rew}
The concept of reweighting transition probabilities is crucial when using enhanced sampling algorithms to build the Markov transition matrix and, thus, CVs. A Markov chain constructed from biased data does not converge to the equilibrium density given by the Boltzmann distribution~\cite{rydzewski2021multiscale,rydzewski2022reweighted}. This bias affects the Markov chain and leads to incorrect density and geometric relations between samples, which can result in reduced space that does not accurately represent the characteristics of the data. Reweighting pairwise probabilities counters the bias from the Markov matrix, yielding the unbiased Markov. While learning biased CVs can still be used to analyze, speed up, and drive the sampling of rare events~\cite{zheng2013rapid,rydzewski2016machine,chiavazzo2017intrinsic}, the necessity of a reweighting algorithm becomes apparent when we seek to restore the equilibrium properties of the system and compute slow CVs. 

The initial approach to learning unbiased CVs from enhanced sampling simulations with the diffusion map algorithm was proposed by Ferguson et al.~\cite{ferguson2011integrating}, in which each configuration is weighted based on its importance in umbrella sampling simulations. A symmetric weighted Gaussian kernel was used by Zhang et al.~\cite{zheng2013molecular} to learn CVs from multiple metadynamics simulations. Building on the local kernels introduced by Berry and Sauer~\cite{berry2016local}, Banisch et al. and Trstanova et al. devised a general approach to reweighting transition probabilities based on target measure reweighting~\cite{banisch2020diffusion,trstanova2020local}. This approach was later employed in works by Evans et al., where diffusion map with the Mahalanobis distance is constructed in $\bz$ space~\cite{banisch2020diffusion,trstanova2020local}.

Zhang and Chen~\cite{zhang2018unfolding} derived an alternative technique for reweighting, which Rydzewski et al.~\cite{rydzewski2022reweighted} later generalized to multiple algorithms employing Markov transition kernels. They demonstrated that the anisotropic diffusion kernel as can be unbiased as:
\begin{equation}
  \label{eq:rew}
  K(\bx_k,\bx_l) = r_{kl} \frac{G_{\varepsilon}(\bx_k,\bx_l)}{{\rho^\alpha(\bx_k)\rho^\alpha(\bx_l)}},
\end{equation}
where a transition reweighting factor $r_{kl}=w_k w_l$ incorporates importance weights from enhanced sampling simulations and $\rho$ are reweighted density estimates:
\begin{equation}
  \rho(\bx_k) = \sum_m w_m G_\varepsilon(\bx_k, \bx_m).
\end{equation}
A detailed derivation with possible approximations is given by Rydzewski et al.~\cite{rydzewski2022reweighted} As explained in \rsct{sec:es}, the form of weight depends on the employed enhanced sampling and reweighting techniques~\cite{bonomi2009reconstructing,tiwary2015time,Sch_fer_2020,giberti2020iterative,kamenik2021enhanced}. 

Several approximate transition reweighting factors can be obtained depending on the scaling of the long-time asymptotics of the kernel with the constant $\alpha$~\cite{rydzewski2022reweighted}. This kind of transition reweighting can be used for diffusion maps~\cite{rydzewski2022reweighted} and deep learning~\cite{zhang2018unfolding,rydzewski2021multiscale,rydzewski2022reweighted}. We refer to the review by Rydzewski et al.~\cite{rydzewski2023manifold} for a detailed discussion.

This idea was recently explored by Rydzewski~\cite{rydzewski2023selecting}, who demonstrated that this form of transition reweighting in diffusion maps can be employed as a feature selection pipeline for further dimensionality reduction. This is done by leveraging the idea that the partial selection of variables should have a similar eigenspectrum to configuration space. This extension can provide an interpretable and explainable description by selecting physically important CVs for the given process~\cite{rydzewski2023selecting}.

For a more general approach to dynamical transition reweighting, not limited to unbiasing transition probabilities in spatial techniques, see reviews by Chen and Chipot~\cite{chen2023chasing}, which discusses many reweighting methods for temporal techniques and Keller and Bolhuis~\cite{keller2024dynamical}, where reweighting is examined from the perspective of Markov state models.

\begin{figure*}[t]
  \includegraphics{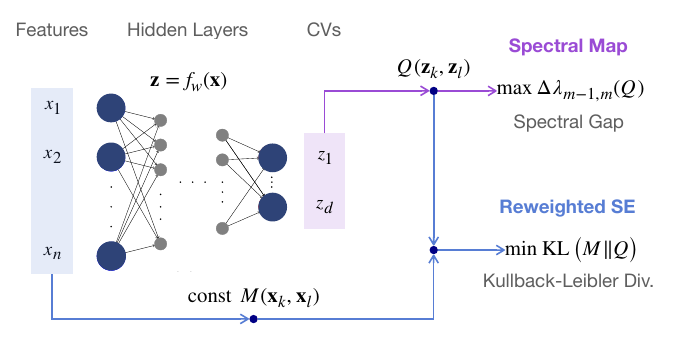}
  \caption{Learning CVs with spatial techniques. Diagram of a neural network showing the difference between reweighted stochastic embedding (RSE) and spectral map. RSE estimates transition matrices $M(\bx_k,\bx_l)$ and $Q(\bz_k,\bz_l)$ in both $\bx$ and $\bz$ spaces, respectively (as $\bx$ space can consist of variables different than the microscopic coordinates, we denote it as features). Then, it uses the Kullback--Leibler (KL) divergence as a loss function to minimize differences between pairs of transition probabilities in $\bx$ and $\bz$ spaces. In contrast, spectral map constructs a transition matrix only in $\bz$ space. Next, it performs an eigendecomposition of $Q$ to calculate the spectral gap between neighboring eigenvalues ($\Delta\lambda_{m-1,m}$ where $m$ is the number of states in $\bz$ space) and maximizes it to improve timescale separation between slow and fast variables.}
  \label{fig:diagram}
\end{figure*}

\subsection{Eigendecomposition}
In learning algorithms that use a few eigenvectors of the Markov transition matrix to span $\bz$ space, a mapping into $\bz$ space is obtained by solving an eigendecomposition problem:
\begin{equation}
  M\psi_k=\lambda_k\psi_k,
\end{equation}
where $\lambda_k$ and $\psi_k$ are the eigenvalues and corresponding eigenvectors of the Markov transition matrix $M$, respectively. As explained in \rsct{sec:ts}, as a result of the existence of the spectral gap between neighboring eigenvalues $\lambda_k$, slow CVs can be approximated by the following truncated mapping:
\begin{equation}
  \bz = \ppar[\big]{\lambda_1\psi_1, \dots, \lambda_d\psi_d},
\end{equation}
where $d$ is the dimension of $\bz$ space. The eigenvalues of the Markov transition matrix $M$ are (sorted in non-ascending order):
\begin{equation}
  \label{eq:eigenvalues}
  \lambda_0 = 1 > \lambda_1 \dots \ge \lambda_N,  
\end{equation}
where the eigenvalue $\lambda_0$ corresponds to the equilibrium distribution of the Markov chain given by the eigenvector $\psi_0$. The dominant eigenvalues related to the slowest relaxation timescales in the system~\cite{bovier2002metastability} and the fast eigenvalues have a negligible contribution to slow CVs. In the case of anisotropic diffusion maps, the eigenvalues ${\lambda_k}$ are related to the eigenvalues of the Fokker--Planck generator ${\mu_k}$ by the relation $\lambda_k = \e^{-\mu_k}$. 

Several techniques use the mapping provided by diffusion maps as an initial guess to improve slow CVs iteratively. For instance, the eigenvectors of the Markov transition matrix $M$ can serve as a basis to approximate kinetic quantities such as the transfer operator. This approach was exploited in works by Boninsegna et al.~\cite{boninsegna2015investigating}, Noe and Clementi~\cite{noe2015kinetic,noe2016commute}, and more recently by Thiede et al. using a Galerkin approximation~\cite{thiede2019galerkin}.

Algorithms that use an eigendecomposition to construct $z$ space require an out-of-sample extension to map samples outside of the dataset. Specifically, for diffusion maps the Nystr\"om extension~\cite{fowlkes2004spectral,long2019landmark}, Laplacian pyramids~\cite{dsilva2013nonlinear}, and geometric harmonics~\cite{coifman2006geometric,evangelou2023double} interpolators were used. A detailed analysis of out-of-sample algorithms was published by Bengio et al~\cite{bengio2003out}.

\subsection{Reweighted Stochastic Embedding}
Reweighted stochastic embedding (RSE) is a recent framework for the parametric learning of slow CVs, introduced by Rydzewski et al.~\cite{rydzewski2022reweighted}, which employs algorithms to construct unbiased Markov transition matrices with transition reweighting (\rsct{sec:rew}), allowing for the estimation of CVs from data collected in enhanced sampling simulations. Building on the work of van Maaten, Hinton, and Roweis~\cite{hinton2002stochastic,maaten2008visualizing,maaten2009learning,van2009dimensionality}, RSE optimizes a loss function to learn the mapping to reduced space. Specifically, it projects samples into $\bz$ space using a neural network, while ensuring that the statistical distance between transition matrices estimated in both configuration space and $\bz$ space is minimized (\rfig{fig:diagram}).

The first technique of this framework is stochastic kinetic embedding (StKE), which was proposed by Zhang and Chen~\cite{zhang2018unfolding}. StKE combines modeling a slow manifold with parametric dimensionality reduction, building upon the reweighted anisotropic diffusion kernel. As such, StKE can learn slow CVs from biased data sampled in enhanced sampling simulations. In addition, it uses an iterative procedure incorporating temperate-accelerated MD~\cite{voter2000temperature,maragliano2006temperature} to alleviate the circular dependency~\cite{zhang2018unfolding}, allowing the use of this algorithm on the fly in atomistic simulations~\cite{zhang2018unfolding,chen2021collective,rydzewski2022reweighted,liu2024unbiasing}. Subsequently, Rydzewski and Valsson introduced a RSE technique called multiscale reweighted stochastic embedding (MRSE)~\cite{rydzewski2021multiscale} that shares similarities with StKE~\cite{rydzewski2022reweighted,rydzewski2023manifold}. The main difference between StKE and MRSE is, as in many methods discussed in this review, boils down to using other kernels to estimate Markov transition matrices. In MRSE, the process of constructing unbiased transition probabilities from enhanced sampling simulations involves adaptively estimating a kernel in $\bx$ space based on information theory principles. In contrast, StKE employs a fixed anisotropic diffusion kernel, as used in diffusion maps (see \rsct{sec:rew}). This topic is discussed in detail in the review by Rydzewski et al~\cite{rydzewski2023manifold}.

RSE employs building transition matrices in both $\bx$ and $\bz$ spaces (\rfig{fig:diagram}). As with many neutral network-based techniques for learning CVs, $\bx$ space can comprise variables other than the microscopic coordinates, which are called features or descriptors. The transition matrix $M$ constructed in $\bx$ space remains constant throughout learning, while the matrix $Q$ in $\bz$ space is adjusted depending on a neural network that performs dimensionality reduction, i.e., $f_w(\bx)=\bz$. Most generally, in RSE, a weighted Gaussian mixture is used to construct the transition matrix in $\bx$ space~\cite{rydzewski2021multiscale}:
\begin{equation}
  M(\bx_k,\bx_l) \propto \sum_{\varepsilon} \frac{w(\bx_l)}{{\rho^\alpha(\bx_l)}} G_{\varepsilon}(\bx_k,\bx_l)  
\end{equation}
where the sum goes over scale parameters. In $\bz$ space, the transition matrix can be given, for example, by a $t$-distribution kernel~\cite{rydzewski2021multiscale}:
\begin{equation}
  Q(\bz_k,\bz_l) \propto \ppar*{1 + \ppar{\bz_k - \bz_l}^2}^{-1}.
\end{equation}
RSE minimizes the Kullback--Leibler divergence~\cite{kullback1951information} to learn CVs, which can be interpreted as a ``distance'' between probability distributions. Thus, after the training converges, the transition probabilities in both spaces should be approximately equal. More details about these algorithms can be found in Refs.~\citenum{rydzewski2022reweighted,rydzewski2021multiscale}. 

\begin{figure*}[t]
  \includegraphics{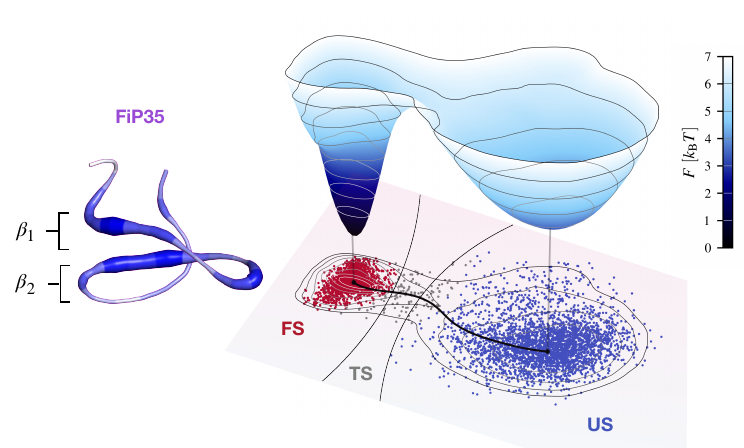}
  \caption{Free energy landscape of the FiP35 protein constructed from slow CVs learned with spectral map (right). The slow CVs discriminate between the folded state (FS) and the unfolded state (US), which are separated by the transition state (TS) near the free energy barrier. The most important physical interactions in the FiP35 consisting of two $\beta$ sheets identified by spectral map are shown in blue (left). [Figure based on Rydzewski, ``Spectral Map for Slow Collective Variables, Markovian Dynamics, and Transition State Ensembles,'' J. Chem. Theory Comput. (2024). Copyright 2024 Author, licensed under Creative Commons Attribution 4.0.]}
  \label{fig:jctc}
\end{figure*}

\subsection{Spectral Map}
The first technique devised to maximize timescale separation to find CVs in complex systems was proposed by Tiwary and Berne~\cite{tiwary2016spectral} and subsequently expanded~\cite{tiwary2017predicting,10.1063/5.0030931,tsai2021sgoop,zou2021toward}. Their technique, called spectral gap optimization of order parameters (SGOOP), is based on constructing a transition matrix using the principles of the maximum caliber framework~\cite{ghosh2020maximum}. As opposed to the techniques reviewed here, SGOOP explicitly uses time information to construct slow CVs.

A recent unsupervised statistical learning technique for learning slow CVs that is also based on maximizing timescale separation is spectral map, developed by Rydzewski~\cite{rydzewski2023spectral}. It is modeled using an overdamped Langevin diffusion in $\bz$ space~\cite{rydzewski2024tse}. Spectral map proceeds by mapping the dynamics into $\bz$ space using a neural network and constructing a Markov transition matrix by row-normalizing the anisotropic diffusion kernel (\req{eq:kernel}), however, from data in $\bz$ space:
\begin{equation}
  \label{eq:reducedmarkov}
  Q(\bz_k,\bz_l) = \frac{K(\bz_k,\bz_l)}{\sum_m K(\bz_k,\bz_m)},
 \end{equation}
where $\bz=f_w(\bx)$ given by the neural network with learnable parameters $w$. The transition matrix is then spectrally decomposed to estimate the degree of timescale separation from the spectral gap in its eigenspectrum. The spectral gap is used as a score function for the neural network and maximized during learning:
\begin{equation}
  \label{eq:sg}
 \Delta\lambda_{m-1,m}(Q) = \lambda_{m-1} - \lambda_m,
\end{equation}
where $\lambda_k$ are the eigenvalues of $Q$ sorted in decreasing order and $m$ is the number of metastable states in the system. As the spectral gap is maximized, $\bz$ space is adjusted accordingly by improving the parameters of the neural network. At the end, $\bz$ space corresponds to slow CVs. A simplified diagram spectral map and comparison to RSE is given in \rfig{fig:diagram}.
  
Rydzewski and Gökdemir~\cite{rydzewski2024learning} showed that maximizing timescale separation in spectral map results in the dynamics in $\bz$ space with Markovian characteristics. In their work, it was shown that it is possible to construct a high-quality Markov state model based on the learned slow CVs and estimate kinetics accurately. In another work, Rydzewski showed that the framework can be easily extended for learning the transition state ensembles~\cite{rydzewski2024tse} (\rfig{fig:jctc}), which is demanding for complex systems due to the scarcity of transitions between states~\cite{hummer2015optimal,martini2017variational,ray2023deep}. 

Using the transition state ensemble to count transition paths~\cite{best2005reaction}, Rydzewski~\cite{rydzewski2024tse} showed that a slow CV learned by spectral map closely approaches a Markovian limit for overdamped Langevin dynamics~\cite{berezhkovskii2018single}. Moreover, it was illustrated that spectral maps can estimate the quality of the reduced representations with commonly used physical descriptors by comparing their spectral gaps. It was demonstrated that spectral map can be used to construct interpretable reaction coordinates for protein folding with a linear model instead of a deep neural network, and they are slower than the fraction of native contacts or end-to-end distance~\cite{rydzewski2024tse}.

\subsection{Enhanced Sampling via Neural Networks}
After the training procedure, a neural network representing CVs can be used for the purposes of ehnanced sampling. To bias such a neural network, a biasing force must be applied in CV space. This force is equal to the negative derivative of the biasing potential with respect to the CVs, which can be estimated using the chain rule:
\begin{equation}
  F(\bx) = - \frac{\dd{V}(\bz)}{\dz} \nabla_{\bx} f(\bx),
\end{equation}
where the second term on the right-hand side is automatically computed through backpropagation. By accumulating the biasing potential in CV space, the neural network can be used to push the system out of local minima. Such CVs, in the form of a neural network, can be integrated into several advanced MD simulation codes, such as PLUMED~\cite{plumed,plumed-nest,plumed-tutorials,bonati2023unified,trizio2024advanced}.

We want to underline that there might be more requirements for slow CVs represented by a neural network (not only limited to spatial techniques). An often overlooked issue that can harm the convergence of biasing methods is that the neural network may learn a function where $\nabla_{\bx} f(\bx)\approx 0$ in basins. According to Darve et al.~\cite{Darve-JCP-2001}, biasing a CV can be imagined in terms of an object that is pulled or pushed, where the CV has a ``mass'' attached to it that is related to the inverse of $\nabla_{\bx} f(\bx)$. Consequently, applying bias to neural networks with $\nabla_{\bx} f(\bx)\approx 0$ in energy minima might be inefficient due to the large mass and lead to numerical stability issues in MD simulations.

\section{Summary}
Overall, we think further research in spatial techniques will follow by carefully incorporating more thermodynamical information into ML. Due to rapid developments in physics-informed algorithms, we expect that the primary effort will be directed toward solving the problem of constructing interpretable and explainable reaction coordinates for complex systems in chemical physics. 

To address this issue, we can examine the theoretical progress in modeling slow dynamics in the context of timescale separation in CV space~\cite{maragliano2006string,lange2006collective,legoll2010effective,zhang2016effective}. By investigating slow dynamics using overdamped Langevin dynamics in a free-energy landscape with configuration-dependent diffusion coefficients, we can propose a Markovian interpretation of the physical process. The diffusion tensors, which depend on the coordinates, are important for reduced dynamics and can impact the free-energy landscape by altering transition states and barrier height~\cite{rhee2005one,krivov2008diffusive,best2010coordinate,dietschreit2022free,nakamura2024derivation}. To account for this in spatial techniques, we can incorporate information about them in anisotropic kernels and transition matrices. Additionally, analyzing spatial techniques from the perspective of spectral graph theory~\cite{chung1997spectral}, especially the long-term behavior of Markov chains, the asymptotic rate of convergence to equilibrium, and mixing rates~\cite{boyd2004fastest,boyd2006convex}, can lead to improvements.

For spatial techniques to learn from enhanced sampling simulations, slow CVs should be computed using unbiased Markov matrices through a transition reweighting algorithm, such as those presented in the review, to capture equilibrium information accurately. It would be interesting to explore the relationship between the reweighting of Markov transition matrices and dynamical path reweighting, for example, based on the Girsanov theorem~\cite{donati2017girsanov,kieninger2020dynamical,donati2022review}. To improve sampling and drive it toward complex physical processes, spatial techniques can be extended with a general iterative learning-sampling framework where rounds of learning slow CVs (including reweighting) are followed by biasing using an enhanced sampling technique. Such iterative approaches have already been implemented using ML to learn from MD simulations~\cite{mccarty2017variational,chen2018molecular,ribeiro2018reweighted,zhang2018unfolding,brotzakis2018enhanced,bonati2021deep,mehdi2022accelerating,bonati2023unified,shmilovich2023girsanov}.

Finally, we underline that apart from spatial techniques, many others can be used to study complex processes in the fields of chemical physics and MD~\cite{molgedey1994separation,wiskott2002slow,ceriotti2011simplifying,naritomi2011slow,ceriotti2013demonstrating,perez2016hierarchical,mcgibbon2017identification,sidky2019high,li2019computing,chen2019capabilities,tribello2019using,wang2019past,zhang2019targeted,bonati2020data,morishita2021time,wang2021state,belkacemi2021chasing,novelli2022characterizing,ketkaew2022deepcv,sun2022multitask,song2022slow,chen2023discovering,jung2023machine}. A detailed introduction to such ML methods can be found in recent reviews~\cite{rohrdanz2013discovering,li2014recent,peters2016reaction,noe2017collective,ceriotti2019unsupervised,wang2020machine,wu2020variational,klus2018data,sidky2020machine,chen2021collective,chen2023chasing,rydzewski2023manifold,mehdi2024enhanced}. We think, however, that recent results in spatial techniques for learning slow CVs are worthy of further development and could provide a valuable alternative to temporal techniques for understanding the physics of complex systems.

\tocless{\section*{Acknowledgements}}
The research was supported by the National Science Center in Poland (Sonata 2021/43/D/ST4/00920, ``Statistical Learning of Slow Collective Variables from Atomistic Simulations''). J. R. acknowledges funding the Ministry of Science and Higher Education in Poland. The authors acknowledge insightful feedback from Haochuan Chen, Luke Evans, Luigi Bonati, and Omar Valsson.

\tocless{\section*{Author Declarations}}

\tocless{\subsection*{Conflict of Interest}}

The authors have no conflicts to disclose.\\[0.1cm]

\tocless{\subsection*{Author Contributions}}

\noindent{\bf Tuğçe Gökdemir}: Conceptualization (equal); Supervision (equal);
Writing - original draft (equal); Writing - review \& editing (equal). {\bf Jakub Rydzewski}: Conceptualization (equal); Supervision (equal); Writing - original draft (equal); Writing - review \& editing (equal).\\[0.1cm]

\tocless{\section*{Data Availability}}

Data sharing is not applicable to this article as no new data were created or analyzed in this study.\\[0.1cm]

\tocless{\section*{References}}

\bibliography{bib/main.bib}

\begin{thebibliography}{212}%
\makeatletter
\providecommand \@ifxundefined [1]{%
 \@ifx{#1\undefined}
}%
\providecommand \@ifnum [1]{%
 \ifnum #1\expandafter \@firstoftwo
 \else \expandafter \@secondoftwo
 \fi
}%
\providecommand \@ifx [1]{%
 \ifx #1\expandafter \@firstoftwo
 \else \expandafter \@secondoftwo
 \fi
}%
\providecommand \natexlab [1]{#1}%
\providecommand \enquote  [1]{``#1''}%
\providecommand \bibnamefont  [1]{#1}%
\providecommand \bibfnamefont [1]{#1}%
\providecommand \citenamefont [1]{#1}%
\providecommand \href@noop [0]{\@secondoftwo}%
\providecommand \href [0]{\begingroup \@sanitize@url \@href}%
\providecommand \@href[1]{\@@startlink{#1}\@@href}%
\providecommand \@@href[1]{\endgroup#1\@@endlink}%
\providecommand \@sanitize@url [0]{\catcode `\\12\catcode `\$12\catcode
  `\&12\catcode `\#12\catcode `\^12\catcode `\_12\catcode `\%12\relax}%
\providecommand \@@startlink[1]{}%
\providecommand \@@endlink[0]{}%
\providecommand \url  [0]{\begingroup\@sanitize@url \@url }%
\providecommand \@url [1]{\endgroup\@href {#1}{\urlprefix }}%
\providecommand \urlprefix  [0]{URL }%
\providecommand \Eprint [0]{\href }%
\providecommand \doibase [0]{https://doi.org/}%
\providecommand \selectlanguage [0]{\@gobble}%
\providecommand \bibinfo  [0]{\@secondoftwo}%
\providecommand \bibfield  [0]{\@secondoftwo}%
\providecommand \translation [1]{[#1]}%
\providecommand \BibitemOpen [0]{}%
\providecommand \bibitemStop [0]{}%
\providecommand \bibitemNoStop [0]{.\EOS\space}%
\providecommand \EOS [0]{\spacefactor3000\relax}%
\providecommand \BibitemShut  [1]{\csname bibitem#1\endcsname}%
\let\auto@bib@innerbib\@empty
\bibitem [{\citenamefont {Chandler}(1987)}]{chandler1987introduction}%
  \BibitemOpen
  \bibfield  {author} {\bibinfo {author} {\bibfnamefont {D.}~\bibnamefont
  {Chandler}},\ }\href@noop {} {\emph {\bibinfo {title} {{Introduction to
  Modern Statistical Mechanics}}}}\ (\bibinfo  {publisher} {Oxford University
  Press, Oxford, UK},\ \bibinfo {year} {1987})\BibitemShut {NoStop}%
\bibitem [{\citenamefont {Lelièvre}, \citenamefont {Rousset},\ and\
  \citenamefont {Stoltz}(2010)}]{stoltz2010free}%
  \BibitemOpen
  \bibfield  {author} {\bibinfo {author} {\bibfnamefont {T.}~\bibnamefont
  {Lelièvre}}, \bibinfo {author} {\bibfnamefont {M.}~\bibnamefont {Rousset}},\
  and\ \bibinfo {author} {\bibfnamefont {G.}~\bibnamefont {Stoltz}},\
  }\href@noop {} {\emph {\bibinfo {title} {{Free Energy Computations: A
  Mathematical Perspective}}}}\ (\bibinfo  {publisher} {Imperial College
  Press},\ \bibinfo {year} {2010})\BibitemShut {NoStop}%
\bibitem [{\citenamefont {Frenkel}\ and\ \citenamefont
  {Smit}(2023)}]{dfrenkel:mc}%
  \BibitemOpen
  \bibfield  {author} {\bibinfo {author} {\bibfnamefont {D.}~\bibnamefont
  {Frenkel}}\ and\ \bibinfo {author} {\bibfnamefont {B.}~\bibnamefont {Smit}},\
  }\href@noop {} {\emph {\bibinfo {title} {{Understanding Molecular Simulation:
  From Algorithms to Applications}}}},\ \bibinfo {edition} {3rd}\ ed.\
  (\bibinfo  {publisher} {Academic Press},\ \bibinfo {year} {2023})\BibitemShut
  {NoStop}%
\bibitem [{\citenamefont {Rogal}(2021)}]{rogal2021reaction}%
  \BibitemOpen
  \bibfield  {author} {\bibinfo {author} {\bibfnamefont {J.}~\bibnamefont
  {Rogal}},\ }\bibfield  {title} {\enquote {\bibinfo {title} {{Reaction
  Coordinates in Complex Systems -- A Perspective}},}\ }\href
  {https://doi.org/https://doi.org/10.1140/epjb/s10051-021-00233-5} {\bibfield
  {journal} {\bibinfo  {journal} {Eur. Phys. J. B}\ }\textbf {\bibinfo {volume}
  {94}},\ \bibinfo {pages} {1--9} (\bibinfo {year} {2021})}\BibitemShut
  {NoStop}%
\bibitem [{\citenamefont {Bolhuis}\ \emph {et~al.}(2002)\citenamefont
  {Bolhuis}, \citenamefont {Chandler}, \citenamefont {Dellago},\ and\
  \citenamefont {Geissler}}]{bolhuis2002transition}%
  \BibitemOpen
  \bibfield  {author} {\bibinfo {author} {\bibfnamefont {P.~G.}\ \bibnamefont
  {Bolhuis}}, \bibinfo {author} {\bibfnamefont {D.}~\bibnamefont {Chandler}},
  \bibinfo {author} {\bibfnamefont {C.}~\bibnamefont {Dellago}},\ and\ \bibinfo
  {author} {\bibfnamefont {P.~L.}\ \bibnamefont {Geissler}},\ }\bibfield
  {title} {\enquote {\bibinfo {title} {{Transition Path Sampling: Throwing
  Ropes over Rough Mountain Passes, in the Dark}},}\ }\href
  {https://doi.org/https://doi.org/10.1146/annurev.physchem.53.082301.113146}
  {\bibfield  {journal} {\bibinfo  {journal} {Annu. Rev. Phys. Chem.}\ }\textbf
  {\bibinfo {volume} {53}},\ \bibinfo {pages} {291--318} (\bibinfo {year}
  {2002})}\BibitemShut {NoStop}%
\bibitem [{\citenamefont {Abrams}\ and\ \citenamefont
  {Bussi}(2014)}]{abrams2014enhanced}%
  \BibitemOpen
  \bibfield  {author} {\bibinfo {author} {\bibfnamefont {C.}~\bibnamefont
  {Abrams}}\ and\ \bibinfo {author} {\bibfnamefont {G.}~\bibnamefont {Bussi}},\
  }\bibfield  {title} {\enquote {\bibinfo {title} {{Enhanced Sampling in
  Molecular Dynamics using Metadynamics, Replica-Exchange, and
  Temperature-Acceleration}},}\ }\href
  {https://doi.org/https://doi.org/10.3390/e16010163} {\bibfield  {journal}
  {\bibinfo  {journal} {Entropy}\ }\textbf {\bibinfo {volume} {16}},\ \bibinfo
  {pages} {163--199} (\bibinfo {year} {2014})}\BibitemShut {NoStop}%
\bibitem [{\citenamefont {Pietrucci}(2017)}]{pietrucci_strategies_2017}%
  \BibitemOpen
  \bibfield  {author} {\bibinfo {author} {\bibfnamefont {F.}~\bibnamefont
  {Pietrucci}},\ }\bibfield  {title} {\enquote {\bibinfo {title} {{Strategies
  for the Exploration of Free Energy Landscapes: Unity in Diversity and
  Challenges Ahead}},}\ }\href
  {https://doi.org/https://doi.org/10.1016/j.revip.2017.05.001} {\bibfield
  {journal} {\bibinfo  {journal} {Rev. Phys.}\ }\textbf {\bibinfo {volume}
  {2}},\ \bibinfo {pages} {32--45} (\bibinfo {year} {2017})}\BibitemShut
  {NoStop}%
\bibitem [{\citenamefont {Valsson}, \citenamefont {Tiwary},\ and\ \citenamefont
  {Parrinello}(2016)}]{valsson2016enhancing}%
  \BibitemOpen
  \bibfield  {author} {\bibinfo {author} {\bibfnamefont {O.}~\bibnamefont
  {Valsson}}, \bibinfo {author} {\bibfnamefont {P.}~\bibnamefont {Tiwary}},\
  and\ \bibinfo {author} {\bibfnamefont {M.}~\bibnamefont {Parrinello}},\
  }\bibfield  {title} {\enquote {\bibinfo {title} {{Enhancing Important
  Fluctuations: Rare Events and Metadynamics from a Conceptual Viewpoint}},}\
  }\href {https://doi.org/10.1146/annurev-physchem-040215-112229} {\bibfield
  {journal} {\bibinfo  {journal} {Annu. Rev. Phys. Chem.}\ }\textbf {\bibinfo
  {volume} {67}},\ \bibinfo {pages} {159--184} (\bibinfo {year}
  {2016})}\BibitemShut {NoStop}%
\bibitem [{\citenamefont {Yang}\ \emph {et~al.}(2019)\citenamefont {Yang},
  \citenamefont {Shao}, \citenamefont {Zhang}, \citenamefont {Yang},\ and\
  \citenamefont {Gao}}]{yang2019enhanced}%
  \BibitemOpen
  \bibfield  {author} {\bibinfo {author} {\bibfnamefont {Y.~I.}\ \bibnamefont
  {Yang}}, \bibinfo {author} {\bibfnamefont {Q.}~\bibnamefont {Shao}}, \bibinfo
  {author} {\bibfnamefont {J.}~\bibnamefont {Zhang}}, \bibinfo {author}
  {\bibfnamefont {L.}~\bibnamefont {Yang}},\ and\ \bibinfo {author}
  {\bibfnamefont {Y.~Q.}\ \bibnamefont {Gao}},\ }\bibfield  {title} {\enquote
  {\bibinfo {title} {{Enhanced Sampling in Molecular Dynamics}},}\ }\href
  {https://doi.org/https://doi.org/10.1063/1.5109531} {\bibfield  {journal}
  {\bibinfo  {journal} {J. Chem. Phys.}\ }\textbf {\bibinfo {volume} {151}},\
  \bibinfo {pages} {070902} (\bibinfo {year} {2019})}\BibitemShut {NoStop}%
\bibitem [{\citenamefont {Bussi}\ and\ \citenamefont
  {Laio}(2020)}]{bussi2020using}%
  \BibitemOpen
  \bibfield  {author} {\bibinfo {author} {\bibfnamefont {G.}~\bibnamefont
  {Bussi}}\ and\ \bibinfo {author} {\bibfnamefont {A.}~\bibnamefont {Laio}},\
  }\bibfield  {title} {\enquote {\bibinfo {title} {{Using Metadynamics to
  Explore Complex Free-Energy Landscapes}},}\ }\href
  {https://doi.org/https://doi.org/10.1038/s42254-020-0153-0} {\bibfield
  {journal} {\bibinfo  {journal} {Nat. Rev. Phys.}\ }\textbf {\bibinfo {volume}
  {2}},\ \bibinfo {pages} {200--2012} (\bibinfo {year} {2020})}\BibitemShut
  {NoStop}%
\bibitem [{\citenamefont {H{\'e}nin}\ \emph {et~al.}(2022)\citenamefont
  {H{\'e}nin}, \citenamefont {Leli{\`e}vre}, \citenamefont {Shirts},
  \citenamefont {Valsson},\ and\ \citenamefont
  {Delemotte}}]{henin2022enhanced}%
  \BibitemOpen
  \bibfield  {author} {\bibinfo {author} {\bibfnamefont {J.}~\bibnamefont
  {H{\'e}nin}}, \bibinfo {author} {\bibfnamefont {T.}~\bibnamefont
  {Leli{\`e}vre}}, \bibinfo {author} {\bibfnamefont {M.~R.}\ \bibnamefont
  {Shirts}}, \bibinfo {author} {\bibfnamefont {O.}~\bibnamefont {Valsson}},\
  and\ \bibinfo {author} {\bibfnamefont {L.}~\bibnamefont {Delemotte}},\
  }\bibfield  {title} {\enquote {\bibinfo {title} {{Enhanced Sampling Methods
  for Molecular Dynamics Simulations [Article v1.0]}},}\ }\href
  {https://doi.org/10.33011/livecoms.4.1.1583} {\bibfield  {journal} {\bibinfo
  {journal} {Living Journal of Computational Molecular Science}\ }\textbf
  {\bibinfo {volume} {4}},\ \bibinfo {pages} {1583} (\bibinfo {year}
  {2022})}\BibitemShut {NoStop}%
\bibitem [{\citenamefont {Chen}\ and\ \citenamefont
  {Chipot}(2022)}]{chen2022enhancing}%
  \BibitemOpen
  \bibfield  {author} {\bibinfo {author} {\bibfnamefont {H.}~\bibnamefont
  {Chen}}\ and\ \bibinfo {author} {\bibfnamefont {C.}~\bibnamefont {Chipot}},\
  }\bibfield  {title} {\enquote {\bibinfo {title} {{Enhancing Sampling with
  Free-Energy Calculations}},}\ }\href
  {https://doi.org/https://doi.org/10.1016/j.sbi.2022.102497} {\bibfield
  {journal} {\bibinfo  {journal} {Curr. Opin. Struct. Biol.}\ }\textbf
  {\bibinfo {volume} {77}},\ \bibinfo {pages} {102497} (\bibinfo {year}
  {2022})}\BibitemShut {NoStop}%
\bibitem [{\citenamefont {Ray}\ and\ \citenamefont
  {Parrinello}(2023)}]{ray2023kinetics}%
  \BibitemOpen
  \bibfield  {author} {\bibinfo {author} {\bibfnamefont {D.}~\bibnamefont
  {Ray}}\ and\ \bibinfo {author} {\bibfnamefont {M.}~\bibnamefont
  {Parrinello}},\ }\bibfield  {title} {\enquote {\bibinfo {title} {{Kinetics
  from Metadynamics: Principles, Applications, and Outlook}},}\ }\href
  {https://doi.org/https://doi.org/10.1021/acs.jctc.3c00660} {\bibfield
  {journal} {\bibinfo  {journal} {J. Chem. Theory Comput.}\ }\textbf {\bibinfo
  {volume} {19}},\ \bibinfo {pages} {5649--5670} (\bibinfo {year}
  {2023})}\BibitemShut {NoStop}%
\bibitem [{\citenamefont {Geissler}, \citenamefont {Dellago},\ and\
  \citenamefont {Chandler}(1999)}]{geissler1999kinetic}%
  \BibitemOpen
  \bibfield  {author} {\bibinfo {author} {\bibfnamefont {P.~L.}\ \bibnamefont
  {Geissler}}, \bibinfo {author} {\bibfnamefont {C.}~\bibnamefont {Dellago}},\
  and\ \bibinfo {author} {\bibfnamefont {D.}~\bibnamefont {Chandler}},\
  }\bibfield  {title} {\enquote {\bibinfo {title} {{Kinetic Pathways of Ion
  Pair Dissociation in Water}},}\ }\href
  {https://doi.org/https://doi.org/10.1021/jp984837g} {\bibfield  {journal}
  {\bibinfo  {journal} {J. Phys. Chem. B}\ }\textbf {\bibinfo {volume} {103}},\
  \bibinfo {pages} {3706--3710} (\bibinfo {year} {1999})}\BibitemShut {NoStop}%
\bibitem [{\citenamefont {Bolhuis}, \citenamefont {Dellago},\ and\
  \citenamefont {Chandler}(2000)}]{bolhuis2000reaction}%
  \BibitemOpen
  \bibfield  {author} {\bibinfo {author} {\bibfnamefont {P.~G.}\ \bibnamefont
  {Bolhuis}}, \bibinfo {author} {\bibfnamefont {C.}~\bibnamefont {Dellago}},\
  and\ \bibinfo {author} {\bibfnamefont {D.}~\bibnamefont {Chandler}},\
  }\bibfield  {title} {\enquote {\bibinfo {title} {{Reaction Coordinates of
  Biomolecular Isomerization}},}\ }\href
  {https://doi.org/https://doi.org/10.1073/pnas.100127697} {\bibfield
  {journal} {\bibinfo  {journal} {Proc. Natl. Acad. Sci. U.S.A.}\ }\textbf
  {\bibinfo {volume} {97}},\ \bibinfo {pages} {5877--5882} (\bibinfo {year}
  {2000})}\BibitemShut {NoStop}%
\bibitem [{\citenamefont {Ma}\ and\ \citenamefont
  {Dinner}(2005)}]{ma2005automatic}%
  \BibitemOpen
  \bibfield  {author} {\bibinfo {author} {\bibfnamefont {A.}~\bibnamefont
  {Ma}}\ and\ \bibinfo {author} {\bibfnamefont {A.~R.}\ \bibnamefont
  {Dinner}},\ }\bibfield  {title} {\enquote {\bibinfo {title} {{Automatic
  Method for Identifying Reaction Coordinates in Complex Systems}},}\ }\href
  {https://doi.org/https://doi.org/10.1021/jp045546c} {\bibfield  {journal}
  {\bibinfo  {journal} {J. Phys. Chem. B}\ }\textbf {\bibinfo {volume} {109}},\
  \bibinfo {pages} {6769--6779} (\bibinfo {year} {2005})}\BibitemShut {NoStop}%
\bibitem [{\citenamefont {H{\"a}nggi}, \citenamefont {Talkner},\ and\
  \citenamefont {Borkovec}(1990)}]{hanggi1990reaction}%
  \BibitemOpen
  \bibfield  {author} {\bibinfo {author} {\bibfnamefont {P.}~\bibnamefont
  {H{\"a}nggi}}, \bibinfo {author} {\bibfnamefont {P.}~\bibnamefont
  {Talkner}},\ and\ \bibinfo {author} {\bibfnamefont {M.}~\bibnamefont
  {Borkovec}},\ }\bibfield  {title} {\enquote {\bibinfo {title} {{Reaction-Rate
  Theory: Fifty Years after Kramers}},}\ }\href
  {https://doi.org/https://doi.org/10.1103/revmodphys.62.251} {\bibfield
  {journal} {\bibinfo  {journal} {Rev. Mod. Phys.}\ }\textbf {\bibinfo {volume}
  {62}},\ \bibinfo {pages} {251} (\bibinfo {year} {1990})}\BibitemShut
  {NoStop}%
\bibitem [{\citenamefont {Shaw}\ \emph {et~al.}(2010)\citenamefont {Shaw},
  \citenamefont {Maragakis}, \citenamefont {Lindorff-Larsen}, \citenamefont
  {Piana}, \citenamefont {Dror}, \citenamefont {Eastwood}, \citenamefont
  {Bank}, \citenamefont {Jumper}, \citenamefont {Salmon},\ and\ \citenamefont
  {Shan}}]{shaw2010atomic}%
  \BibitemOpen
  \bibfield  {author} {\bibinfo {author} {\bibfnamefont {D.~E.}\ \bibnamefont
  {Shaw}}, \bibinfo {author} {\bibfnamefont {P.}~\bibnamefont {Maragakis}},
  \bibinfo {author} {\bibfnamefont {K.}~\bibnamefont {Lindorff-Larsen}},
  \bibinfo {author} {\bibfnamefont {S.}~\bibnamefont {Piana}}, \bibinfo
  {author} {\bibfnamefont {R.~O.}\ \bibnamefont {Dror}}, \bibinfo {author}
  {\bibfnamefont {M.~P.}\ \bibnamefont {Eastwood}}, \bibinfo {author}
  {\bibfnamefont {J.~A.}\ \bibnamefont {Bank}}, \bibinfo {author}
  {\bibfnamefont {J.~M.}\ \bibnamefont {Jumper}}, \bibinfo {author}
  {\bibfnamefont {J.~K.}\ \bibnamefont {Salmon}},\ and\ \bibinfo {author}
  {\bibfnamefont {Y.}~\bibnamefont {Shan}},\ }\bibfield  {title} {\enquote
  {\bibinfo {title} {{Atomic-Level Characterization of the Structural Dynamics
  of Proteins}},}\ }\href
  {https://doi.org/https://doi.org/10.1126/science.1187409} {\bibfield
  {journal} {\bibinfo  {journal} {Science}\ }\textbf {\bibinfo {volume}
  {330}},\ \bibinfo {pages} {341--346} (\bibinfo {year} {2010})}\BibitemShut
  {NoStop}%
\bibitem [{\citenamefont {Lindorff-Larsen}\ \emph {et~al.}(2011)\citenamefont
  {Lindorff-Larsen}, \citenamefont {Piana}, \citenamefont {Dror},\ and\
  \citenamefont {Shaw}}]{lindorff2011fast}%
  \BibitemOpen
  \bibfield  {author} {\bibinfo {author} {\bibfnamefont {K.}~\bibnamefont
  {Lindorff-Larsen}}, \bibinfo {author} {\bibfnamefont {S.}~\bibnamefont
  {Piana}}, \bibinfo {author} {\bibfnamefont {R.~O.}\ \bibnamefont {Dror}},\
  and\ \bibinfo {author} {\bibfnamefont {D.~E.}\ \bibnamefont {Shaw}},\
  }\bibfield  {title} {\enquote {\bibinfo {title} {{How Fast-Folding Proteins
  Fold}},}\ }\href {https://doi.org/https://doi.org/10.1126/science.1208351}
  {\bibfield  {journal} {\bibinfo  {journal} {Science}\ }\textbf {\bibinfo
  {volume} {334}},\ \bibinfo {pages} {517--520} (\bibinfo {year}
  {2011})}\BibitemShut {NoStop}%
\bibitem [{\citenamefont {Neha}\ \emph {et~al.}(2023)\citenamefont {Neha},
  \citenamefont {Tiwari}, \citenamefont {Mondal}, \citenamefont {Kumari},\ and\
  \citenamefont {Karmakar}}]{neha2022collective}%
  \BibitemOpen
  \bibfield  {author} {\bibinfo {author} {\bibnamefont {Neha}}, \bibinfo
  {author} {\bibfnamefont {V.}~\bibnamefont {Tiwari}}, \bibinfo {author}
  {\bibfnamefont {S.}~\bibnamefont {Mondal}}, \bibinfo {author} {\bibfnamefont
  {N.}~\bibnamefont {Kumari}},\ and\ \bibinfo {author} {\bibfnamefont
  {T.}~\bibnamefont {Karmakar}},\ }\bibfield  {title} {\enquote {\bibinfo
  {title} {{Collective Variables for Crystallization Simulations--from Early
  Developments to Recent Advances}},}\ }\href
  {https://doi.org/https://doi.org/10.1021/acsomega.2c06310} {\bibfield
  {journal} {\bibinfo  {journal} {ACS Omega}\ }\textbf {\bibinfo {volume}
  {8}},\ \bibinfo {pages} {127--146} (\bibinfo {year} {2023})}\BibitemShut
  {NoStop}%
\bibitem [{\citenamefont {Beyerle}, \citenamefont {Zou},\ and\ \citenamefont
  {Tiwary}(2023)}]{beyerle2023recent}%
  \BibitemOpen
  \bibfield  {author} {\bibinfo {author} {\bibfnamefont {E.~R.}\ \bibnamefont
  {Beyerle}}, \bibinfo {author} {\bibfnamefont {Z.}~\bibnamefont {Zou}},\ and\
  \bibinfo {author} {\bibfnamefont {P.}~\bibnamefont {Tiwary}},\ }\bibfield
  {title} {\enquote {\bibinfo {title} {{Recent Advances in Describing and
  Driving Crystal Nucleation using Machine Learning and Artificial
  Intelligence}},}\ }\href
  {https://doi.org/https://doi.org/10.1016/j.cossms.2023.101093} {\bibfield
  {journal} {\bibinfo  {journal} {Curr. Opin. Solid State Mater. Sci.}\
  }\textbf {\bibinfo {volume} {27}},\ \bibinfo {pages} {101093} (\bibinfo
  {year} {2023})}\BibitemShut {NoStop}%
\bibitem [{\citenamefont {Berthier}\ and\ \citenamefont
  {Biroli}(2011)}]{berthier2011theoretical}%
  \BibitemOpen
  \bibfield  {author} {\bibinfo {author} {\bibfnamefont {L.}~\bibnamefont
  {Berthier}}\ and\ \bibinfo {author} {\bibfnamefont {G.}~\bibnamefont
  {Biroli}},\ }\bibfield  {title} {\enquote {\bibinfo {title} {{Theoretical
  Perspective on the Glass Transition and Amorphous Materials}},}\ }\href
  {https://doi.org/https://doi.org/10.1103/revmodphys.83.587} {\bibfield
  {journal} {\bibinfo  {journal} {Rev. Mod. Phys.}\ }\textbf {\bibinfo {volume}
  {83}},\ \bibinfo {pages} {587} (\bibinfo {year} {2011})}\BibitemShut
  {NoStop}%
\bibitem [{\citenamefont {Hohenberg}\ and\ \citenamefont
  {Krekhov}(2015)}]{hohenberg2015introduction}%
  \BibitemOpen
  \bibfield  {author} {\bibinfo {author} {\bibfnamefont {P.~C.}\ \bibnamefont
  {Hohenberg}}\ and\ \bibinfo {author} {\bibfnamefont {A.~P.}\ \bibnamefont
  {Krekhov}},\ }\bibfield  {title} {\enquote {\bibinfo {title} {{An
  Introduction to the Ginzburg--Landau Theory of Phase Transitions and
  Nonequilibrium Patterns}},}\ }\href
  {https://doi.org/https://doi.org/10.1016/j.physrep.2015.01.001} {\bibfield
  {journal} {\bibinfo  {journal} {Phys. Rep.}\ }\textbf {\bibinfo {volume}
  {572}},\ \bibinfo {pages} {1--42} (\bibinfo {year} {2015})}\BibitemShut
  {NoStop}%
\bibitem [{\citenamefont {Banerjee}\ \emph {et~al.}(2024)\citenamefont
  {Banerjee}, \citenamefont {Azizi}, \citenamefont {Egan}, \citenamefont
  {Donkor}, \citenamefont {Malosso}, \citenamefont {Pino}, \citenamefont
  {Mirón}, \citenamefont {Stella}, \citenamefont {Sormani}, \citenamefont
  {Hozana}, \citenamefont {Monti}, \citenamefont {Morzan}, \citenamefont
  {Rodriguez}, \citenamefont {Cassone}, \citenamefont {Jelic}, \citenamefont
  {Scherlis},\ and\ \citenamefont {Hassanali}}]{10.1063/5.0207567}%
  \BibitemOpen
  \bibfield  {author} {\bibinfo {author} {\bibfnamefont {D.}~\bibnamefont
  {Banerjee}}, \bibinfo {author} {\bibfnamefont {K.}~\bibnamefont {Azizi}},
  \bibinfo {author} {\bibfnamefont {C.~K.}\ \bibnamefont {Egan}}, \bibinfo
  {author} {\bibfnamefont {E.~D.}\ \bibnamefont {Donkor}}, \bibinfo {author}
  {\bibfnamefont {C.}~\bibnamefont {Malosso}}, \bibinfo {author} {\bibfnamefont
  {S.~D.}\ \bibnamefont {Pino}}, \bibinfo {author} {\bibfnamefont {G.~D.}\
  \bibnamefont {Mirón}}, \bibinfo {author} {\bibfnamefont {M.}~\bibnamefont
  {Stella}}, \bibinfo {author} {\bibfnamefont {G.}~\bibnamefont {Sormani}},
  \bibinfo {author} {\bibfnamefont {G.~N.}\ \bibnamefont {Hozana}}, \bibinfo
  {author} {\bibfnamefont {M.}~\bibnamefont {Monti}}, \bibinfo {author}
  {\bibfnamefont {U.~N.}\ \bibnamefont {Morzan}}, \bibinfo {author}
  {\bibfnamefont {A.}~\bibnamefont {Rodriguez}}, \bibinfo {author}
  {\bibfnamefont {G.}~\bibnamefont {Cassone}}, \bibinfo {author} {\bibfnamefont
  {A.}~\bibnamefont {Jelic}}, \bibinfo {author} {\bibfnamefont
  {D.}~\bibnamefont {Scherlis}},\ and\ \bibinfo {author} {\bibfnamefont
  {A.}~\bibnamefont {Hassanali}},\ }\bibfield  {title} {\enquote {\bibinfo
  {title} {{Aqueous Solution Chemistry in Silico and the Role of Data-Driven
  Approaches}},}\ }\href {https://doi.org/10.1063/5.0207567} {\bibfield
  {journal} {\bibinfo  {journal} {Chem. Phys. Rev.}\ }\textbf {\bibinfo
  {volume} {5}},\ \bibinfo {pages} {021308} (\bibinfo {year}
  {2024})}\BibitemShut {NoStop}%
\bibitem [{\citenamefont {Piccini}\ \emph {et~al.}(2022)\citenamefont
  {Piccini}, \citenamefont {Lee}, \citenamefont {Yuk}, \citenamefont {Zhang},
  \citenamefont {Collinge}, \citenamefont {Kollias}, \citenamefont {Nguyen},
  \citenamefont {Glezakou},\ and\ \citenamefont {Rousseau}}]{piccini2022ab}%
  \BibitemOpen
  \bibfield  {author} {\bibinfo {author} {\bibfnamefont {G.}~\bibnamefont
  {Piccini}}, \bibinfo {author} {\bibfnamefont {M.-S.}\ \bibnamefont {Lee}},
  \bibinfo {author} {\bibfnamefont {S.~F.}\ \bibnamefont {Yuk}}, \bibinfo
  {author} {\bibfnamefont {D.}~\bibnamefont {Zhang}}, \bibinfo {author}
  {\bibfnamefont {G.}~\bibnamefont {Collinge}}, \bibinfo {author}
  {\bibfnamefont {L.}~\bibnamefont {Kollias}}, \bibinfo {author} {\bibfnamefont
  {M.-T.}\ \bibnamefont {Nguyen}}, \bibinfo {author} {\bibfnamefont {V.-A.}\
  \bibnamefont {Glezakou}},\ and\ \bibinfo {author} {\bibfnamefont
  {R.}~\bibnamefont {Rousseau}},\ }\bibfield  {title} {\enquote {\bibinfo
  {title} {{Ab Initio Molecular Dynamics with Enhanced Sampling in
  Heterogeneous Catalysis}},}\ }\href
  {https://doi.org/https://doi.org/10.1039/D1CY01329G} {\bibfield  {journal}
  {\bibinfo  {journal} {Catal. Sci. Technol.}\ }\textbf {\bibinfo {volume}
  {12}},\ \bibinfo {pages} {12--37} (\bibinfo {year} {2022})}\BibitemShut
  {NoStop}%
\bibitem [{\citenamefont {Baron}\ and\ \citenamefont
  {McCammon}(2013)}]{baron2013molecular}%
  \BibitemOpen
  \bibfield  {author} {\bibinfo {author} {\bibfnamefont {R.}~\bibnamefont
  {Baron}}\ and\ \bibinfo {author} {\bibfnamefont {J.~A.}\ \bibnamefont
  {McCammon}},\ }\bibfield  {title} {\enquote {\bibinfo {title} {{Molecular
  Recognition and Ligand Association}},}\ }\href
  {https://doi.org/https://doi.org/10.1146/annurev-physchem-040412-110047}
  {\bibfield  {journal} {\bibinfo  {journal} {Annu. Rev. Phys. Chem.}\ }\textbf
  {\bibinfo {volume} {64}},\ \bibinfo {pages} {151--175} (\bibinfo {year}
  {2013})}\BibitemShut {NoStop}%
\bibitem [{\citenamefont {Chipot}(2014)}]{chipot2014frontiers}%
  \BibitemOpen
  \bibfield  {author} {\bibinfo {author} {\bibfnamefont {C.}~\bibnamefont
  {Chipot}},\ }\bibfield  {title} {\enquote {\bibinfo {title} {{Frontiers in
  Free-Energy Calculations of Biological Systems}},}\ }\href
  {https://doi.org/https://doi.org/10.1002/wcms.1157} {\bibfield  {journal}
  {\bibinfo  {journal} {Wiley Interdiscip. Rev. Comput. Mol. Sci.}\ }\textbf
  {\bibinfo {volume} {4}},\ \bibinfo {pages} {71--89} (\bibinfo {year}
  {2014})}\BibitemShut {NoStop}%
\bibitem [{\citenamefont {Rydzewski}\ and\ \citenamefont
  {Nowak}(2017)}]{rydzewski2017ligand}%
  \BibitemOpen
  \bibfield  {author} {\bibinfo {author} {\bibfnamefont {J.}~\bibnamefont
  {Rydzewski}}\ and\ \bibinfo {author} {\bibfnamefont {W.}~\bibnamefont
  {Nowak}},\ }\bibfield  {title} {\enquote {\bibinfo {title} {{Ligand Diffusion
  in Proteins via Enhanced Sampling in Molecular Dynamics}},}\ }\href
  {https://doi.org/https://doi.org/10.1016/j.plrev.2017.03.003} {\bibfield
  {journal} {\bibinfo  {journal} {Phys. Life Rev.}\ }\textbf {\bibinfo {volume}
  {22}},\ \bibinfo {pages} {58--74} (\bibinfo {year} {2017})}\BibitemShut
  {NoStop}%
\bibitem [{\citenamefont {Bernetti}\ \emph {et~al.}(2019)\citenamefont
  {Bernetti}, \citenamefont {Masetti}, \citenamefont {Rocchia},\ and\
  \citenamefont {Cavalli}}]{bernetti2019kinetics}%
  \BibitemOpen
  \bibfield  {author} {\bibinfo {author} {\bibfnamefont {M.}~\bibnamefont
  {Bernetti}}, \bibinfo {author} {\bibfnamefont {M.}~\bibnamefont {Masetti}},
  \bibinfo {author} {\bibfnamefont {W.}~\bibnamefont {Rocchia}},\ and\ \bibinfo
  {author} {\bibfnamefont {A.}~\bibnamefont {Cavalli}},\ }\bibfield  {title}
  {\enquote {\bibinfo {title} {{Kinetics of Drug Binding and Residence
  Time}},}\ }\href
  {https://doi.org/https://doi.org/10.1146/annurev-physchem-042018-052340}
  {\bibfield  {journal} {\bibinfo  {journal} {Annu. Rev. Phys. Chem.}\ }\textbf
  {\bibinfo {volume} {70}},\ \bibinfo {pages} {143--171} (\bibinfo {year}
  {2019})}\BibitemShut {NoStop}%
\bibitem [{\citenamefont {Pedregosa}\ \emph {et~al.}(2011)\citenamefont
  {Pedregosa}, \citenamefont {Varoquaux}, \citenamefont {Gramfort},
  \citenamefont {Michel}, \citenamefont {Thirion}, \citenamefont {Grisel},
  \citenamefont {Blondel}, \citenamefont {Prettenhofer}, \citenamefont {Weiss},
  \citenamefont {Dubourg}, \citenamefont {Vanderplas}, \citenamefont {Passos},
  \citenamefont {Cournapeau}, \citenamefont {Brucher}, \citenamefont {Perrot},\
  and\ \citenamefont {Duchesnay}}]{scikit-learn}%
  \BibitemOpen
  \bibfield  {author} {\bibinfo {author} {\bibfnamefont {F.}~\bibnamefont
  {Pedregosa}}, \bibinfo {author} {\bibfnamefont {G.}~\bibnamefont
  {Varoquaux}}, \bibinfo {author} {\bibfnamefont {A.}~\bibnamefont {Gramfort}},
  \bibinfo {author} {\bibfnamefont {V.}~\bibnamefont {Michel}}, \bibinfo
  {author} {\bibfnamefont {B.}~\bibnamefont {Thirion}}, \bibinfo {author}
  {\bibfnamefont {O.}~\bibnamefont {Grisel}}, \bibinfo {author} {\bibfnamefont
  {M.}~\bibnamefont {Blondel}}, \bibinfo {author} {\bibfnamefont
  {P.}~\bibnamefont {Prettenhofer}}, \bibinfo {author} {\bibfnamefont
  {R.}~\bibnamefont {Weiss}}, \bibinfo {author} {\bibfnamefont
  {V.}~\bibnamefont {Dubourg}}, \bibinfo {author} {\bibfnamefont
  {J.}~\bibnamefont {Vanderplas}}, \bibinfo {author} {\bibfnamefont
  {A.}~\bibnamefont {Passos}}, \bibinfo {author} {\bibfnamefont
  {D.}~\bibnamefont {Cournapeau}}, \bibinfo {author} {\bibfnamefont
  {M.}~\bibnamefont {Brucher}}, \bibinfo {author} {\bibfnamefont
  {M.}~\bibnamefont {Perrot}},\ and\ \bibinfo {author} {\bibfnamefont
  {E.}~\bibnamefont {Duchesnay}},\ }\bibfield  {title} {\enquote {\bibinfo
  {title} {{Scikit-learn: Machine Learning in Python}},}\ }\href
  {https://doi.org/https://www.jmlr.org/papers/v12/pedregosa11a.html}
  {\bibfield  {journal} {\bibinfo  {journal} {J. Mach. Learn. Res.}\ }\textbf
  {\bibinfo {volume} {12}},\ \bibinfo {pages} {2825--2830} (\bibinfo {year}
  {2011})}\BibitemShut {NoStop}%
\bibitem [{\citenamefont {Paszke}\ \emph {et~al.}(2019)\citenamefont {Paszke},
  \citenamefont {Gross}, \citenamefont {Massa}, \citenamefont {Lerer},
  \citenamefont {Bradbury}, \citenamefont {Chanan}, \citenamefont {Killeen},
  \citenamefont {Lin}, \citenamefont {Gimelshein}, \citenamefont {Antiga} \emph
  {et~al.}}]{paszke2019pytorch}%
  \BibitemOpen
  \bibfield  {author} {\bibinfo {author} {\bibfnamefont {A.}~\bibnamefont
  {Paszke}}, \bibinfo {author} {\bibfnamefont {S.}~\bibnamefont {Gross}},
  \bibinfo {author} {\bibfnamefont {F.}~\bibnamefont {Massa}}, \bibinfo
  {author} {\bibfnamefont {A.}~\bibnamefont {Lerer}}, \bibinfo {author}
  {\bibfnamefont {J.}~\bibnamefont {Bradbury}}, \bibinfo {author}
  {\bibfnamefont {G.}~\bibnamefont {Chanan}}, \bibinfo {author} {\bibfnamefont
  {T.}~\bibnamefont {Killeen}}, \bibinfo {author} {\bibfnamefont
  {Z.}~\bibnamefont {Lin}}, \bibinfo {author} {\bibfnamefont {N.}~\bibnamefont
  {Gimelshein}}, \bibinfo {author} {\bibfnamefont {L.}~\bibnamefont {Antiga}},
  \emph {et~al.},\ }\bibfield  {title} {\enquote {\bibinfo {title} {{PyTorch:
  An Imperative Style, High-Performance Deep Learning Library}},}\ }\href
  {https://doi.org/https://papers.neurips.cc/paper/9015-pytorch-an-imperative-style-high-performance-deep-learning-library.pdf}
  {\bibfield  {journal} {\bibinfo  {journal} {Adv. Neural Inf. Process. Syst.}\
  }\textbf {\bibinfo {volume} {32}},\ \bibinfo {pages} {8026--8037} (\bibinfo
  {year} {2019})}\BibitemShut {NoStop}%
\bibitem [{\citenamefont {Hastie}\ \emph {et~al.}(2009)\citenamefont {Hastie},
  \citenamefont {Tibshirani}, \citenamefont {Friedman},\ and\ \citenamefont
  {Friedman}}]{hastie2009elements}%
  \BibitemOpen
  \bibfield  {author} {\bibinfo {author} {\bibfnamefont {T.}~\bibnamefont
  {Hastie}}, \bibinfo {author} {\bibfnamefont {R.}~\bibnamefont {Tibshirani}},
  \bibinfo {author} {\bibfnamefont {J.~H.}\ \bibnamefont {Friedman}},\ and\
  \bibinfo {author} {\bibfnamefont {J.~H.}\ \bibnamefont {Friedman}},\
  }\href@noop {} {\emph {\bibinfo {title} {{The Elements of Statistical
  Learning: Data Mining, Inference, and Prediction}}}},\ \bibinfo {edition}
  {2nd}\ ed.\ (\bibinfo  {publisher} {Springer},\ \bibinfo {year}
  {2009})\BibitemShut {NoStop}%
\bibitem [{\citenamefont {Fukunaga}(2013)}]{fukunaga2013introduction}%
  \BibitemOpen
  \bibfield  {author} {\bibinfo {author} {\bibfnamefont {K.}~\bibnamefont
  {Fukunaga}},\ }\href@noop {} {\emph {\bibinfo {title} {{Introduction to
  Statistical Pattern Recognition}}}},\ \bibinfo {edition} {2nd}\ ed.\
  (\bibinfo  {publisher} {Elsevier},\ \bibinfo {year} {2013})\BibitemShut
  {NoStop}%
\bibitem [{\citenamefont {Bengio}, \citenamefont {Courville},\ and\
  \citenamefont {Vincent}(2013)}]{bengio2013representation}%
  \BibitemOpen
  \bibfield  {author} {\bibinfo {author} {\bibfnamefont {Y.}~\bibnamefont
  {Bengio}}, \bibinfo {author} {\bibfnamefont {A.}~\bibnamefont {Courville}},\
  and\ \bibinfo {author} {\bibfnamefont {P.}~\bibnamefont {Vincent}},\
  }\bibfield  {title} {\enquote {\bibinfo {title} {{Representation Learning: A
  Review and New Perspectives}},}\ }\href
  {https://doi.org/https://doi.org/10.1109/tpami.2013.50} {\bibfield  {journal}
  {\bibinfo  {journal} {IEEE Trans. Pattern Anal. Mach. Intell.}\ }\textbf
  {\bibinfo {volume} {35}},\ \bibinfo {pages} {1798--1828} (\bibinfo {year}
  {2013})}\BibitemShut {NoStop}%
\bibitem [{\citenamefont {Xie}\ \emph {et~al.}(2020)\citenamefont {Xie},
  \citenamefont {Gao}, \citenamefont {Nijkamp}, \citenamefont {Zhu},\ and\
  \citenamefont {Wu}}]{xie2020representation}%
  \BibitemOpen
  \bibfield  {author} {\bibinfo {author} {\bibfnamefont {J.}~\bibnamefont
  {Xie}}, \bibinfo {author} {\bibfnamefont {R.}~\bibnamefont {Gao}}, \bibinfo
  {author} {\bibfnamefont {E.}~\bibnamefont {Nijkamp}}, \bibinfo {author}
  {\bibfnamefont {S.-C.}\ \bibnamefont {Zhu}},\ and\ \bibinfo {author}
  {\bibfnamefont {Y.~N.}\ \bibnamefont {Wu}},\ }\bibfield  {title} {\enquote
  {\bibinfo {title} {{Representation Learning: A Statistical Perspective}},}\
  }\href
  {https://doi.org/https://doi.org/10.1146/annurev-statistics-031219-041131}
  {\bibfield  {journal} {\bibinfo  {journal} {Annu. Rev. Stat. Appl.}\ }\textbf
  {\bibinfo {volume} {7}},\ \bibinfo {pages} {303--335} (\bibinfo {year}
  {2020})}\BibitemShut {NoStop}%
\bibitem [{\citenamefont {Karniadakis}\ \emph {et~al.}(2021)\citenamefont
  {Karniadakis}, \citenamefont {Kevrekidis}, \citenamefont {Lu}, \citenamefont
  {Perdikaris}, \citenamefont {Wang},\ and\ \citenamefont
  {Yang}}]{karniadakis2021physics}%
  \BibitemOpen
  \bibfield  {author} {\bibinfo {author} {\bibfnamefont {G.~E.}\ \bibnamefont
  {Karniadakis}}, \bibinfo {author} {\bibfnamefont {I.~G.}\ \bibnamefont
  {Kevrekidis}}, \bibinfo {author} {\bibfnamefont {L.}~\bibnamefont {Lu}},
  \bibinfo {author} {\bibfnamefont {P.}~\bibnamefont {Perdikaris}}, \bibinfo
  {author} {\bibfnamefont {S.}~\bibnamefont {Wang}},\ and\ \bibinfo {author}
  {\bibfnamefont {L.}~\bibnamefont {Yang}},\ }\bibfield  {title} {\enquote
  {\bibinfo {title} {{Physics-Informed Machine Learning}},}\ }\href
  {https://doi.org/https://doi.org/10.1038/s42254-021-00314-5} {\bibfield
  {journal} {\bibinfo  {journal} {Nat. Rev. Phys.}\ }\textbf {\bibinfo {volume}
  {3}},\ \bibinfo {pages} {422--440} (\bibinfo {year} {2021})}\BibitemShut
  {NoStop}%
\bibitem [{\citenamefont {Meil{\u{a}}}\ and\ \citenamefont
  {Zhang}(2024)}]{meilua2024manifold}%
  \BibitemOpen
  \bibfield  {author} {\bibinfo {author} {\bibfnamefont {M.}~\bibnamefont
  {Meil{\u{a}}}}\ and\ \bibinfo {author} {\bibfnamefont {H.}~\bibnamefont
  {Zhang}},\ }\bibfield  {title} {\enquote {\bibinfo {title} {{Manifold
  Learning: What, How, and Why}},}\ }\href
  {https://doi.org/https://doi.org/10.1146/annurev-statistics-040522-115238}
  {\bibfield  {journal} {\bibinfo  {journal} {Annu. Rev. Stat. Appl.}\ }\textbf
  {\bibinfo {volume} {11}},\ \bibinfo {pages} {393--417} (\bibinfo {year}
  {2024})}\BibitemShut {NoStop}%
\bibitem [{\citenamefont {Glielmo}\ \emph {et~al.}(2021)\citenamefont
  {Glielmo}, \citenamefont {Husic}, \citenamefont {Rodriguez}, \citenamefont
  {Clementi}, \citenamefont {Noé},\ and\ \citenamefont
  {Laio}}]{glielmo2021unsupervised}%
  \BibitemOpen
  \bibfield  {author} {\bibinfo {author} {\bibfnamefont {A.}~\bibnamefont
  {Glielmo}}, \bibinfo {author} {\bibfnamefont {B.~E.}\ \bibnamefont {Husic}},
  \bibinfo {author} {\bibfnamefont {A.}~\bibnamefont {Rodriguez}}, \bibinfo
  {author} {\bibfnamefont {C.}~\bibnamefont {Clementi}}, \bibinfo {author}
  {\bibfnamefont {F.}~\bibnamefont {Noé}},\ and\ \bibinfo {author}
  {\bibfnamefont {A.}~\bibnamefont {Laio}},\ }\bibfield  {title} {\enquote
  {\bibinfo {title} {{Unsupervised Learning Methods for Molecular Simulation
  Data}},}\ }\href {https://doi.org/10.1021/acs.chemrev.0c01195} {\bibfield
  {journal} {\bibinfo  {journal} {Chem. Rev.}\ }\textbf {\bibinfo {volume}
  {121}},\ \bibinfo {pages} {9722--9758} (\bibinfo {year} {2021})}\BibitemShut
  {NoStop}%
\bibitem [{\citenamefont {Brunton}\ \emph {et~al.}(2022)\citenamefont
  {Brunton}, \citenamefont {Budišić}, \citenamefont {Kaiser},\ and\
  \citenamefont {Kutz}}]{brunton2021modern}%
  \BibitemOpen
  \bibfield  {author} {\bibinfo {author} {\bibfnamefont {S.~L.}\ \bibnamefont
  {Brunton}}, \bibinfo {author} {\bibfnamefont {M.}~\bibnamefont {Budišić}},
  \bibinfo {author} {\bibfnamefont {E.}~\bibnamefont {Kaiser}},\ and\ \bibinfo
  {author} {\bibfnamefont {J.~N.}\ \bibnamefont {Kutz}},\ }\bibfield  {title}
  {\enquote {\bibinfo {title} {{Modern Koopman Theory for Dynamical
  Systems}},}\ }\href {https://doi.org/10.1137/21m1401243} {\bibfield
  {journal} {\bibinfo  {journal} {SIAM Rev.}\ }\textbf {\bibinfo {volume}
  {64}},\ \bibinfo {pages} {229–340} (\bibinfo {year} {2022})}\BibitemShut
  {NoStop}%
\bibitem [{\citenamefont {Rohrdanz}, \citenamefont {Zheng},\ and\ \citenamefont
  {Clementi}(2013)}]{rohrdanz2013discovering}%
  \BibitemOpen
  \bibfield  {author} {\bibinfo {author} {\bibfnamefont {M.~A.}\ \bibnamefont
  {Rohrdanz}}, \bibinfo {author} {\bibfnamefont {W.}~\bibnamefont {Zheng}},\
  and\ \bibinfo {author} {\bibfnamefont {C.}~\bibnamefont {Clementi}},\
  }\bibfield  {title} {\enquote {\bibinfo {title} {{Discovering Mountain Passes
  via Torchlight: Methods for the Definition of Reaction Coordinates and
  Pathways in Complex Macromolecular Reactions}},}\ }\href
  {https://doi.org/https://doi.org/10.1146/annurev-physchem-040412-110006}
  {\bibfield  {journal} {\bibinfo  {journal} {Annu. Rev. Phys. Chem.}\ }\textbf
  {\bibinfo {volume} {64}},\ \bibinfo {pages} {295--316} (\bibinfo {year}
  {2013})}\BibitemShut {NoStop}%
\bibitem [{\citenamefont {Li}\ and\ \citenamefont {Ma}(2014)}]{li2014recent}%
  \BibitemOpen
  \bibfield  {author} {\bibinfo {author} {\bibfnamefont {W.}~\bibnamefont
  {Li}}\ and\ \bibinfo {author} {\bibfnamefont {A.}~\bibnamefont {Ma}},\
  }\bibfield  {title} {\enquote {\bibinfo {title} {{Recent Developments in
  Methods for Identifying Reaction Coordinates}},}\ }\href
  {https://doi.org/https://doi.org/10.1080/08927022.2014.907898} {\bibfield
  {journal} {\bibinfo  {journal} {Mol. Sim.}\ }\textbf {\bibinfo {volume}
  {40}},\ \bibinfo {pages} {784--793} (\bibinfo {year} {2014})}\BibitemShut
  {NoStop}%
\bibitem [{\citenamefont {Peters}(2016)}]{peters2016reaction}%
  \BibitemOpen
  \bibfield  {author} {\bibinfo {author} {\bibfnamefont {B.}~\bibnamefont
  {Peters}},\ }\bibfield  {title} {\enquote {\bibinfo {title} {{Reaction
  Coordinates and Mechanistic Hypothesis Tests}},}\ }\href
  {https://doi.org/https://doi.org/10.1146/annurev-physchem-040215-112215}
  {\bibfield  {journal} {\bibinfo  {journal} {Annu. Rev. Phys. Chem.}\ }\textbf
  {\bibinfo {volume} {67}},\ \bibinfo {pages} {669--690} (\bibinfo {year}
  {2016})}\BibitemShut {NoStop}%
\bibitem [{\citenamefont {Noé}\ and\ \citenamefont
  {Clementi}(2017)}]{noe2017collective}%
  \BibitemOpen
  \bibfield  {author} {\bibinfo {author} {\bibfnamefont {F.}~\bibnamefont
  {Noé}}\ and\ \bibinfo {author} {\bibfnamefont {C.}~\bibnamefont
  {Clementi}},\ }\bibfield  {title} {\enquote {\bibinfo {title} {{Collective
  Variables for the Study of Long-Time Kinetics from Molecular Trajectories:
  Theory and Methods}},}\ }\href
  {https://doi.org/https://doi.org/10.1016/j.sbi.2017.02.006} {\bibfield
  {journal} {\bibinfo  {journal} {Curr. Opin. Struct. Biol.}\ }\textbf
  {\bibinfo {volume} {43}},\ \bibinfo {pages} {141--147} (\bibinfo {year}
  {2017})}\BibitemShut {NoStop}%
\bibitem [{\citenamefont {Ceriotti}(2019)}]{ceriotti2019unsupervised}%
  \BibitemOpen
  \bibfield  {author} {\bibinfo {author} {\bibfnamefont {M.}~\bibnamefont
  {Ceriotti}},\ }\bibfield  {title} {\enquote {\bibinfo {title} {{Unsupervised
  Machine Learning in Atomistic Simulations, between Predictions and
  Understanding}},}\ }\href {https://doi.org/https://doi.org/10.1063/1.5091842}
  {\bibfield  {journal} {\bibinfo  {journal} {J. Chem. Phys.}\ }\textbf
  {\bibinfo {volume} {150}},\ \bibinfo {pages} {150901} (\bibinfo {year}
  {2019})}\BibitemShut {NoStop}%
\bibitem [{\citenamefont {Wang}, \citenamefont {Ribeiro},\ and\ \citenamefont
  {Tiwary}(2020)}]{wang2020machine}%
  \BibitemOpen
  \bibfield  {author} {\bibinfo {author} {\bibfnamefont {Y.}~\bibnamefont
  {Wang}}, \bibinfo {author} {\bibfnamefont {J.~M.~L.}\ \bibnamefont
  {Ribeiro}},\ and\ \bibinfo {author} {\bibfnamefont {P.}~\bibnamefont
  {Tiwary}},\ }\bibfield  {title} {\enquote {\bibinfo {title} {{Machine
  Learning Approaches for Analyzing and Enhancing Molecular Dynamics
  Simulations}},}\ }\href
  {https://doi.org/https://doi.org/10.1016/j.sbi.2019.12.016} {\bibfield
  {journal} {\bibinfo  {journal} {Curr. Opin. Struct. Biol.}\ }\textbf
  {\bibinfo {volume} {61}},\ \bibinfo {pages} {139--145} (\bibinfo {year}
  {2020})}\BibitemShut {NoStop}%
\bibitem [{\citenamefont {Wu}\ and\ \citenamefont
  {No{\'e}}(2020)}]{wu2020variational}%
  \BibitemOpen
  \bibfield  {author} {\bibinfo {author} {\bibfnamefont {H.}~\bibnamefont
  {Wu}}\ and\ \bibinfo {author} {\bibfnamefont {F.}~\bibnamefont {No{\'e}}},\
  }\bibfield  {title} {\enquote {\bibinfo {title} {{Variational Approach for
  Learning Markov Processes from Time Series Data}},}\ }\href
  {https://doi.org/10.1007/s00332-019-09567-y} {\bibfield  {journal} {\bibinfo
  {journal} {J. Nonlinear Sci.}\ }\textbf {\bibinfo {volume} {30}},\ \bibinfo
  {pages} {23--66} (\bibinfo {year} {2020})}\BibitemShut {NoStop}%
\bibitem [{\citenamefont {Klus}\ \emph {et~al.}(2018)\citenamefont {Klus},
  \citenamefont {Nüske}, \citenamefont {Koltai}, \citenamefont {Wu},
  \citenamefont {Kevrekidis}, \citenamefont {Schütte},\ and\ \citenamefont
  {Noé}}]{klus2018data}%
  \BibitemOpen
  \bibfield  {author} {\bibinfo {author} {\bibfnamefont {S.}~\bibnamefont
  {Klus}}, \bibinfo {author} {\bibfnamefont {F.}~\bibnamefont {Nüske}},
  \bibinfo {author} {\bibfnamefont {P.}~\bibnamefont {Koltai}}, \bibinfo
  {author} {\bibfnamefont {H.}~\bibnamefont {Wu}}, \bibinfo {author}
  {\bibfnamefont {I.}~\bibnamefont {Kevrekidis}}, \bibinfo {author}
  {\bibfnamefont {C.}~\bibnamefont {Schütte}},\ and\ \bibinfo {author}
  {\bibfnamefont {F.}~\bibnamefont {Noé}},\ }\bibfield  {title} {\enquote
  {\bibinfo {title} {{Data-Driven Model Reduction and Transfer Operator
  Approximation}},}\ }\href {https://doi.org/10.1007/s00332-017-9437-7}
  {\bibfield  {journal} {\bibinfo  {journal} {J. Nonlinear Sci.}\ }\textbf
  {\bibinfo {volume} {28}},\ \bibinfo {pages} {985--1010} (\bibinfo {year}
  {2018})}\BibitemShut {NoStop}%
\bibitem [{\citenamefont {Sidky}, \citenamefont {Chen},\ and\ \citenamefont
  {Ferguson}(2020)}]{sidky2020machine}%
  \BibitemOpen
  \bibfield  {author} {\bibinfo {author} {\bibfnamefont {H.}~\bibnamefont
  {Sidky}}, \bibinfo {author} {\bibfnamefont {W.}~\bibnamefont {Chen}},\ and\
  \bibinfo {author} {\bibfnamefont {A.~L.}\ \bibnamefont {Ferguson}},\
  }\bibfield  {title} {\enquote {\bibinfo {title} {{Machine Learning for
  Collective Variable Discovery and Enhanced Sampling in Biomolecular
  Simulation}},}\ }\href
  {https://doi.org/https://doi.org/10.1080/00268976.2020.1737742} {\bibfield
  {journal} {\bibinfo  {journal} {Mol. Phys.}\ }\textbf {\bibinfo {volume}
  {118}},\ \bibinfo {pages} {e1737742} (\bibinfo {year} {2020})}\BibitemShut
  {NoStop}%
\bibitem [{\citenamefont {Chen}(2021)}]{chen2021collective}%
  \BibitemOpen
  \bibfield  {author} {\bibinfo {author} {\bibfnamefont {M.}~\bibnamefont
  {Chen}},\ }\bibfield  {title} {\enquote {\bibinfo {title} {{Collective
  Variable-Based Enhanced Sampling and Machine Learning}},}\ }\href
  {https://doi.org/10.1140/epjb/s10051-021-00220-w} {\bibfield  {journal}
  {\bibinfo  {journal} {Eur. Phys. J. B}\ }\textbf {\bibinfo {volume} {94}},\
  \bibinfo {pages} {1--17} (\bibinfo {year} {2021})}\BibitemShut {NoStop}%
\bibitem [{\citenamefont {Chen}\ and\ \citenamefont
  {Chipot}(2023)}]{chen2023chasing}%
  \BibitemOpen
  \bibfield  {author} {\bibinfo {author} {\bibfnamefont {H.}~\bibnamefont
  {Chen}}\ and\ \bibinfo {author} {\bibfnamefont {C.}~\bibnamefont {Chipot}},\
  }\bibfield  {title} {\enquote {\bibinfo {title} {{Chasing Collective
  Variables using Temporal Data-Driven Strategies}},}\ }\href
  {https://doi.org/https://doi.org/10.1017/qrd.2022.23} {\bibfield  {journal}
  {\bibinfo  {journal} {QRB Discovery}\ }\textbf {\bibinfo {volume} {4}},\
  \bibinfo {pages} {e2} (\bibinfo {year} {2023})}\BibitemShut {NoStop}%
\bibitem [{\citenamefont {Rydzewski}, \citenamefont {Chen},\ and\ \citenamefont
  {Valsson}(2023)}]{rydzewski2023manifold}%
  \BibitemOpen
  \bibfield  {author} {\bibinfo {author} {\bibfnamefont {J.}~\bibnamefont
  {Rydzewski}}, \bibinfo {author} {\bibfnamefont {M.}~\bibnamefont {Chen}},\
  and\ \bibinfo {author} {\bibfnamefont {O.}~\bibnamefont {Valsson}},\
  }\bibfield  {title} {\enquote {\bibinfo {title} {{Manifold Learning in
  Atomistic Simulations: A Conceptual Review}},}\ }\href
  {https://doi.org/10.1088/2632-2153/ace81a} {\bibfield  {journal} {\bibinfo
  {journal} {Mach. Learn.: Sci. Technol.}\ }\textbf {\bibinfo {volume} {4}},\
  \bibinfo {pages} {031001} (\bibinfo {year} {2023})}\BibitemShut {NoStop}%
\bibitem [{\citenamefont {Mehdi}\ \emph {et~al.}(2024)\citenamefont {Mehdi},
  \citenamefont {Smith}, \citenamefont {Herron}, \citenamefont {Zou},\ and\
  \citenamefont {Tiwary}}]{mehdi2024enhanced}%
  \BibitemOpen
  \bibfield  {author} {\bibinfo {author} {\bibfnamefont {S.}~\bibnamefont
  {Mehdi}}, \bibinfo {author} {\bibfnamefont {Z.}~\bibnamefont {Smith}},
  \bibinfo {author} {\bibfnamefont {L.}~\bibnamefont {Herron}}, \bibinfo
  {author} {\bibfnamefont {Z.}~\bibnamefont {Zou}},\ and\ \bibinfo {author}
  {\bibfnamefont {P.}~\bibnamefont {Tiwary}},\ }\bibfield  {title} {\enquote
  {\bibinfo {title} {{Enhanced Sampling with Machine Learning}},}\ }\href
  {https://doi.org/https://doi.org/10.1146/annurev-physchem-083122-125941}
  {\bibfield  {journal} {\bibinfo  {journal} {Annu. Rev. Phys. Chem.}\ }\textbf
  {\bibinfo {volume} {75}},\ \bibinfo {pages} {347–370} (\bibinfo {year}
  {2024})}\BibitemShut {NoStop}%
\bibitem [{\citenamefont {Zwanzig}(2001)}]{zwanzig2001nonequilibrium}%
  \BibitemOpen
  \bibfield  {author} {\bibinfo {author} {\bibfnamefont {R.}~\bibnamefont
  {Zwanzig}},\ }\href {https://doi.org/10.1093/oso/9780195140187.001.0001}
  {\emph {\bibinfo {title} {{Nonequilibrium Statistical Mechanics}}}},\
  \bibinfo {edition} {1st}\ ed.\ (\bibinfo  {publisher} {Oxford University
  Press},\ \bibinfo {year} {2001})\BibitemShut {NoStop}%
\bibitem [{\citenamefont {Peters}\ \emph {et~al.}(2013)\citenamefont {Peters},
  \citenamefont {Bolhuis}, \citenamefont {Mullen},\ and\ \citenamefont
  {Shea}}]{peters2013reaction}%
  \BibitemOpen
  \bibfield  {author} {\bibinfo {author} {\bibfnamefont {B.}~\bibnamefont
  {Peters}}, \bibinfo {author} {\bibfnamefont {P.~G.}\ \bibnamefont {Bolhuis}},
  \bibinfo {author} {\bibfnamefont {R.~G.}\ \bibnamefont {Mullen}},\ and\
  \bibinfo {author} {\bibfnamefont {J.-E.}\ \bibnamefont {Shea}},\ }\bibfield
  {title} {\enquote {\bibinfo {title} {{Reaction Coordinates, One-Dimensional
  Smoluchowski Equations, and a Test for Dynamical Self-Consistency}},}\ }\href
  {https://doi.org/https://doi.org/10.1063/1.4775807} {\bibfield  {journal}
  {\bibinfo  {journal} {J. Chem. Phys.}\ }\textbf {\bibinfo {volume} {138}},\
  \bibinfo {pages} {054106} (\bibinfo {year} {2013})}\BibitemShut {NoStop}%
\bibitem [{\citenamefont {Hummer}(2004)}]{hummer2004transition}%
  \BibitemOpen
  \bibfield  {author} {\bibinfo {author} {\bibfnamefont {G.}~\bibnamefont
  {Hummer}},\ }\bibfield  {title} {\enquote {\bibinfo {title} {{From Transition
  Paths to Transition States and Rate Coefficients}},}\ }\href
  {https://doi.org/https://doi.org/10.1063/1.1630572} {\bibfield  {journal}
  {\bibinfo  {journal} {J. Chem. Phys.}\ }\textbf {\bibinfo {volume} {120}},\
  \bibinfo {pages} {516--523} (\bibinfo {year} {2004})}\BibitemShut {NoStop}%
\bibitem [{\citenamefont {Coifman}\ \emph {et~al.}(2008)\citenamefont
  {Coifman}, \citenamefont {Kevrekidis}, \citenamefont {Lafon}, \citenamefont
  {Maggioni},\ and\ \citenamefont {Nadler}}]{coifman2008diffusion}%
  \BibitemOpen
  \bibfield  {author} {\bibinfo {author} {\bibfnamefont {R.~R.}\ \bibnamefont
  {Coifman}}, \bibinfo {author} {\bibfnamefont {I.~G.}\ \bibnamefont
  {Kevrekidis}}, \bibinfo {author} {\bibfnamefont {S.}~\bibnamefont {Lafon}},
  \bibinfo {author} {\bibfnamefont {M.}~\bibnamefont {Maggioni}},\ and\
  \bibinfo {author} {\bibfnamefont {B.}~\bibnamefont {Nadler}},\ }\bibfield
  {title} {\enquote {\bibinfo {title} {{Diffusion Maps, Reduction Coordinates,
  and Low Dimensional Representation of Stochastic Systems}},}\ }\href
  {https://doi.org/https://doi.org/10.1137/070696325} {\bibfield  {journal}
  {\bibinfo  {journal} {Multiscale Model. Simul.}\ }\textbf {\bibinfo {volume}
  {7}},\ \bibinfo {pages} {842--864} (\bibinfo {year} {2008})}\BibitemShut
  {NoStop}%
\bibitem [{\citenamefont {Berezhkovskii}\ and\ \citenamefont
  {Szabo}(2005)}]{berezhkovskii2005one}%
  \BibitemOpen
  \bibfield  {author} {\bibinfo {author} {\bibfnamefont {A.}~\bibnamefont
  {Berezhkovskii}}\ and\ \bibinfo {author} {\bibfnamefont {A.}~\bibnamefont
  {Szabo}},\ }\bibfield  {title} {\enquote {\bibinfo {title} {{One-Dimensional
  Reaction Coordinates for Diffusive Activated Rate Processes in Many
  Dimensions}},}\ }\href {https://doi.org/https://doi.org/10.1063/1.1818091}
  {\bibfield  {journal} {\bibinfo  {journal} {J. Chem. Phys.}\ }\textbf
  {\bibinfo {volume} {122}},\ \bibinfo {pages} {014503} (\bibinfo {year}
  {2005})}\BibitemShut {NoStop}%
\bibitem [{\citenamefont {Berezhkovskii}\ and\ \citenamefont
  {Szabo}(2011)}]{berezhkovskii2011time}%
  \BibitemOpen
  \bibfield  {author} {\bibinfo {author} {\bibfnamefont {A.}~\bibnamefont
  {Berezhkovskii}}\ and\ \bibinfo {author} {\bibfnamefont {A.}~\bibnamefont
  {Szabo}},\ }\bibfield  {title} {\enquote {\bibinfo {title} {{Time Scale
  Separation Leads to Position-Dependent Diffusion along a Slow Coordinate}},}\
  }\href {https://doi.org/https://doi.org/10.1063/1.3626215} {\bibfield
  {journal} {\bibinfo  {journal} {J. Chem. Phys.}\ }\textbf {\bibinfo {volume}
  {135}},\ \bibinfo {pages} {074108} (\bibinfo {year} {2011})}\BibitemShut
  {NoStop}%
\bibitem [{\citenamefont {Berezhkovskii}\ and\ \citenamefont
  {Makarov}(2018)}]{berezhkovskii2018single}%
  \BibitemOpen
  \bibfield  {author} {\bibinfo {author} {\bibfnamefont {A.~M.}\ \bibnamefont
  {Berezhkovskii}}\ and\ \bibinfo {author} {\bibfnamefont {D.~E.}\ \bibnamefont
  {Makarov}},\ }\bibfield  {title} {\enquote {\bibinfo {title}
  {{Single-Molecule Test for Markovianity of the Dynamics along a Reaction
  Coordinate}},}\ }\href
  {https://doi.org/https://doi.org/10.1021/acs.jpclett.8b00956} {\bibfield
  {journal} {\bibinfo  {journal} {J. Phys. Chem. Lett.}\ }\textbf {\bibinfo
  {volume} {9}},\ \bibinfo {pages} {2190--2195} (\bibinfo {year}
  {2018})}\BibitemShut {NoStop}%
\bibitem [{\citenamefont {Torrie}\ and\ \citenamefont
  {Valleau}(1977)}]{torrie1977nonphysical}%
  \BibitemOpen
  \bibfield  {author} {\bibinfo {author} {\bibfnamefont {G.~M.}\ \bibnamefont
  {Torrie}}\ and\ \bibinfo {author} {\bibfnamefont {J.~P.}\ \bibnamefont
  {Valleau}},\ }\bibfield  {title} {\enquote {\bibinfo {title} {{Nonphysical
  Sampling Distributions in Monte Carlo Free-Energy Estimation: Umbrella
  Sampling}},}\ }\href
  {https://doi.org/https://doi.org/10.1016/0021-9991(77)90121-8} {\bibfield
  {journal} {\bibinfo  {journal} {J. Comp. Phys.}\ }\textbf {\bibinfo {volume}
  {23}},\ \bibinfo {pages} {187--199} (\bibinfo {year} {1977})}\BibitemShut
  {NoStop}%
\bibitem [{\citenamefont {Mezei}(1987)}]{mezei1987adaptive}%
  \BibitemOpen
  \bibfield  {author} {\bibinfo {author} {\bibfnamefont {M.}~\bibnamefont
  {Mezei}},\ }\bibfield  {title} {\enquote {\bibinfo {title} {{Adaptive
  Umbrella Sampling: Self-Consistent Determination of the Non-Boltzmann
  Bias}},}\ }\href
  {https://doi.org/https://doi.org/10.1016/0021-9991(87)90054-4} {\bibfield
  {journal} {\bibinfo  {journal} {J. Comput. Phys.}\ }\textbf {\bibinfo
  {volume} {68}},\ \bibinfo {pages} {237--248} (\bibinfo {year}
  {1987})}\BibitemShut {NoStop}%
\bibitem [{\citenamefont {Maragakis}, \citenamefont {van~der Vaart},\ and\
  \citenamefont {Karplus}(2009)}]{Maragakis-JPCB-2009}%
  \BibitemOpen
  \bibfield  {author} {\bibinfo {author} {\bibfnamefont {P.}~\bibnamefont
  {Maragakis}}, \bibinfo {author} {\bibfnamefont {A.}~\bibnamefont {van~der
  Vaart}},\ and\ \bibinfo {author} {\bibfnamefont {M.}~\bibnamefont
  {Karplus}},\ }\bibfield  {title} {\enquote {\bibinfo {title}
  {{Gaussian-Mixture Umbrella Sampling}},}\ }\href
  {https://doi.org/https://doi.org/10.1021/jp808381s} {\bibfield  {journal}
  {\bibinfo  {journal} {J. Phys. Chem. B}\ }\textbf {\bibinfo {volume} {113}},\
  \bibinfo {pages} {4664--4673} (\bibinfo {year} {2009})}\BibitemShut {NoStop}%
\bibitem [{\citenamefont {K\"{a}stner}(2011)}]{Kastner2011umbreallsampling}%
  \BibitemOpen
  \bibfield  {author} {\bibinfo {author} {\bibfnamefont {J.}~\bibnamefont
  {K\"{a}stner}},\ }\bibfield  {title} {\enquote {\bibinfo {title} {{Umbrella
  Sampling}},}\ }\href {https://doi.org/https://doi.org/10.1002/wcms.66}
  {\bibfield  {journal} {\bibinfo  {journal} {Wiley Interdiscip. Rev. Comput.
  Mol. Sci.}\ }\textbf {\bibinfo {volume} {1}},\ \bibinfo {pages} {932--942}
  (\bibinfo {year} {2011})}\BibitemShut {NoStop}%
\bibitem [{\citenamefont {Laio}\ and\ \citenamefont
  {Parrinello}(2002)}]{laio2002escaping}%
  \BibitemOpen
  \bibfield  {author} {\bibinfo {author} {\bibfnamefont {A.}~\bibnamefont
  {Laio}}\ and\ \bibinfo {author} {\bibfnamefont {M.}~\bibnamefont
  {Parrinello}},\ }\bibfield  {title} {\enquote {\bibinfo {title} {{Escaping
  Free-Energy Minima}},}\ }\href
  {https://doi.org/https://doi.org/10.1073/pnas.202427399} {\bibfield
  {journal} {\bibinfo  {journal} {Proc. Natl. Acad. Sci. U.S.A.}\ }\textbf
  {\bibinfo {volume} {99}},\ \bibinfo {pages} {12562--12566} (\bibinfo {year}
  {2002})}\BibitemShut {NoStop}%
\bibitem [{\citenamefont {Barducci}, \citenamefont {Bussi},\ and\ \citenamefont
  {Parrinello}(2008)}]{barducci2008well}%
  \BibitemOpen
  \bibfield  {author} {\bibinfo {author} {\bibfnamefont {A.}~\bibnamefont
  {Barducci}}, \bibinfo {author} {\bibfnamefont {G.}~\bibnamefont {Bussi}},\
  and\ \bibinfo {author} {\bibfnamefont {M.}~\bibnamefont {Parrinello}},\
  }\bibfield  {title} {\enquote {\bibinfo {title} {{Well-Tempered Metadynamics:
  A Smoothly Converging and Tunable Free-Energy Method}},}\ }\href
  {https://doi.org/https://doi.org/10.1103/PhysRevLett.100.020603} {\bibfield
  {journal} {\bibinfo  {journal} {Phys. Rev. Lett.}\ }\textbf {\bibinfo
  {volume} {100}},\ \bibinfo {pages} {020603} (\bibinfo {year}
  {2008})}\BibitemShut {NoStop}%
\bibitem [{\citenamefont {Invernizzi}\ and\ \citenamefont
  {Parrinello}(2020)}]{Invernizzi2020opus}%
  \BibitemOpen
  \bibfield  {author} {\bibinfo {author} {\bibfnamefont {M.}~\bibnamefont
  {Invernizzi}}\ and\ \bibinfo {author} {\bibfnamefont {M.}~\bibnamefont
  {Parrinello}},\ }\bibfield  {title} {\enquote {\bibinfo {title} {{Rethinking
  Metadynamics: From Bias Potentials to Probability Distributions}},}\ }\href
  {https://doi.org/10.1021/acs.jpclett.0c00497} {\bibfield  {journal} {\bibinfo
   {journal} {J. Phys. Chem. Lett.}\ }\textbf {\bibinfo {volume} {11}},\
  \bibinfo {pages} {2731--2736} (\bibinfo {year} {2020})}\BibitemShut {NoStop}%
\bibitem [{\citenamefont {Invernizzi}, \citenamefont {Piaggi},\ and\
  \citenamefont {Parrinello}(2020)}]{invernizzi2020unified}%
  \BibitemOpen
  \bibfield  {author} {\bibinfo {author} {\bibfnamefont {M.}~\bibnamefont
  {Invernizzi}}, \bibinfo {author} {\bibfnamefont {P.~M.}\ \bibnamefont
  {Piaggi}},\ and\ \bibinfo {author} {\bibfnamefont {M.}~\bibnamefont
  {Parrinello}},\ }\bibfield  {title} {\enquote {\bibinfo {title} {{Unified
  Approach to Enhanced Sampling}},}\ }\href
  {https://doi.org/https://doi.org/10.1103/PhysRevX.10.041034} {\bibfield
  {journal} {\bibinfo  {journal} {Phys. Rev. X}\ }\textbf {\bibinfo {volume}
  {10}},\ \bibinfo {pages} {041034} (\bibinfo {year} {2020})}\BibitemShut
  {NoStop}%
\bibitem [{\citenamefont {Valsson}\ and\ \citenamefont
  {Parrinello}(2014)}]{valsson2014variational}%
  \BibitemOpen
  \bibfield  {author} {\bibinfo {author} {\bibfnamefont {O.}~\bibnamefont
  {Valsson}}\ and\ \bibinfo {author} {\bibfnamefont {M.}~\bibnamefont
  {Parrinello}},\ }\bibfield  {title} {\enquote {\bibinfo {title} {{Variational
  Approach to Enhanced Sampling and Free Energy Calculations}},}\ }\href
  {https://doi.org/10.1103/PhysRevLett.113.090601} {\bibfield  {journal}
  {\bibinfo  {journal} {Phys. Rev. Lett.}\ }\textbf {\bibinfo {volume} {113}},\
  \bibinfo {pages} {090601} (\bibinfo {year} {2014})}\BibitemShut {NoStop}%
\bibitem [{\citenamefont {Yang}\ and\ \citenamefont
  {Parrinello}(2018)}]{yang2018refining}%
  \BibitemOpen
  \bibfield  {author} {\bibinfo {author} {\bibfnamefont {Y.~I.}\ \bibnamefont
  {Yang}}\ and\ \bibinfo {author} {\bibfnamefont {M.}~\bibnamefont
  {Parrinello}},\ }\bibfield  {title} {\enquote {\bibinfo {title} {{Refining
  Collective Coordinates and Improving Free Energy Representation in
  Variational Enhanced Sampling}},}\ }\href
  {https://doi.org/https://doi.org/10.1021/acs.jctc.8b00231} {\bibfield
  {journal} {\bibinfo  {journal} {J. Chem. Theory Comput.}\ }\textbf {\bibinfo
  {volume} {14}},\ \bibinfo {pages} {2889--2894} (\bibinfo {year}
  {2018})}\BibitemShut {NoStop}%
\bibitem [{\citenamefont {Bonati}, \citenamefont {Zhang},\ and\ \citenamefont
  {Parrinello}(2019)}]{bonati2019neural}%
  \BibitemOpen
  \bibfield  {author} {\bibinfo {author} {\bibfnamefont {L.}~\bibnamefont
  {Bonati}}, \bibinfo {author} {\bibfnamefont {Y.-Y.}\ \bibnamefont {Zhang}},\
  and\ \bibinfo {author} {\bibfnamefont {M.}~\bibnamefont {Parrinello}},\
  }\bibfield  {title} {\enquote {\bibinfo {title} {{Neural Networks-Based
  Variationally Enhanced Sampling}},}\ }\href
  {https://doi.org/https://doi.org/10.1073/pnas.1907975116} {\bibfield
  {journal} {\bibinfo  {journal} {Proc. Natl. Acad. Sci. U.S.A.}\ }\textbf
  {\bibinfo {volume} {116}},\ \bibinfo {pages} {17641--17647} (\bibinfo {year}
  {2019})}\BibitemShut {NoStop}%
\bibitem [{\citenamefont {Roux}(2022)}]{roux2022transition}%
  \BibitemOpen
  \bibfield  {author} {\bibinfo {author} {\bibfnamefont {B.}~\bibnamefont
  {Roux}},\ }\bibfield  {title} {\enquote {\bibinfo {title} {{Transition Rate
  Theory, Spectral Analysis, and Reactive Paths}},}\ }\href
  {https://doi.org/https://doi.org/10.1063/5.0084209} {\bibfield  {journal}
  {\bibinfo  {journal} {J. Chem. Phys.}\ }\textbf {\bibinfo {volume} {156}},\
  \bibinfo {pages} {134111} (\bibinfo {year} {2022})}\BibitemShut {NoStop}%
\bibitem [{\citenamefont {Shuler}(1959)}]{shuler1959relaxation}%
  \BibitemOpen
  \bibfield  {author} {\bibinfo {author} {\bibfnamefont {K.~E.}\ \bibnamefont
  {Shuler}},\ }\bibfield  {title} {\enquote {\bibinfo {title} {{Relaxation
  Processes in Multistate Systems}},}\ }\href
  {https://doi.org/https://doi.org/10.1063/1.1724416} {\bibfield  {journal}
  {\bibinfo  {journal} {Phys. Fluids}\ }\textbf {\bibinfo {volume} {2}},\
  \bibinfo {pages} {442--448} (\bibinfo {year} {1959})}\BibitemShut {NoStop}%
\bibitem [{\citenamefont {Risken}(1996)}]{risken1996fokker}%
  \BibitemOpen
  \bibfield  {author} {\bibinfo {author} {\bibfnamefont {H.}~\bibnamefont
  {Risken}},\ }\href@noop {} {\emph {\bibinfo {title} {{Fokker--Planck
  Equation}}}}\ (\bibinfo  {publisher} {Springer},\ \bibinfo {year}
  {1996})\BibitemShut {NoStop}%
\bibitem [{\citenamefont {Bovier}\ \emph {et~al.}(2002)\citenamefont {Bovier},
  \citenamefont {Eckhoff}, \citenamefont {Gayrard},\ and\ \citenamefont
  {Klein}}]{bovier2002metastability}%
  \BibitemOpen
  \bibfield  {author} {\bibinfo {author} {\bibfnamefont {A.}~\bibnamefont
  {Bovier}}, \bibinfo {author} {\bibfnamefont {M.}~\bibnamefont {Eckhoff}},
  \bibinfo {author} {\bibfnamefont {V.}~\bibnamefont {Gayrard}},\ and\ \bibinfo
  {author} {\bibfnamefont {M.}~\bibnamefont {Klein}},\ }\bibfield  {title}
  {\enquote {\bibinfo {title} {{Metastability and Low Lying Spectra in
  Reversible Markov Chains}},}\ }\href
  {https://doi.org/https://doi.org/10.1007/s002200200609} {\bibfield  {journal}
  {\bibinfo  {journal} {Commun. Math. Phys.}\ }\textbf {\bibinfo {volume}
  {228}},\ \bibinfo {pages} {219--255} (\bibinfo {year} {2002})}\BibitemShut
  {NoStop}%
\bibitem [{\citenamefont {Gaveau}\ and\ \citenamefont
  {Schulman}(1996)}]{gaveau1996master}%
  \BibitemOpen
  \bibfield  {author} {\bibinfo {author} {\bibfnamefont {B.}~\bibnamefont
  {Gaveau}}\ and\ \bibinfo {author} {\bibfnamefont {L.~S.}\ \bibnamefont
  {Schulman}},\ }\bibfield  {title} {\enquote {\bibinfo {title} {{Master
  Equation based Formulation of Nonequilibrium Statistical Mechanics}},}\
  }\href {https://doi.org/https://doi.org/10.1063/1.531608} {\bibfield
  {journal} {\bibinfo  {journal} {J. Math. Phys.}\ }\textbf {\bibinfo {volume}
  {37}},\ \bibinfo {pages} {3897--3932} (\bibinfo {year} {1996})}\BibitemShut
  {NoStop}%
\bibitem [{\citenamefont {Gaveau}\ and\ \citenamefont
  {Schulman}(1998)}]{gaveau1998theory}%
  \BibitemOpen
  \bibfield  {author} {\bibinfo {author} {\bibfnamefont {B.}~\bibnamefont
  {Gaveau}}\ and\ \bibinfo {author} {\bibfnamefont {L.~S.}\ \bibnamefont
  {Schulman}},\ }\bibfield  {title} {\enquote {\bibinfo {title} {{Theory of
  Nonequilibrium First-Order Phase Transitions for Stochastic Dynamics}},}\
  }\href {https://doi.org/https://doi.org/10.1063/1.532394} {\bibfield
  {journal} {\bibinfo  {journal} {J. Math. Phys.}\ }\textbf {\bibinfo {volume}
  {39}},\ \bibinfo {pages} {1517--1533} (\bibinfo {year} {1998})}\BibitemShut
  {NoStop}%
\bibitem [{\citenamefont {Davies}(1983)}]{davies1983spectral}%
  \BibitemOpen
  \bibfield  {author} {\bibinfo {author} {\bibfnamefont {E.~B.}\ \bibnamefont
  {Davies}},\ }\bibfield  {title} {\enquote {\bibinfo {title} {{Spectral
  Properties of Metastable Markov Semigroups}},}\ }\href@noop {} {\bibfield
  {journal} {\bibinfo  {journal} {J. Funct. Anal.}\ }\textbf {\bibinfo {volume}
  {52}},\ \bibinfo {pages} {315--329} (\bibinfo {year} {1983})}\BibitemShut
  {NoStop}%
\bibitem [{\citenamefont
  {Davies}(1982{\natexlab{a}})}]{davies1982metastable-1}%
  \BibitemOpen
  \bibfield  {author} {\bibinfo {author} {\bibfnamefont {E.~B.}\ \bibnamefont
  {Davies}},\ }\bibfield  {title} {\enquote {\bibinfo {title} {{Metastable
  States of Symmetric Markov Semigroups I}},}\ }\href@noop {} {\bibfield
  {journal} {\bibinfo  {journal} {Proc. London Math. Soc.}\ }\textbf {\bibinfo
  {volume} {3}},\ \bibinfo {pages} {133--150} (\bibinfo {year}
  {1982}{\natexlab{a}})}\BibitemShut {NoStop}%
\bibitem [{\citenamefont
  {Davies}(1982{\natexlab{b}})}]{davies1982metastable-2}%
  \BibitemOpen
  \bibfield  {author} {\bibinfo {author} {\bibfnamefont {E.~B.}\ \bibnamefont
  {Davies}},\ }\bibfield  {title} {\enquote {\bibinfo {title} {{Metastable
  States of Symmetric Markov Semigroups II}},}\ }\href@noop {} {\bibfield
  {journal} {\bibinfo  {journal} {J. London Math. Soc.}\ }\textbf {\bibinfo
  {volume} {2}},\ \bibinfo {pages} {541--556} (\bibinfo {year}
  {1982}{\natexlab{b}})}\BibitemShut {NoStop}%
\bibitem [{\citenamefont {Bovier}\ and\ \citenamefont
  {Den~Hollander}(2016)}]{bovier2016metastability}%
  \BibitemOpen
  \bibfield  {author} {\bibinfo {author} {\bibfnamefont {A.}~\bibnamefont
  {Bovier}}\ and\ \bibinfo {author} {\bibfnamefont {F.}~\bibnamefont
  {Den~Hollander}},\ }\href@noop {} {\emph {\bibinfo {title} {{Metastability: A
  Potential-Theoretic Approach}}}},\ Vol.\ \bibinfo {volume} {351}\ (\bibinfo
  {publisher} {Springer},\ \bibinfo {year} {2016})\BibitemShut {NoStop}%
\bibitem [{\citenamefont {Darve}\ and\ \citenamefont
  {Pohorille}(2001)}]{Darve-JCP-2001}%
  \BibitemOpen
  \bibfield  {author} {\bibinfo {author} {\bibfnamefont {E.}~\bibnamefont
  {Darve}}\ and\ \bibinfo {author} {\bibfnamefont {A.}~\bibnamefont
  {Pohorille}},\ }\bibfield  {title} {\enquote {\bibinfo {title} {{Calculating
  Free Energies using Average Force}},}\ }\href
  {https://doi.org/https://doi.org/10.1063/1.1410978} {\bibfield  {journal}
  {\bibinfo  {journal} {J. Chem. Phys.}\ }\textbf {\bibinfo {volume} {115}},\
  \bibinfo {pages} {9169} (\bibinfo {year} {2001})}\BibitemShut {NoStop}%
\bibitem [{\citenamefont {Rosso}\ \emph {et~al.}(2002)\citenamefont {Rosso},
  \citenamefont {Min{\'a}ry}, \citenamefont {Zhu},\ and\ \citenamefont
  {Tuckerman}}]{rosso2002use}%
  \BibitemOpen
  \bibfield  {author} {\bibinfo {author} {\bibfnamefont {L.}~\bibnamefont
  {Rosso}}, \bibinfo {author} {\bibfnamefont {P.}~\bibnamefont {Min{\'a}ry}},
  \bibinfo {author} {\bibfnamefont {Z.}~\bibnamefont {Zhu}},\ and\ \bibinfo
  {author} {\bibfnamefont {M.~E.}\ \bibnamefont {Tuckerman}},\ }\bibfield
  {title} {\enquote {\bibinfo {title} {{On the Use of the Adiabatic Molecular
  Dynamics Technique in the Calculation of Free Energy Profiles}},}\ }\href
  {https://doi.org/https://doi.org/10.1063/1.1448491} {\bibfield  {journal}
  {\bibinfo  {journal} {J. Chem. Phys.}\ }\textbf {\bibinfo {volume} {116}},\
  \bibinfo {pages} {4389--4402} (\bibinfo {year} {2002})}\BibitemShut {NoStop}%
\bibitem [{\citenamefont {Morishita}\ \emph {et~al.}(2012)\citenamefont
  {Morishita}, \citenamefont {Itoh}, \citenamefont {Okumura},\ and\
  \citenamefont {Mikami}}]{morishita2012free}%
  \BibitemOpen
  \bibfield  {author} {\bibinfo {author} {\bibfnamefont {T.}~\bibnamefont
  {Morishita}}, \bibinfo {author} {\bibfnamefont {S.~G.}\ \bibnamefont {Itoh}},
  \bibinfo {author} {\bibfnamefont {H.}~\bibnamefont {Okumura}},\ and\ \bibinfo
  {author} {\bibfnamefont {M.}~\bibnamefont {Mikami}},\ }\bibfield  {title}
  {\enquote {\bibinfo {title} {{Free-Energy Calculation via Mean-Force Dynamics
  using a Logarithmic Energy Landscape}},}\ }\href
  {https://doi.org/http://dx.doi.org/10.1103/PhysRevE.85.066702} {\bibfield
  {journal} {\bibinfo  {journal} {Phys. Rev. E}\ }\textbf {\bibinfo {volume}
  {85}},\ \bibinfo {pages} {066702} (\bibinfo {year} {2012})}\BibitemShut
  {NoStop}%
\bibitem [{\citenamefont {Giberti}\ \emph {et~al.}(2020)\citenamefont
  {Giberti}, \citenamefont {Cheng}, \citenamefont {Tribello},\ and\
  \citenamefont {Ceriotti}}]{giberti2020iterative}%
  \BibitemOpen
  \bibfield  {author} {\bibinfo {author} {\bibfnamefont {F.}~\bibnamefont
  {Giberti}}, \bibinfo {author} {\bibfnamefont {B.}~\bibnamefont {Cheng}},
  \bibinfo {author} {\bibfnamefont {G.~A.}\ \bibnamefont {Tribello}},\ and\
  \bibinfo {author} {\bibfnamefont {M.}~\bibnamefont {Ceriotti}},\ }\bibfield
  {title} {\enquote {\bibinfo {title} {{Iterative Unbiasing of
  Quasi-Equilibrium Sampling}},}\ }\href
  {https://doi.org/10.1021/acs.jctc.9b00907} {\bibfield  {journal} {\bibinfo
  {journal} {J. Chem. Theory Comput.}\ }\textbf {\bibinfo {volume} {16}},\
  \bibinfo {pages} {100--107} (\bibinfo {year} {2020})}\BibitemShut {NoStop}%
\bibitem [{\citenamefont {Tiwary}\ and\ \citenamefont
  {Parrinello}(2015)}]{tiwary2015time}%
  \BibitemOpen
  \bibfield  {author} {\bibinfo {author} {\bibfnamefont {P.}~\bibnamefont
  {Tiwary}}\ and\ \bibinfo {author} {\bibfnamefont {M.}~\bibnamefont
  {Parrinello}},\ }\bibfield  {title} {\enquote {\bibinfo {title} {{A
  Time-Independent Free Energy Estimator for Metadynamics}},}\ }\href
  {https://doi.org/10.1021/jp504920s} {\bibfield  {journal} {\bibinfo
  {journal} {J. Phys. Chem. B}\ }\textbf {\bibinfo {volume} {119}},\ \bibinfo
  {pages} {736--742} (\bibinfo {year} {2015})}\BibitemShut {NoStop}%
\bibitem [{\citenamefont {Bonomi}, \citenamefont {Barducci},\ and\
  \citenamefont {Parrinello}(2009)}]{bonomi2009reconstructing}%
  \BibitemOpen
  \bibfield  {author} {\bibinfo {author} {\bibfnamefont {M.}~\bibnamefont
  {Bonomi}}, \bibinfo {author} {\bibfnamefont {A.}~\bibnamefont {Barducci}},\
  and\ \bibinfo {author} {\bibfnamefont {M.}~\bibnamefont {Parrinello}},\
  }\bibfield  {title} {\enquote {\bibinfo {title} {{Reconstructing the
  Equilibrium Boltzmann Distribution from Well-Tempered Metadynamics}},}\
  }\href {https://doi.org/https://doi.org/10.1002/jcc.21305} {\bibfield
  {journal} {\bibinfo  {journal} {J. Comput. Chem.}\ }\textbf {\bibinfo
  {volume} {30}},\ \bibinfo {pages} {1615--1621} (\bibinfo {year}
  {2009})}\BibitemShut {NoStop}%
\bibitem [{\citenamefont {Schäfer}\ and\ \citenamefont
  {Settanni}(2020)}]{Sch_fer_2020}%
  \BibitemOpen
  \bibfield  {author} {\bibinfo {author} {\bibfnamefont {T.~M.}\ \bibnamefont
  {Schäfer}}\ and\ \bibinfo {author} {\bibfnamefont {G.}~\bibnamefont
  {Settanni}},\ }\bibfield  {title} {\enquote {\bibinfo {title} {{Data
  Reweighting in Metadynamics Simulations}},}\ }\href
  {https://doi.org/10.1021/acs.jctc.9b00867} {\bibfield  {journal} {\bibinfo
  {journal} {J. Chem. Theory Comput.}\ }\textbf {\bibinfo {volume} {16}},\
  \bibinfo {pages} {2042--2052} (\bibinfo {year} {2020})}\BibitemShut {NoStop}%
\bibitem [{\citenamefont {Linker}, \citenamefont {Wei{\ss}},\ and\
  \citenamefont {Riniker}(2020)}]{linker2020connecting}%
  \BibitemOpen
  \bibfield  {author} {\bibinfo {author} {\bibfnamefont {S.~M.}\ \bibnamefont
  {Linker}}, \bibinfo {author} {\bibfnamefont {R.~G.}\ \bibnamefont
  {Wei{\ss}}},\ and\ \bibinfo {author} {\bibfnamefont {S.}~\bibnamefont
  {Riniker}},\ }\bibfield  {title} {\enquote {\bibinfo {title} {{Connecting
  Dynamic Reweighting Algorithms: Derivation of the Dynamic Reweighting Family
  Tree}},}\ }\href {https://doi.org/https://doi.org/10.1063/5.0019687}
  {\bibfield  {journal} {\bibinfo  {journal} {J. Chem. Phys.}\ }\textbf
  {\bibinfo {volume} {153}},\ \bibinfo {pages} {234106} (\bibinfo {year}
  {2020})}\BibitemShut {NoStop}%
\bibitem [{\citenamefont {Kamenik}, \citenamefont {Linker},\ and\ \citenamefont
  {Riniker}(2022)}]{kamenik2021enhanced}%
  \BibitemOpen
  \bibfield  {author} {\bibinfo {author} {\bibfnamefont {A.~S.}\ \bibnamefont
  {Kamenik}}, \bibinfo {author} {\bibfnamefont {S.~M.}\ \bibnamefont
  {Linker}},\ and\ \bibinfo {author} {\bibfnamefont {S.}~\bibnamefont
  {Riniker}},\ }\bibfield  {title} {\enquote {\bibinfo {title} {{Enhanced
  Sampling without Borders: On Global Biasing Functions and how to Reweight
  them}},}\ }\href {https://doi.org/https://doi.org/10.1039/D1CP04809K}
  {\bibfield  {journal} {\bibinfo  {journal} {Phys. Chem. Chem. Phys.}\
  }\textbf {\bibinfo {volume} {24}},\ \bibinfo {pages} {1225--1236} (\bibinfo
  {year} {2022})}\BibitemShut {NoStop}%
\bibitem [{\citenamefont {Pérez-Hernández}\ \emph {et~al.}(2013)\citenamefont
  {Pérez-Hernández}, \citenamefont {Paul}, \citenamefont {Giorgino},
  \citenamefont {De~Fabritiis},\ and\ \citenamefont
  {Noé}}]{hernandez2013identification}%
  \BibitemOpen
  \bibfield  {author} {\bibinfo {author} {\bibfnamefont {G.}~\bibnamefont
  {Pérez-Hernández}}, \bibinfo {author} {\bibfnamefont {F.}~\bibnamefont
  {Paul}}, \bibinfo {author} {\bibfnamefont {T.}~\bibnamefont {Giorgino}},
  \bibinfo {author} {\bibfnamefont {G.}~\bibnamefont {De~Fabritiis}},\ and\
  \bibinfo {author} {\bibfnamefont {F.}~\bibnamefont {Noé}},\ }\bibfield
  {title} {\enquote {\bibinfo {title} {{Identification of Slow Molecular Order
  Parameters for Markov Model Construction}},}\ }\href
  {https://doi.org/10.1063/1.4811489} {\bibfield  {journal} {\bibinfo
  {journal} {J. Chem. Phys.}\ }\textbf {\bibinfo {volume} {139}},\ \bibinfo
  {pages} {015102} (\bibinfo {year} {2013})}\BibitemShut {NoStop}%
\bibitem [{\citenamefont {Wehmeyer}\ and\ \citenamefont
  {No{\'e}}(2018)}]{wehmeyer2018time}%
  \BibitemOpen
  \bibfield  {author} {\bibinfo {author} {\bibfnamefont {C.}~\bibnamefont
  {Wehmeyer}}\ and\ \bibinfo {author} {\bibfnamefont {F.}~\bibnamefont
  {No{\'e}}},\ }\bibfield  {title} {\enquote {\bibinfo {title} {{Time-Lagged
  Autoencoders: Deep Learning of Slow Collective Variables for Molecular
  Kinetics}},}\ }\href {https://doi.org/https://doi.org/10.1063/1.5011399}
  {\bibfield  {journal} {\bibinfo  {journal} {J. Chem. Phys.}\ }\textbf
  {\bibinfo {volume} {148}},\ \bibinfo {pages} {241703} (\bibinfo {year}
  {2018})}\BibitemShut {NoStop}%
\bibitem [{\citenamefont {McCarty}\ and\ \citenamefont
  {Parrinello}(2017)}]{mccarty2017variational}%
  \BibitemOpen
  \bibfield  {author} {\bibinfo {author} {\bibfnamefont {J.}~\bibnamefont
  {McCarty}}\ and\ \bibinfo {author} {\bibfnamefont {M.}~\bibnamefont
  {Parrinello}},\ }\bibfield  {title} {\enquote {\bibinfo {title} {{A
  Variational Conformational Dynamics Approach to the Selection of Collective
  Variables in Metadynamics}},}\ }\href
  {https://doi.org/https://doi.org/10.1063/1.4998598} {\bibfield  {journal}
  {\bibinfo  {journal} {J. Chem. Phys.}\ }\textbf {\bibinfo {volume} {147}},\
  \bibinfo {pages} {204109} (\bibinfo {year} {2017})}\BibitemShut {NoStop}%
\bibitem [{\citenamefont {Bonati}, \citenamefont {Piccini},\ and\ \citenamefont
  {Parrinello}(2021)}]{bonati2021deep}%
  \BibitemOpen
  \bibfield  {author} {\bibinfo {author} {\bibfnamefont {L.}~\bibnamefont
  {Bonati}}, \bibinfo {author} {\bibfnamefont {G.}~\bibnamefont {Piccini}},\
  and\ \bibinfo {author} {\bibfnamefont {M.}~\bibnamefont {Parrinello}},\
  }\bibfield  {title} {\enquote {\bibinfo {title} {{Deep Learning the Slow
  Modes for Rare Events Sampling}},}\ }\href
  {https://doi.org/10.1073/pnas.2113533118} {\bibfield  {journal} {\bibinfo
  {journal} {Proc. Natl. Acad. Sci. U.S.A.}\ }\textbf {\bibinfo {volume}
  {118}},\ \bibinfo {pages} {e2113533118} (\bibinfo {year} {2021})}\BibitemShut
  {NoStop}%
\bibitem [{\citenamefont {Mardt}\ \emph {et~al.}(2018)\citenamefont {Mardt},
  \citenamefont {Pasquali}, \citenamefont {Wu},\ and\ \citenamefont
  {No{\'e}}}]{mardt2018vampnets}%
  \BibitemOpen
  \bibfield  {author} {\bibinfo {author} {\bibfnamefont {A.}~\bibnamefont
  {Mardt}}, \bibinfo {author} {\bibfnamefont {L.}~\bibnamefont {Pasquali}},
  \bibinfo {author} {\bibfnamefont {H.}~\bibnamefont {Wu}},\ and\ \bibinfo
  {author} {\bibfnamefont {F.}~\bibnamefont {No{\'e}}},\ }\bibfield  {title}
  {\enquote {\bibinfo {title} {{VAMPnets for Deep Learning of Molecular
  Kinetics}},}\ }\href {https://doi.org/10.1038/s41467-017-02388-1} {\bibfield
  {journal} {\bibinfo  {journal} {Nat. Commun.}\ }\textbf {\bibinfo {volume}
  {9}},\ \bibinfo {pages} {5} (\bibinfo {year} {2018})}\BibitemShut {NoStop}%
\bibitem [{\citenamefont {Chen}, \citenamefont {Sidky},\ and\ \citenamefont
  {Ferguson}(2019{\natexlab{a}})}]{chen2019nonlinear}%
  \BibitemOpen
  \bibfield  {author} {\bibinfo {author} {\bibfnamefont {W.}~\bibnamefont
  {Chen}}, \bibinfo {author} {\bibfnamefont {H.}~\bibnamefont {Sidky}},\ and\
  \bibinfo {author} {\bibfnamefont {A.}~\bibnamefont {Ferguson}},\ }\bibfield
  {title} {\enquote {\bibinfo {title} {{Nonlinear Discovery of Slow Molecular
  Modes using State-Free Reversible VAMPnets}},}\ }\href
  {https://doi.org/https://doi.org/10.1063/1.5092521} {\bibfield  {journal}
  {\bibinfo  {journal} {J. Chem. Phys.}\ }\textbf {\bibinfo {volume} {150}},\
  \bibinfo {pages} {214114} (\bibinfo {year} {2019}{\natexlab{a}})}\BibitemShut
  {NoStop}%
\bibitem [{\citenamefont {Shi}\ and\ \citenamefont
  {Malik}(2000)}]{shi2000normalized}%
  \BibitemOpen
  \bibfield  {author} {\bibinfo {author} {\bibfnamefont {J.}~\bibnamefont
  {Shi}}\ and\ \bibinfo {author} {\bibfnamefont {J.}~\bibnamefont {Malik}},\
  }\bibfield  {title} {\enquote {\bibinfo {title} {{Normalized Cuts and Image
  Segmentation}},}\ }\href
  {https://doi.org/https://doi.org/10.1109/cvpr.1997.609407} {\bibfield
  {journal} {\bibinfo  {journal} {IEEE Trans. Pattern Anal. Mach. Intell.}\
  }\textbf {\bibinfo {volume} {22}},\ \bibinfo {pages} {888--905} (\bibinfo
  {year} {2000})}\BibitemShut {NoStop}%
\bibitem [{\citenamefont {Belkin}\ and\ \citenamefont
  {Niyogi}(2001)}]{belkin2001laplacian}%
  \BibitemOpen
  \bibfield  {author} {\bibinfo {author} {\bibfnamefont {M.}~\bibnamefont
  {Belkin}}\ and\ \bibinfo {author} {\bibfnamefont {P.}~\bibnamefont
  {Niyogi}},\ }\bibfield  {title} {\enquote {\bibinfo {title} {{Laplacian
  Eigenmaps and Spectral Techniques for Embedding and Clustering}},}\
  }\bibfield  {booktitle} {\emph {\bibinfo {booktitle} {Adv. Neural Inf.
  Process. Syst.}},\ }\href
  {https://doi.org/https://proceedings.neurips.cc/paper_files/paper/2001/file/f106b7f99d2cb30c3db1c3cc0fde9ccb-Paper.pdf}
  {\ \textbf {\bibinfo {volume} {14}},\ \bibinfo {pages} {585--591} (\bibinfo
  {year} {2001})}\BibitemShut {NoStop}%
\bibitem [{\citenamefont {Belkin}\ and\ \citenamefont
  {Niyogi}(2003)}]{belkin2003laplacian}%
  \BibitemOpen
  \bibfield  {author} {\bibinfo {author} {\bibfnamefont {M.}~\bibnamefont
  {Belkin}}\ and\ \bibinfo {author} {\bibfnamefont {P.}~\bibnamefont
  {Niyogi}},\ }\bibfield  {title} {\enquote {\bibinfo {title} {{Laplacian
  Eigenmaps for Dimensionality Reduction and Data Representation}},}\ }\href
  {https://doi.org/https://doi.org/10.1162/089976603321780317} {\bibfield
  {journal} {\bibinfo  {journal} {Neural Comput.}\ }\textbf {\bibinfo {volume}
  {15}},\ \bibinfo {pages} {1373--1396} (\bibinfo {year} {2003})}\BibitemShut
  {NoStop}%
\bibitem [{\citenamefont {Belkin}\ and\ \citenamefont
  {Niyogi}(2004)}]{belkin2004semi}%
  \BibitemOpen
  \bibfield  {author} {\bibinfo {author} {\bibfnamefont {M.}~\bibnamefont
  {Belkin}}\ and\ \bibinfo {author} {\bibfnamefont {P.}~\bibnamefont
  {Niyogi}},\ }\bibfield  {title} {\enquote {\bibinfo {title} {{Semi-Supervised
  Learning on Riemannian Manifolds}},}\ }\href
  {https://doi.org/https://doi.org/10.1023/b:mach.0000033120.25363.1e}
  {\bibfield  {journal} {\bibinfo  {journal} {Mach. Learn.}\ }\textbf {\bibinfo
  {volume} {56}},\ \bibinfo {pages} {209--239} (\bibinfo {year}
  {2004})}\BibitemShut {NoStop}%
\bibitem [{\citenamefont {Belkin}\ and\ \citenamefont
  {Niyogi}(2008)}]{belkin2008towards}%
  \BibitemOpen
  \bibfield  {author} {\bibinfo {author} {\bibfnamefont {M.}~\bibnamefont
  {Belkin}}\ and\ \bibinfo {author} {\bibfnamefont {P.}~\bibnamefont
  {Niyogi}},\ }\bibfield  {title} {\enquote {\bibinfo {title} {{Towards a
  Theoretical Foundation for Laplacian-based Manifold Methods}},}\ }\href
  {https://doi.org/https://doi.org/10.1016/j.jcss.2007.08.006} {\bibfield
  {journal} {\bibinfo  {journal} {J. Comput. Syst. Sci.}\ }\textbf {\bibinfo
  {volume} {74}},\ \bibinfo {pages} {1289--1308} (\bibinfo {year}
  {2008})}\BibitemShut {NoStop}%
\bibitem [{\citenamefont {Chung}(1997)}]{chung1997spectral}%
  \BibitemOpen
  \bibfield  {author} {\bibinfo {author} {\bibfnamefont {F.~R.~K.}\
  \bibnamefont {Chung}},\ }\href@noop {} {\emph {\bibinfo {title} {{Spectral
  Graph Theory}}}},\ \bibinfo {number} {92}\ (\bibinfo  {publisher} {American
  Mathematical Society},\ \bibinfo {year} {1997})\BibitemShut {NoStop}%
\bibitem [{\citenamefont {Sch{\"o}lkopf}, \citenamefont {Smola},\ and\
  \citenamefont {M{\"u}ller}(1998)}]{scholkopf1998nonlinear}%
  \BibitemOpen
  \bibfield  {author} {\bibinfo {author} {\bibfnamefont {B.}~\bibnamefont
  {Sch{\"o}lkopf}}, \bibinfo {author} {\bibfnamefont {A.}~\bibnamefont
  {Smola}},\ and\ \bibinfo {author} {\bibfnamefont {K.-R.}\ \bibnamefont
  {M{\"u}ller}},\ }\bibfield  {title} {\enquote {\bibinfo {title} {{Nonlinear
  Component Analysis as a Kernel Eigenvalue Problem}},}\ }\href
  {https://doi.org/https://doi.org/10.1162/089976698300017467} {\bibfield
  {journal} {\bibinfo  {journal} {Neural Comput.}\ }\textbf {\bibinfo {volume}
  {10}},\ \bibinfo {pages} {1299--1319} (\bibinfo {year} {1998})}\BibitemShut
  {NoStop}%
\bibitem [{\citenamefont {Szummer}\ and\ \citenamefont
  {Jaakkola}(2001)}]{szummer2001partially}%
  \BibitemOpen
  \bibfield  {author} {\bibinfo {author} {\bibfnamefont {M.}~\bibnamefont
  {Szummer}}\ and\ \bibinfo {author} {\bibfnamefont {T.}~\bibnamefont
  {Jaakkola}},\ }\bibfield  {title} {\enquote {\bibinfo {title} {{Partially
  Labeled Classification with Markov Random Walks}},}\ }\href
  {https://doi.org/http://papers.neurips.cc/paper/1967-partially-labeled-classification-with-markov-random-walks.pdf}
  {\bibfield  {journal} {\bibinfo  {journal} {Adv. Neural Inf. Process. Syst.}\
  }\textbf {\bibinfo {volume} {14}},\ \bibinfo {pages} {945–952} (\bibinfo
  {year} {2001})}\BibitemShut {NoStop}%
\bibitem [{\citenamefont {Kondor}\ and\ \citenamefont
  {Lafferty}(2002)}]{10.5555/645531.655996}%
  \BibitemOpen
  \bibfield  {author} {\bibinfo {author} {\bibfnamefont {R.~I.}\ \bibnamefont
  {Kondor}}\ and\ \bibinfo {author} {\bibfnamefont {J.~D.}\ \bibnamefont
  {Lafferty}},\ }\bibfield  {title} {\enquote {\bibinfo {title} {{Diffusion
  Kernels on Graphs and Other Discrete Input Spaces}},}\ }\href
  {https://doi.org/https://www.ml.cmu.edu/research/dap-papers/kondor-diffusion-kernels.pdf}
  {\bibfield  {journal} {\bibinfo  {journal} {Proc. ICML}\ ,\ \bibinfo {pages}
  {315–322}} (\bibinfo {year} {2002})}\BibitemShut {NoStop}%
\bibitem [{\citenamefont {Izenman}(2012)}]{izenman2012introduction}%
  \BibitemOpen
  \bibfield  {author} {\bibinfo {author} {\bibfnamefont {A.~J.}\ \bibnamefont
  {Izenman}},\ }\bibfield  {title} {\enquote {\bibinfo {title} {{Introduction
  to Manifold Learning}},}\ }\href
  {https://doi.org/https://doi.org/10.1002/wics.1222} {\bibfield  {journal}
  {\bibinfo  {journal} {Wiley Interdiscip. Rev. Comput. Stat.}\ }\textbf
  {\bibinfo {volume} {4}},\ \bibinfo {pages} {439--446} (\bibinfo {year}
  {2012})}\BibitemShut {NoStop}%
\bibitem [{\citenamefont {Coifman}\ \emph {et~al.}(2005)\citenamefont
  {Coifman}, \citenamefont {Lafon}, \citenamefont {Lee}, \citenamefont
  {Maggioni}, \citenamefont {Nadler}, \citenamefont {Warner},\ and\
  \citenamefont {Zucker}}]{coifman2005geometric}%
  \BibitemOpen
  \bibfield  {author} {\bibinfo {author} {\bibfnamefont {R.~R.}\ \bibnamefont
  {Coifman}}, \bibinfo {author} {\bibfnamefont {S.}~\bibnamefont {Lafon}},
  \bibinfo {author} {\bibfnamefont {A.~B.}\ \bibnamefont {Lee}}, \bibinfo
  {author} {\bibfnamefont {M.}~\bibnamefont {Maggioni}}, \bibinfo {author}
  {\bibfnamefont {B.}~\bibnamefont {Nadler}}, \bibinfo {author} {\bibfnamefont
  {F.}~\bibnamefont {Warner}},\ and\ \bibinfo {author} {\bibfnamefont {S.~W.}\
  \bibnamefont {Zucker}},\ }\bibfield  {title} {\enquote {\bibinfo {title}
  {{Geometric Diffusions as a Tool for Harmonic Analysis and Structure
  Definition of Data: Diffusion Maps}},}\ }\href
  {https://doi.org/https://doi.org/10.1073/pnas.0500334102} {\bibfield
  {journal} {\bibinfo  {journal} {Proc. Natl. Acad. Sci. U.S.A.}\ }\textbf
  {\bibinfo {volume} {102}},\ \bibinfo {pages} {7426--7431} (\bibinfo {year}
  {2005})}\BibitemShut {NoStop}%
\bibitem [{\citenamefont {Nadler}\ \emph
  {et~al.}(2006{\natexlab{a}})\citenamefont {Nadler}, \citenamefont {Lafon},
  \citenamefont {Coifman},\ and\ \citenamefont
  {Kevrekidis}}]{nadler2006diffusion}%
  \BibitemOpen
  \bibfield  {author} {\bibinfo {author} {\bibfnamefont {B.}~\bibnamefont
  {Nadler}}, \bibinfo {author} {\bibfnamefont {S.}~\bibnamefont {Lafon}},
  \bibinfo {author} {\bibfnamefont {R.~R.}\ \bibnamefont {Coifman}},\ and\
  \bibinfo {author} {\bibfnamefont {I.~G.}\ \bibnamefont {Kevrekidis}},\
  }\bibfield  {title} {\enquote {\bibinfo {title} {{Diffusion Maps, Spectral
  Clustering and Reaction Coordinates of Dynamical Systems}},}\ }\href
  {https://doi.org/https://doi.org/10.1016/j.acha.2005.07.004} {\bibfield
  {journal} {\bibinfo  {journal} {Appl. Comput. Harmon. Anal.}\ }\textbf
  {\bibinfo {volume} {21}},\ \bibinfo {pages} {113--127} (\bibinfo {year}
  {2006}{\natexlab{a}})}\BibitemShut {NoStop}%
\bibitem [{\citenamefont {Nadler}\ \emph
  {et~al.}(2006{\natexlab{b}})\citenamefont {Nadler}, \citenamefont {Lafon},
  \citenamefont {Kevrekidis},\ and\ \citenamefont
  {Coifman}}]{NIPS2005_2a0f97f8}%
  \BibitemOpen
  \bibfield  {author} {\bibinfo {author} {\bibfnamefont {B.}~\bibnamefont
  {Nadler}}, \bibinfo {author} {\bibfnamefont {S.}~\bibnamefont {Lafon}},
  \bibinfo {author} {\bibfnamefont {I.}~\bibnamefont {Kevrekidis}},\ and\
  \bibinfo {author} {\bibfnamefont {R.}~\bibnamefont {Coifman}},\ }\bibfield
  {title} {\enquote {\bibinfo {title} {{Diffusion Maps, Spectral Clustering and
  Eigenfunctions of Fokker-Planck Operators}},}\ }\href
  {https://doi.org/https://proceedings.neurips.cc/paper/2005/file/2a0f97f81755e2878b264adf39cba68e-Paper.pdf}
  {\bibfield  {journal} {\bibinfo  {journal} {Adv. Neural Inf. Process. Syst.}\
  }\textbf {\bibinfo {volume} {18}},\ \bibinfo {pages} {955--962} (\bibinfo
  {year} {2006}{\natexlab{b}})}\BibitemShut {NoStop}%
\bibitem [{\citenamefont {Nadler}\ and\ \citenamefont
  {Galun}(2006)}]{nadler2006fundamental}%
  \BibitemOpen
  \bibfield  {author} {\bibinfo {author} {\bibfnamefont {B.}~\bibnamefont
  {Nadler}}\ and\ \bibinfo {author} {\bibfnamefont {M.}~\bibnamefont {Galun}},\
  }\bibfield  {title} {\enquote {\bibinfo {title} {{Fundamental Limitations of
  Spectral Clustering}},}\ }\href
  {https://doi.org/https://doi.org/10.7551/mitpress/7503.003.0132} {\bibfield
  {journal} {\bibinfo  {journal} {Adv. Neural Inf. Process. Syst.}\ }\textbf
  {\bibinfo {volume} {19}},\ \bibinfo {pages} {1017--1024} (\bibinfo {year}
  {2006})}\BibitemShut {NoStop}%
\bibitem [{\citenamefont {Coifman}\ and\ \citenamefont
  {Lafon}(2006{\natexlab{a}})}]{coifman2006diffusion}%
  \BibitemOpen
  \bibfield  {author} {\bibinfo {author} {\bibfnamefont {R.~R.}\ \bibnamefont
  {Coifman}}\ and\ \bibinfo {author} {\bibfnamefont {S.}~\bibnamefont
  {Lafon}},\ }\bibfield  {title} {\enquote {\bibinfo {title} {{Diffusion
  Maps}},}\ }\href {https://doi.org/https://doi.org/10.1016/j.acha.2006.04.006}
  {\bibfield  {journal} {\bibinfo  {journal} {Appl. Comput. Harmon. Anal.}\
  }\textbf {\bibinfo {volume} {21}},\ \bibinfo {pages} {5--30} (\bibinfo {year}
  {2006}{\natexlab{a}})}\BibitemShut {NoStop}%
\bibitem [{\citenamefont {Zelnik-Manor}\ and\ \citenamefont
  {Perona}(2004)}]{zelnik2004self}%
  \BibitemOpen
  \bibfield  {author} {\bibinfo {author} {\bibfnamefont {L.}~\bibnamefont
  {Zelnik-Manor}}\ and\ \bibinfo {author} {\bibfnamefont {P.}~\bibnamefont
  {Perona}},\ }\bibfield  {title} {\enquote {\bibinfo {title} {{Self-Tuning
  Spectral Clustering}},}\ }\href
  {https://doi.org/http://papers.neurips.cc/paper/2619-self-tuning-spectral-clustering.pdf}
  {\bibfield  {journal} {\bibinfo  {journal} {Adv. Neural Inf. Process. Syst.}\
  }\textbf {\bibinfo {volume} {17}},\ \bibinfo {pages} {1601--1608} (\bibinfo
  {year} {2004})}\BibitemShut {NoStop}%
\bibitem [{\citenamefont {Rohrdanz}\ \emph {et~al.}(2011)\citenamefont
  {Rohrdanz}, \citenamefont {Zheng}, \citenamefont {Maggioni},\ and\
  \citenamefont {Clementi}}]{rohrdanz2011determination}%
  \BibitemOpen
  \bibfield  {author} {\bibinfo {author} {\bibfnamefont {M.~A.}\ \bibnamefont
  {Rohrdanz}}, \bibinfo {author} {\bibfnamefont {W.}~\bibnamefont {Zheng}},
  \bibinfo {author} {\bibfnamefont {M.}~\bibnamefont {Maggioni}},\ and\
  \bibinfo {author} {\bibfnamefont {C.}~\bibnamefont {Clementi}},\ }\bibfield
  {title} {\enquote {\bibinfo {title} {{Determination of Reaction Coordinates
  via Locally Scaled Diffusion Map}},}\ }\href
  {https://doi.org/https://doi.org/10.1063/1.3569857} {\bibfield  {journal}
  {\bibinfo  {journal} {J. Chem. Phys.}\ }\textbf {\bibinfo {volume} {134}},\
  \bibinfo {pages} {03B624} (\bibinfo {year} {2011})}\BibitemShut {NoStop}%
\bibitem [{\citenamefont {Zheng}, \citenamefont {Rohrdanz},\ and\ \citenamefont
  {Clementi}(2013)}]{zheng2013rapid}%
  \BibitemOpen
  \bibfield  {author} {\bibinfo {author} {\bibfnamefont {W.}~\bibnamefont
  {Zheng}}, \bibinfo {author} {\bibfnamefont {M.~A.}\ \bibnamefont
  {Rohrdanz}},\ and\ \bibinfo {author} {\bibfnamefont {C.}~\bibnamefont
  {Clementi}},\ }\bibfield  {title} {\enquote {\bibinfo {title} {{Rapid
  Exploration of Configuration Space with Diffusion-Map-Directed Molecular
  Dynamics}},}\ }\href {https://doi.org/https://doi.org/10.1021/jp401911h}
  {\bibfield  {journal} {\bibinfo  {journal} {J. Phys. Chem. B}\ }\textbf
  {\bibinfo {volume} {117}},\ \bibinfo {pages} {12769--12776} (\bibinfo {year}
  {2013})}\BibitemShut {NoStop}%
\bibitem [{\citenamefont {Zheng}\ \emph {et~al.}(2013)\citenamefont {Zheng},
  \citenamefont {Vargiu}, \citenamefont {Rohrdanz}, \citenamefont {Carloni},\
  and\ \citenamefont {Clementi}}]{zheng2013molecular}%
  \BibitemOpen
  \bibfield  {author} {\bibinfo {author} {\bibfnamefont {W.}~\bibnamefont
  {Zheng}}, \bibinfo {author} {\bibfnamefont {A.~V.}\ \bibnamefont {Vargiu}},
  \bibinfo {author} {\bibfnamefont {M.~A.}\ \bibnamefont {Rohrdanz}}, \bibinfo
  {author} {\bibfnamefont {P.}~\bibnamefont {Carloni}},\ and\ \bibinfo {author}
  {\bibfnamefont {C.}~\bibnamefont {Clementi}},\ }\bibfield  {title} {\enquote
  {\bibinfo {title} {{Molecular Recognition of DNA by Ligands: Roughness and
  Complexity of the Free Energy Profile}},}\ }\href
  {https://doi.org/https://doi.org/10.1063/1.4824106} {\bibfield  {journal}
  {\bibinfo  {journal} {J. Chem. Phys.}\ }\textbf {\bibinfo {volume} {139}},\
  \bibinfo {pages} {10B612\_1} (\bibinfo {year} {2013})}\BibitemShut {NoStop}%
\bibitem [{\citenamefont {Berry}, \citenamefont {Giannakis},\ and\
  \citenamefont {Harlim}(2015)}]{berry2015nonparametric}%
  \BibitemOpen
  \bibfield  {author} {\bibinfo {author} {\bibfnamefont {T.}~\bibnamefont
  {Berry}}, \bibinfo {author} {\bibfnamefont {D.}~\bibnamefont {Giannakis}},\
  and\ \bibinfo {author} {\bibfnamefont {J.}~\bibnamefont {Harlim}},\
  }\bibfield  {title} {\enquote {\bibinfo {title} {{Nonparametric Forecasting
  of Low-Dimensional Dynamical Systems}},}\ }\href
  {https://doi.org/https://doi.org/10.1103/PhysRevE.91.032915} {\bibfield
  {journal} {\bibinfo  {journal} {Phys. Rev. E}\ }\textbf {\bibinfo {volume}
  {91}},\ \bibinfo {pages} {032915} (\bibinfo {year} {2015})}\BibitemShut
  {NoStop}%
\bibitem [{\citenamefont {Berry}\ and\ \citenamefont
  {Harlim}(2016)}]{berry2016variable}%
  \BibitemOpen
  \bibfield  {author} {\bibinfo {author} {\bibfnamefont {T.}~\bibnamefont
  {Berry}}\ and\ \bibinfo {author} {\bibfnamefont {J.}~\bibnamefont {Harlim}},\
  }\bibfield  {title} {\enquote {\bibinfo {title} {{Variable Bandwidth
  Diffusion Kernels}},}\ }\href
  {https://doi.org/https://doi.org/10.1016/j.acha.2015.01.001} {\bibfield
  {journal} {\bibinfo  {journal} {Appl. Comput. Harmon. Anal.}\ }\textbf
  {\bibinfo {volume} {40}},\ \bibinfo {pages} {68--96} (\bibinfo {year}
  {2016})}\BibitemShut {NoStop}%
\bibitem [{\citenamefont {Dsilva}\ \emph {et~al.}(2013)\citenamefont {Dsilva},
  \citenamefont {Talmon}, \citenamefont {Rabin}, \citenamefont {Coifman},\ and\
  \citenamefont {Kevrekidis}}]{dsilva2013nonlinear}%
  \BibitemOpen
  \bibfield  {author} {\bibinfo {author} {\bibfnamefont {C.~J.}\ \bibnamefont
  {Dsilva}}, \bibinfo {author} {\bibfnamefont {R.}~\bibnamefont {Talmon}},
  \bibinfo {author} {\bibfnamefont {N.}~\bibnamefont {Rabin}}, \bibinfo
  {author} {\bibfnamefont {R.~R.}\ \bibnamefont {Coifman}},\ and\ \bibinfo
  {author} {\bibfnamefont {I.~G.}\ \bibnamefont {Kevrekidis}},\ }\bibfield
  {title} {\enquote {\bibinfo {title} {{Nonlinear Intrinsic Variables and State
  Reconstruction in Multiscale Simulations}},}\ }\href
  {https://doi.org/https://doi.org/10.1063/1.4828457} {\bibfield  {journal}
  {\bibinfo  {journal} {J. Chem. Phys.}\ }\textbf {\bibinfo {volume} {139}},\
  \bibinfo {pages} {184109} (\bibinfo {year} {2013})}\BibitemShut {NoStop}%
\bibitem [{\citenamefont {Dsilva}\ \emph {et~al.}(2016)\citenamefont {Dsilva},
  \citenamefont {Talmon}, \citenamefont {Gear}, \citenamefont {Coifman},\ and\
  \citenamefont {Kevrekidis}}]{dsilva2015data}%
  \BibitemOpen
  \bibfield  {author} {\bibinfo {author} {\bibfnamefont {C.~J.}\ \bibnamefont
  {Dsilva}}, \bibinfo {author} {\bibfnamefont {R.}~\bibnamefont {Talmon}},
  \bibinfo {author} {\bibfnamefont {C.~W.}\ \bibnamefont {Gear}}, \bibinfo
  {author} {\bibfnamefont {R.~R.}\ \bibnamefont {Coifman}},\ and\ \bibinfo
  {author} {\bibfnamefont {I.~G.}\ \bibnamefont {Kevrekidis}},\ }\bibfield
  {title} {\enquote {\bibinfo {title} {{Data-Driven Reduction for a Class of
  Multiscale Fast-Slow Stochastic Dynamical Systems}},}\ }\href
  {https://doi.org/10.1137/151004896} {\bibfield  {journal} {\bibinfo
  {journal} {SIAM J. Appl. Dyn. Syst.}\ }\textbf {\bibinfo {volume} {15}},\
  \bibinfo {pages} {1327--1351} (\bibinfo {year} {2016})}\BibitemShut {NoStop}%
\bibitem [{\citenamefont {Singer}\ \emph {et~al.}(2009)\citenamefont {Singer},
  \citenamefont {Erban}, \citenamefont {Kevrekidis},\ and\ \citenamefont
  {Coifman}}]{singer2009detecting}%
  \BibitemOpen
  \bibfield  {author} {\bibinfo {author} {\bibfnamefont {A.}~\bibnamefont
  {Singer}}, \bibinfo {author} {\bibfnamefont {R.}~\bibnamefont {Erban}},
  \bibinfo {author} {\bibfnamefont {I.~G.}\ \bibnamefont {Kevrekidis}},\ and\
  \bibinfo {author} {\bibfnamefont {R.~R.}\ \bibnamefont {Coifman}},\
  }\bibfield  {title} {\enquote {\bibinfo {title} {{Detecting Intrinsic Slow
  Variables in Stochastic Dynamical Systems by Anisotropic Diffusion Maps}},}\
  }\href {https://doi.org/https://doi.org/10.1073/pnas.0905547106} {\bibfield
  {journal} {\bibinfo  {journal} {Proc. Natl. Acad. Sci. U.S.A.}\ }\textbf
  {\bibinfo {volume} {106}},\ \bibinfo {pages} {16090--16095} (\bibinfo {year}
  {2009})}\BibitemShut {NoStop}%
\bibitem [{\citenamefont {Singer}\ and\ \citenamefont
  {Coifman}(2008)}]{singer2008non}%
  \BibitemOpen
  \bibfield  {author} {\bibinfo {author} {\bibfnamefont {A.}~\bibnamefont
  {Singer}}\ and\ \bibinfo {author} {\bibfnamefont {R.~R.}\ \bibnamefont
  {Coifman}},\ }\bibfield  {title} {\enquote {\bibinfo {title} {{Non-Linear
  Independent Component Analysis with Diffusion Maps}},}\ }\href
  {https://doi.org/https://doi.org/10.1016/j.acha.2007.11.001} {\bibfield
  {journal} {\bibinfo  {journal} {Appl. Comput. Harmon. Anal.}\ }\textbf
  {\bibinfo {volume} {25}},\ \bibinfo {pages} {226--239} (\bibinfo {year}
  {2008})}\BibitemShut {NoStop}%
\bibitem [{\citenamefont {Berry}\ and\ \citenamefont
  {Sauer}(2016)}]{berry2016local}%
  \BibitemOpen
  \bibfield  {author} {\bibinfo {author} {\bibfnamefont {T.}~\bibnamefont
  {Berry}}\ and\ \bibinfo {author} {\bibfnamefont {T.}~\bibnamefont {Sauer}},\
  }\bibfield  {title} {\enquote {\bibinfo {title} {{Local Kernels and the
  Geometric Structure of Data}},}\ }\href
  {https://doi.org/https://doi.org/10.1016/j.acha.2015.03.002} {\bibfield
  {journal} {\bibinfo  {journal} {Appl. Comput. Harmon. Anal.}\ }\textbf
  {\bibinfo {volume} {40}},\ \bibinfo {pages} {439--469} (\bibinfo {year}
  {2016})}\BibitemShut {NoStop}%
\bibitem [{\citenamefont {Mugnai}\ and\ \citenamefont
  {Elber}(2015)}]{mugnai2015extracting}%
  \BibitemOpen
  \bibfield  {author} {\bibinfo {author} {\bibfnamefont {M.~L.}\ \bibnamefont
  {Mugnai}}\ and\ \bibinfo {author} {\bibfnamefont {R.}~\bibnamefont {Elber}},\
  }\bibfield  {title} {\enquote {\bibinfo {title} {{Extracting the Diffusion
  Tensor from Molecular Dynamics Simulation with Milestoning}},}\ }\href
  {https://doi.org/https://doi.org/10.1063/1.4904882} {\bibfield  {journal}
  {\bibinfo  {journal} {J. Chem. Phys.}\ }\textbf {\bibinfo {volume} {142}},\
  \bibinfo {pages} {014105} (\bibinfo {year} {2015})}\BibitemShut {NoStop}%
\bibitem [{\citenamefont {Domingues}, \citenamefont {Coifman},\ and\
  \citenamefont {Haji-Akbari}(2024)}]{domingues2024estimating}%
  \BibitemOpen
  \bibfield  {author} {\bibinfo {author} {\bibfnamefont {T.~S.}\ \bibnamefont
  {Domingues}}, \bibinfo {author} {\bibfnamefont {R.}~\bibnamefont {Coifman}},\
  and\ \bibinfo {author} {\bibfnamefont {A.}~\bibnamefont {Haji-Akbari}},\
  }\bibfield  {title} {\enquote {\bibinfo {title} {{Estimating
  Position-Dependent and Anisotropic Diffusivity Tensors from Molecular
  Dynamics Trajectories: Existing Methods and Future Outlook}},}\ }\href
  {https://doi.org/https://doi.org/10.1021/acs.jctc.4c00148} {\bibfield
  {journal} {\bibinfo  {journal} {J. Chem. Theory Comput.}\ }\textbf {\bibinfo
  {volume} {11}},\ \bibinfo {pages} {4427--4455} (\bibinfo {year}
  {2024})}\BibitemShut {NoStop}%
\bibitem [{\citenamefont {Berry}\ \emph {et~al.}(2013)\citenamefont {Berry},
  \citenamefont {Cressman}, \citenamefont {Greguric-Ferencek},\ and\
  \citenamefont {Sauer}}]{berry2013time}%
  \BibitemOpen
  \bibfield  {author} {\bibinfo {author} {\bibfnamefont {T.}~\bibnamefont
  {Berry}}, \bibinfo {author} {\bibfnamefont {J.~R.}\ \bibnamefont {Cressman}},
  \bibinfo {author} {\bibfnamefont {Z.}~\bibnamefont {Greguric-Ferencek}},\
  and\ \bibinfo {author} {\bibfnamefont {T.}~\bibnamefont {Sauer}},\ }\bibfield
   {title} {\enquote {\bibinfo {title} {{Time-Scale Separation from
  Diffusion-Mapped Delay Coordinates}},}\ }\href
  {https://doi.org/https://doi.org/10.1137/12088183x} {\bibfield  {journal}
  {\bibinfo  {journal} {SIAM J. Appl. Dyn. Syst.}\ }\textbf {\bibinfo {volume}
  {12}},\ \bibinfo {pages} {618--649} (\bibinfo {year} {2013})}\BibitemShut
  {NoStop}%
\bibitem [{\citenamefont {Lafon}\ and\ \citenamefont
  {Lee}(2006)}]{lafon2006diffusion}%
  \BibitemOpen
  \bibfield  {author} {\bibinfo {author} {\bibfnamefont {S.}~\bibnamefont
  {Lafon}}\ and\ \bibinfo {author} {\bibfnamefont {A.~B.}\ \bibnamefont
  {Lee}},\ }\bibfield  {title} {\enquote {\bibinfo {title} {{Diffusion Maps and
  Coarse-Graining: A Unified Framework for Dimensionality Reduction, Graph
  Partitioning, and Data Set Parameterization}},}\ }\href
  {https://doi.org/https://doi.org/10.1109/tpami.2006.184} {\bibfield
  {journal} {\bibinfo  {journal} {IEEE Trans. Pattern Anal. Mach. Intel.}\
  }\textbf {\bibinfo {volume} {28}},\ \bibinfo {pages} {1393--1403} (\bibinfo
  {year} {2006})}\BibitemShut {NoStop}%
\bibitem [{\citenamefont {Rydzewski}\ and\ \citenamefont
  {Valsson}(2021)}]{rydzewski2021multiscale}%
  \BibitemOpen
  \bibfield  {author} {\bibinfo {author} {\bibfnamefont {J.}~\bibnamefont
  {Rydzewski}}\ and\ \bibinfo {author} {\bibfnamefont {O.}~\bibnamefont
  {Valsson}},\ }\bibfield  {title} {\enquote {\bibinfo {title} {{Multiscale
  Reweighted Stochastic Embedding: Deep Learning of Collective Variables for
  Enhanced Sampling}},}\ }\href
  {https://doi.org/https://doi.org/10.1021/acs.jpca.1c02869} {\bibfield
  {journal} {\bibinfo  {journal} {J. Phys. Chem. A}\ }\textbf {\bibinfo
  {volume} {125}},\ \bibinfo {pages} {6286--6302} (\bibinfo {year}
  {2021})}\BibitemShut {NoStop}%
\bibitem [{\citenamefont {Rydzewski}\ \emph {et~al.}(2022)\citenamefont
  {Rydzewski}, \citenamefont {Chen}, \citenamefont {Ghosh},\ and\ \citenamefont
  {Valsson}}]{rydzewski2022reweighted}%
  \BibitemOpen
  \bibfield  {author} {\bibinfo {author} {\bibfnamefont {J.}~\bibnamefont
  {Rydzewski}}, \bibinfo {author} {\bibfnamefont {M.}~\bibnamefont {Chen}},
  \bibinfo {author} {\bibfnamefont {T.~K.}\ \bibnamefont {Ghosh}},\ and\
  \bibinfo {author} {\bibfnamefont {O.}~\bibnamefont {Valsson}},\ }\bibfield
  {title} {\enquote {\bibinfo {title} {{Reweighted Manifold Learning of
  Collective Variables from Enhanced Sampling Simulations}},}\ }\href
  {https://doi.org/https://doi.org/10.1021/acs.jctc.2c00873} {\bibfield
  {journal} {\bibinfo  {journal} {J. Chem. Theory Comput.}\ }\textbf {\bibinfo
  {volume} {18}},\ \bibinfo {pages} {7179--7192} (\bibinfo {year}
  {2022})}\BibitemShut {NoStop}%
\bibitem [{\citenamefont {Rydzewski}\ and\ \citenamefont
  {Nowak}(2016)}]{rydzewski2016machine}%
  \BibitemOpen
  \bibfield  {author} {\bibinfo {author} {\bibfnamefont {J.}~\bibnamefont
  {Rydzewski}}\ and\ \bibinfo {author} {\bibfnamefont {W.}~\bibnamefont
  {Nowak}},\ }\bibfield  {title} {\enquote {\bibinfo {title} {{Machine Learning
  Based Dimensionality Reduction Facilitates Ligand Diffusion Paths Assessment:
  A Case of Cytochrome P450cam}},}\ }\href
  {https://doi.org/https://doi.org/10.1021/acs.jctc.6b00212} {\bibfield
  {journal} {\bibinfo  {journal} {J. Chem. Theory Comput.}\ }\textbf {\bibinfo
  {volume} {12}},\ \bibinfo {pages} {2110--2120} (\bibinfo {year}
  {2016})}\BibitemShut {NoStop}%
\bibitem [{\citenamefont {Chiavazzo}\ \emph {et~al.}(2017)\citenamefont
  {Chiavazzo}, \citenamefont {Covino}, \citenamefont {Coifman}, \citenamefont
  {Gear}, \citenamefont {Georgiou}, \citenamefont {Hummer},\ and\ \citenamefont
  {Kevrekidis}}]{chiavazzo2017intrinsic}%
  \BibitemOpen
  \bibfield  {author} {\bibinfo {author} {\bibfnamefont {E.}~\bibnamefont
  {Chiavazzo}}, \bibinfo {author} {\bibfnamefont {R.}~\bibnamefont {Covino}},
  \bibinfo {author} {\bibfnamefont {R.~R.}\ \bibnamefont {Coifman}}, \bibinfo
  {author} {\bibfnamefont {C.~W.}\ \bibnamefont {Gear}}, \bibinfo {author}
  {\bibfnamefont {A.~S.}\ \bibnamefont {Georgiou}}, \bibinfo {author}
  {\bibfnamefont {G.}~\bibnamefont {Hummer}},\ and\ \bibinfo {author}
  {\bibfnamefont {I.~G.}\ \bibnamefont {Kevrekidis}},\ }\bibfield  {title}
  {\enquote {\bibinfo {title} {{Intrinsic Map Dynamics Exploration for
  Uncharted Effective Free-Energy Landscapes}},}\ }\href
  {https://doi.org/https://doi.org/10.1073/pnas.1621481114} {\bibfield
  {journal} {\bibinfo  {journal} {Proc. Natl Acad. Sci. U.S.A.}\ }\textbf
  {\bibinfo {volume} {114}},\ \bibinfo {pages} {E5494--E5503} (\bibinfo {year}
  {2017})}\BibitemShut {NoStop}%
\bibitem [{\citenamefont {Ferguson}\ \emph {et~al.}(2011)\citenamefont
  {Ferguson}, \citenamefont {Panagiotopoulos}, \citenamefont {Debenedetti},\
  and\ \citenamefont {Kevrekidis}}]{ferguson2011integrating}%
  \BibitemOpen
  \bibfield  {author} {\bibinfo {author} {\bibfnamefont {A.~L.}\ \bibnamefont
  {Ferguson}}, \bibinfo {author} {\bibfnamefont {A.~Z.}\ \bibnamefont
  {Panagiotopoulos}}, \bibinfo {author} {\bibfnamefont {P.~G.}\ \bibnamefont
  {Debenedetti}},\ and\ \bibinfo {author} {\bibfnamefont {I.~G.}\ \bibnamefont
  {Kevrekidis}},\ }\bibfield  {title} {\enquote {\bibinfo {title} {{Integrating
  Diffusion Maps with Umbrella Sampling: Application to Alanine Dipeptide}},}\
  }\href {https://doi.org/https://doi.org/10.1063/1.3574394} {\bibfield
  {journal} {\bibinfo  {journal} {J. Chem. Phys.}\ }\textbf {\bibinfo {volume}
  {134}},\ \bibinfo {pages} {04B606} (\bibinfo {year} {2011})}\BibitemShut
  {NoStop}%
\bibitem [{\citenamefont {Banisch}\ \emph {et~al.}(2020)\citenamefont
  {Banisch}, \citenamefont {Trstanova}, \citenamefont {Bittracher},
  \citenamefont {Klus},\ and\ \citenamefont {Koltai}}]{banisch2020diffusion}%
  \BibitemOpen
  \bibfield  {author} {\bibinfo {author} {\bibfnamefont {R.}~\bibnamefont
  {Banisch}}, \bibinfo {author} {\bibfnamefont {Z.}~\bibnamefont {Trstanova}},
  \bibinfo {author} {\bibfnamefont {A.}~\bibnamefont {Bittracher}}, \bibinfo
  {author} {\bibfnamefont {S.}~\bibnamefont {Klus}},\ and\ \bibinfo {author}
  {\bibfnamefont {P.}~\bibnamefont {Koltai}},\ }\bibfield  {title} {\enquote
  {\bibinfo {title} {{Diffusion Maps Tailored to Arbitrary Non-Degenerate Itô
  Processes}},}\ }\href
  {https://doi.org/https://doi.org/10.1016/j.acha.2018.05.001} {\bibfield
  {journal} {\bibinfo  {journal} {Appl. Comput. Harmon. Anal.}\ }\textbf
  {\bibinfo {volume} {48}},\ \bibinfo {pages} {242--265} (\bibinfo {year}
  {2020})}\BibitemShut {NoStop}%
\bibitem [{\citenamefont {Trstanova}, \citenamefont {Leimkuhler},\ and\
  \citenamefont {Leli{\`e}vre}(2020)}]{trstanova2020local}%
  \BibitemOpen
  \bibfield  {author} {\bibinfo {author} {\bibfnamefont {Z.}~\bibnamefont
  {Trstanova}}, \bibinfo {author} {\bibfnamefont {B.}~\bibnamefont
  {Leimkuhler}},\ and\ \bibinfo {author} {\bibfnamefont {T.}~\bibnamefont
  {Leli{\`e}vre}},\ }\bibfield  {title} {\enquote {\bibinfo {title} {{Local and
  Global Perspectives on Diffusion Maps in the Analysis of Molecular
  Systems}},}\ }\href {https://doi.org/https://doi.org/10.1098/rspa.2019.0036}
  {\bibfield  {journal} {\bibinfo  {journal} {Proc. Royal Soc. A}\ }\textbf
  {\bibinfo {volume} {476}},\ \bibinfo {pages} {20190036} (\bibinfo {year}
  {2020})}\BibitemShut {NoStop}%
\bibitem [{\citenamefont {Zhang}\ and\ \citenamefont
  {Chen}(2018)}]{zhang2018unfolding}%
  \BibitemOpen
  \bibfield  {author} {\bibinfo {author} {\bibfnamefont {J.}~\bibnamefont
  {Zhang}}\ and\ \bibinfo {author} {\bibfnamefont {M.}~\bibnamefont {Chen}},\
  }\bibfield  {title} {\enquote {\bibinfo {title} {{Unfolding Hidden Barriers
  by Active Enhanced Sampling}},}\ }\href
  {https://doi.org/https://doi.org/10.1103/PhysRevLett.121.010601} {\bibfield
  {journal} {\bibinfo  {journal} {Phys. Rev. Lett.}\ }\textbf {\bibinfo
  {volume} {121}},\ \bibinfo {pages} {010601} (\bibinfo {year}
  {2018})}\BibitemShut {NoStop}%
\bibitem [{\citenamefont
  {Rydzewski}(2023{\natexlab{a}})}]{rydzewski2023selecting}%
  \BibitemOpen
  \bibfield  {author} {\bibinfo {author} {\bibfnamefont {J.}~\bibnamefont
  {Rydzewski}},\ }\bibfield  {title} {\enquote {\bibinfo {title} {{Selecting
  High-Dimensional Representations of Physical Systems by Reweighted Diffusion
  Maps}},}\ }\href
  {https://doi.org/https://doi.org/10.1021/acs.jpclett.3c00265} {\bibfield
  {journal} {\bibinfo  {journal} {J. Phys. Chem. Lett.}\ }\textbf {\bibinfo
  {volume} {14}},\ \bibinfo {pages} {2778--2783} (\bibinfo {year}
  {2023}{\natexlab{a}})}\BibitemShut {NoStop}%
\bibitem [{\citenamefont {Keller}\ and\ \citenamefont
  {Bolhuis}(2024)}]{keller2024dynamical}%
  \BibitemOpen
  \bibfield  {author} {\bibinfo {author} {\bibfnamefont {B.~G.}\ \bibnamefont
  {Keller}}\ and\ \bibinfo {author} {\bibfnamefont {P.~G.}\ \bibnamefont
  {Bolhuis}},\ }\bibfield  {title} {\enquote {\bibinfo {title} {{Dynamical
  Reweighting for Biased Rare Event Simulations}},}\ }\href
  {https://doi.org/https://doi.org/10.1146/annurev-physchem-083122-124538}
  {\bibfield  {journal} {\bibinfo  {journal} {Annu. Rev. Phys. Chem.}\ }\textbf
  {\bibinfo {volume} {75}},\ \bibinfo {pages} {137--162} (\bibinfo {year}
  {2024})}\BibitemShut {NoStop}%
\bibitem [{\citenamefont {Boninsegna}\ \emph {et~al.}(2015)\citenamefont
  {Boninsegna}, \citenamefont {Gobbo}, \citenamefont {No{\'e}},\ and\
  \citenamefont {Clementi}}]{boninsegna2015investigating}%
  \BibitemOpen
  \bibfield  {author} {\bibinfo {author} {\bibfnamefont {L.}~\bibnamefont
  {Boninsegna}}, \bibinfo {author} {\bibfnamefont {G.}~\bibnamefont {Gobbo}},
  \bibinfo {author} {\bibfnamefont {F.}~\bibnamefont {No{\'e}}},\ and\ \bibinfo
  {author} {\bibfnamefont {C.}~\bibnamefont {Clementi}},\ }\bibfield  {title}
  {\enquote {\bibinfo {title} {{Investigating Molecular Kinetics by
  Variationally Optimized Diffusion Maps}},}\ }\href
  {https://doi.org/https://doi.org/10.1021/acs.jctc.5b00749} {\bibfield
  {journal} {\bibinfo  {journal} {J. Chem. Theory Comput.}\ }\textbf {\bibinfo
  {volume} {11}},\ \bibinfo {pages} {5947--5960} (\bibinfo {year}
  {2015})}\BibitemShut {NoStop}%
\bibitem [{\citenamefont {No{\'e}}\ and\ \citenamefont
  {Clementi}(2015)}]{noe2015kinetic}%
  \BibitemOpen
  \bibfield  {author} {\bibinfo {author} {\bibfnamefont {F.}~\bibnamefont
  {No{\'e}}}\ and\ \bibinfo {author} {\bibfnamefont {C.}~\bibnamefont
  {Clementi}},\ }\bibfield  {title} {\enquote {\bibinfo {title} {{Kinetic
  Distance and Kinetic Maps from Molecular Dynamics Simulation}},}\ }\href
  {https://doi.org/https://doi.org/10.1021/acs.jctc.5b00553} {\bibfield
  {journal} {\bibinfo  {journal} {J. Chem. Theory Comput.}\ }\textbf {\bibinfo
  {volume} {11}},\ \bibinfo {pages} {5002--5011} (\bibinfo {year}
  {2015})}\BibitemShut {NoStop}%
\bibitem [{\citenamefont {No{\'e}}, \citenamefont {Banisch},\ and\
  \citenamefont {Clementi}(2016)}]{noe2016commute}%
  \BibitemOpen
  \bibfield  {author} {\bibinfo {author} {\bibfnamefont {F.}~\bibnamefont
  {No{\'e}}}, \bibinfo {author} {\bibfnamefont {R.}~\bibnamefont {Banisch}},\
  and\ \bibinfo {author} {\bibfnamefont {C.}~\bibnamefont {Clementi}},\
  }\bibfield  {title} {\enquote {\bibinfo {title} {{Commute Maps: Separating
  Slowly Mixing Molecular Configurations for Kinetic Modeling}},}\ }\href
  {https://doi.org/https://doi.org/10.1021/acs.jctc.6b00762} {\bibfield
  {journal} {\bibinfo  {journal} {J. Chem. Theory Comput.}\ }\textbf {\bibinfo
  {volume} {12}},\ \bibinfo {pages} {5620--5630} (\bibinfo {year}
  {2016})}\BibitemShut {NoStop}%
\bibitem [{\citenamefont {Thiede}\ \emph {et~al.}(2019)\citenamefont {Thiede},
  \citenamefont {Giannakis}, \citenamefont {Dinner},\ and\ \citenamefont
  {Weare}}]{thiede2019galerkin}%
  \BibitemOpen
  \bibfield  {author} {\bibinfo {author} {\bibfnamefont {E.~H.}\ \bibnamefont
  {Thiede}}, \bibinfo {author} {\bibfnamefont {D.}~\bibnamefont {Giannakis}},
  \bibinfo {author} {\bibfnamefont {A.~R.}\ \bibnamefont {Dinner}},\ and\
  \bibinfo {author} {\bibfnamefont {J.}~\bibnamefont {Weare}},\ }\bibfield
  {title} {\enquote {\bibinfo {title} {{Galerkin Approximation of Dynamical
  Quantities using Trajectory Data}},}\ }\href
  {https://doi.org/https://doi.org/10.1063/1.5063730} {\bibfield  {journal}
  {\bibinfo  {journal} {J. Chem. Phys.}\ }\textbf {\bibinfo {volume} {150}},\
  \bibinfo {pages} {244111} (\bibinfo {year} {2019})}\BibitemShut {NoStop}%
\bibitem [{\citenamefont {Fowlkes}\ \emph {et~al.}(2004)\citenamefont
  {Fowlkes}, \citenamefont {Belongie}, \citenamefont {Chung},\ and\
  \citenamefont {Malik}}]{fowlkes2004spectral}%
  \BibitemOpen
  \bibfield  {author} {\bibinfo {author} {\bibfnamefont {C.}~\bibnamefont
  {Fowlkes}}, \bibinfo {author} {\bibfnamefont {S.}~\bibnamefont {Belongie}},
  \bibinfo {author} {\bibfnamefont {F.}~\bibnamefont {Chung}},\ and\ \bibinfo
  {author} {\bibfnamefont {J.}~\bibnamefont {Malik}},\ }\bibfield  {title}
  {\enquote {\bibinfo {title} {{Spectral Grouping using the Nystrom Method}},}\
  }\href {https://doi.org/https://doi.org/10.1109/tpami.2004.1262185}
  {\bibfield  {journal} {\bibinfo  {journal} {IEEE Trans. Pattern Anal. Mach.
  Intell.}\ }\textbf {\bibinfo {volume} {26}},\ \bibinfo {pages} {214--225}
  (\bibinfo {year} {2004})}\BibitemShut {NoStop}%
\bibitem [{\citenamefont {Long}\ and\ \citenamefont
  {Ferguson}(2019)}]{long2019landmark}%
  \BibitemOpen
  \bibfield  {author} {\bibinfo {author} {\bibfnamefont {A.~W.}\ \bibnamefont
  {Long}}\ and\ \bibinfo {author} {\bibfnamefont {A.~L.}\ \bibnamefont
  {Ferguson}},\ }\bibfield  {title} {\enquote {\bibinfo {title} {{Landmark
  Diffusion Maps (L-dMaps): Accelerated Manifold Learning Out-of-Sample
  Extension}},}\ }\href
  {https://doi.org/https://doi.org/10.1016/j.acha.2017.08.004} {\bibfield
  {journal} {\bibinfo  {journal} {Appl. Comput. Harmon. Anal.}\ }\textbf
  {\bibinfo {volume} {47}},\ \bibinfo {pages} {190--211} (\bibinfo {year}
  {2019})}\BibitemShut {NoStop}%
\bibitem [{\citenamefont {Coifman}\ and\ \citenamefont
  {Lafon}(2006{\natexlab{b}})}]{coifman2006geometric}%
  \BibitemOpen
  \bibfield  {author} {\bibinfo {author} {\bibfnamefont {R.~R.}\ \bibnamefont
  {Coifman}}\ and\ \bibinfo {author} {\bibfnamefont {S.}~\bibnamefont
  {Lafon}},\ }\bibfield  {title} {\enquote {\bibinfo {title} {{Geometric
  Harmonics: A Novel Tool for Multiscale Out-of-Sample Extension of Empirical
  Functions}},}\ }\href
  {https://doi.org/https://doi.org/10.1016/j.acha.2005.07.005} {\bibfield
  {journal} {\bibinfo  {journal} {Appl. Comput. Harmon. Anal.}\ }\textbf
  {\bibinfo {volume} {21}},\ \bibinfo {pages} {31--52} (\bibinfo {year}
  {2006}{\natexlab{b}})}\BibitemShut {NoStop}%
\bibitem [{\citenamefont {Evangelou}\ \emph {et~al.}(2023)\citenamefont
  {Evangelou}, \citenamefont {Dietrich}, \citenamefont {Chiavazzo},
  \citenamefont {Lehmberg}, \citenamefont {Meila},\ and\ \citenamefont
  {Kevrekidis}}]{evangelou2023double}%
  \BibitemOpen
  \bibfield  {author} {\bibinfo {author} {\bibfnamefont {N.}~\bibnamefont
  {Evangelou}}, \bibinfo {author} {\bibfnamefont {F.}~\bibnamefont {Dietrich}},
  \bibinfo {author} {\bibfnamefont {E.}~\bibnamefont {Chiavazzo}}, \bibinfo
  {author} {\bibfnamefont {D.}~\bibnamefont {Lehmberg}}, \bibinfo {author}
  {\bibfnamefont {M.}~\bibnamefont {Meila}},\ and\ \bibinfo {author}
  {\bibfnamefont {I.~G.}\ \bibnamefont {Kevrekidis}},\ }\bibfield  {title}
  {\enquote {\bibinfo {title} {{Double Diffusion Maps and Their Latent
  Harmonics for Scientific Computations in Latent Space}},}\ }\href
  {https://doi.org/https://doi.org/10.1016/j.jcp.2023.112072} {\bibfield
  {journal} {\bibinfo  {journal} {J. Comput. Phys.}\ }\textbf {\bibinfo
  {volume} {485}},\ \bibinfo {pages} {112072} (\bibinfo {year}
  {2023})}\BibitemShut {NoStop}%
\bibitem [{\citenamefont {Bengio}\ \emph {et~al.}(2003)\citenamefont {Bengio},
  \citenamefont {Paiement}, \citenamefont {Vincent}, \citenamefont {Delalleau},
  \citenamefont {Roux},\ and\ \citenamefont {Ouimet}}]{bengio2003out}%
  \BibitemOpen
  \bibfield  {author} {\bibinfo {author} {\bibfnamefont {Y.}~\bibnamefont
  {Bengio}}, \bibinfo {author} {\bibfnamefont {J.-F.}\ \bibnamefont
  {Paiement}}, \bibinfo {author} {\bibfnamefont {P.}~\bibnamefont {Vincent}},
  \bibinfo {author} {\bibfnamefont {O.}~\bibnamefont {Delalleau}}, \bibinfo
  {author} {\bibfnamefont {N.}~\bibnamefont {Roux}},\ and\ \bibinfo {author}
  {\bibfnamefont {M.}~\bibnamefont {Ouimet}},\ }\bibfield  {title} {\enquote
  {\bibinfo {title} {{Out-of-Sample Extensions for LLE, Isomap, MDS, Eigenmaps,
  and Spectral Clustering}},}\ }\href
  {https://doi.org/https://proceedings.neurips.cc/paper/2003/file/cf05968255451bdefe3c5bc64d550517-Paper.pdf}
  {\bibfield  {journal} {\bibinfo  {journal} {Adv. Neural Inf. Process. Syst.}\
  }\textbf {\bibinfo {volume} {16}},\ \bibinfo {pages} {177--184} (\bibinfo
  {year} {2003})}\BibitemShut {NoStop}%
\bibitem [{\citenamefont {Hinton}\ and\ \citenamefont
  {Roweis}(2002)}]{hinton2002stochastic}%
  \BibitemOpen
  \bibfield  {author} {\bibinfo {author} {\bibfnamefont {G.~E.}\ \bibnamefont
  {Hinton}}\ and\ \bibinfo {author} {\bibfnamefont {S.}~\bibnamefont
  {Roweis}},\ }\bibfield  {title} {\enquote {\bibinfo {title} {{Stochastic
  Neighbor Embedding}},}\ }\href
  {https://doi.org/https://proceedings.neurips.cc/paper_files/paper/2002/file/6150ccc6069bea6b5716254057a194ef-Paper.pdf}
  {\bibfield  {journal} {\bibinfo  {journal} {Adv. Neural Inf. Process. Syst.}\
  }\textbf {\bibinfo {volume} {15}},\ \bibinfo {pages} {833--864} (\bibinfo
  {year} {2002})}\BibitemShut {NoStop}%
\bibitem [{\citenamefont {van~der Maaten}\ and\ \citenamefont
  {Hinton}(2008)}]{maaten2008visualizing}%
  \BibitemOpen
  \bibfield  {author} {\bibinfo {author} {\bibfnamefont {L.}~\bibnamefont
  {van~der Maaten}}\ and\ \bibinfo {author} {\bibfnamefont {G.}~\bibnamefont
  {Hinton}},\ }\bibfield  {title} {\enquote {\bibinfo {title} {{Visualizing
  Data using $t$-SNE}},}\ }\href
  {https://doi.org/http://www.jmlr.org/papers/v9/vandermaaten08a.html}
  {\bibfield  {journal} {\bibinfo  {journal} {J. Mach. Learn. Res.}\ }\textbf
  {\bibinfo {volume} {9}},\ \bibinfo {pages} {2579--2605} (\bibinfo {year}
  {2008})}\BibitemShut {NoStop}%
\bibitem [{\citenamefont {van~der Maaten}(2009)}]{maaten2009learning}%
  \BibitemOpen
  \bibfield  {author} {\bibinfo {author} {\bibfnamefont {L.}~\bibnamefont
  {van~der Maaten}},\ }\bibfield  {title} {\enquote {\bibinfo {title}
  {{Learning a Parametric Embedding by Preserving Local Structure}},}\ }\href
  {https://doi.org/http://proceedings.mlr.press/v5/maaten09a.html} {\bibfield
  {journal} {\bibinfo  {journal} {J. Mach. Learn. Res.}\ }\textbf {\bibinfo
  {volume} {5}},\ \bibinfo {pages} {384--391} (\bibinfo {year}
  {2009})}\BibitemShut {NoStop}%
\bibitem [{\citenamefont {van Der~Maaten}, \citenamefont {Postma},\ and\
  \citenamefont {van~den Herik}(2009)}]{van2009dimensionality}%
  \BibitemOpen
  \bibfield  {author} {\bibinfo {author} {\bibfnamefont {L.}~\bibnamefont {van
  Der~Maaten}}, \bibinfo {author} {\bibfnamefont {E.}~\bibnamefont {Postma}},\
  and\ \bibinfo {author} {\bibfnamefont {J.}~\bibnamefont {van~den Herik}},\
  }\bibfield  {title} {\enquote {\bibinfo {title} {{Dimensionality Reduction: A
  Comparative Review}},}\ }\href
  {https://lvdmaaten.github.io/publications/papers/TR_Dimensionality_Reduction_Review_2009.pdf}
  {\bibfield  {journal} {\bibinfo  {journal} {J. Mach. Learn. Res.}\ }\textbf
  {\bibinfo {volume} {10}},\ \bibinfo {pages} {66--71} (\bibinfo {year}
  {2009})}\BibitemShut {NoStop}%
\bibitem [{\citenamefont {Sorensen}\ and\ \citenamefont
  {Voter}(2000)}]{voter2000temperature}%
  \BibitemOpen
  \bibfield  {author} {\bibinfo {author} {\bibfnamefont {M.~R.}\ \bibnamefont
  {Sorensen}}\ and\ \bibinfo {author} {\bibfnamefont {A.~F.}\ \bibnamefont
  {Voter}},\ }\bibfield  {title} {\enquote {\bibinfo {title}
  {{Temperature-Accelerated Dynamics for Simulation of Infrequent Events}},}\
  }\href {https://doi.org/https://doi.org/10.1063/1.481576} {\bibfield
  {journal} {\bibinfo  {journal} {J. Chem. Phys.}\ }\textbf {\bibinfo {volume}
  {112}},\ \bibinfo {pages} {9599--9606} (\bibinfo {year} {2000})}\BibitemShut
  {NoStop}%
\bibitem [{\citenamefont {Maragliano}\ and\ \citenamefont
  {Vanden-Eijnden}(2006)}]{maragliano2006temperature}%
  \BibitemOpen
  \bibfield  {author} {\bibinfo {author} {\bibfnamefont {L.}~\bibnamefont
  {Maragliano}}\ and\ \bibinfo {author} {\bibfnamefont {E.}~\bibnamefont
  {Vanden-Eijnden}},\ }\bibfield  {title} {\enquote {\bibinfo {title} {{A
  Temperature Accelerated Method for Sampling Free Energy and Determining
  Reaction Pathways in Rare Events Simulations}},}\ }\href
  {https://doi.org/https://doi.org/10.1016/j.cplett.2006.05.062} {\bibfield
  {journal} {\bibinfo  {journal} {Chem. Phys. Lett.}\ }\textbf {\bibinfo
  {volume} {426}},\ \bibinfo {pages} {168--175} (\bibinfo {year}
  {2006})}\BibitemShut {NoStop}%
\bibitem [{\citenamefont {Liu}\ \emph {et~al.}(2024)\citenamefont {Liu},
  \citenamefont {Ghosh}, \citenamefont {Lin},\ and\ \citenamefont
  {Chen}}]{liu2024unbiasing}%
  \BibitemOpen
  \bibfield  {author} {\bibinfo {author} {\bibfnamefont {Y.}~\bibnamefont
  {Liu}}, \bibinfo {author} {\bibfnamefont {T.~K.}\ \bibnamefont {Ghosh}},
  \bibinfo {author} {\bibfnamefont {G.}~\bibnamefont {Lin}},\ and\ \bibinfo
  {author} {\bibfnamefont {M.}~\bibnamefont {Chen}},\ }\bibfield  {title}
  {\enquote {\bibinfo {title} {{Unbiasing Enhanced Sampling on a
  High-Dimensional Free Energy Surface with a Deep Generative Model}},}\ }\href
  {https://doi.org/https://doi.org/10.1021/acs.jpclett.3c03515} {\bibfield
  {journal} {\bibinfo  {journal} {J. Phys. Chem. Lett.}\ }\textbf {\bibinfo
  {volume} {15}},\ \bibinfo {pages} {3938--3945} (\bibinfo {year}
  {2024})}\BibitemShut {NoStop}%
\bibitem [{\citenamefont {Kullback}\ and\ \citenamefont
  {Leibler}(1951)}]{kullback1951information}%
  \BibitemOpen
  \bibfield  {author} {\bibinfo {author} {\bibfnamefont {S.}~\bibnamefont
  {Kullback}}\ and\ \bibinfo {author} {\bibfnamefont {R.~A.}\ \bibnamefont
  {Leibler}},\ }\bibfield  {title} {\enquote {\bibinfo {title} {{On Information
  and Sufficiency}},}\ }\href
  {https://doi.org/https://doi.org/10.1214/aoms/1177729694} {\bibfield
  {journal} {\bibinfo  {journal} {Ann. Math. Stat.}\ }\textbf {\bibinfo
  {volume} {22}},\ \bibinfo {pages} {79--86} (\bibinfo {year}
  {1951})}\BibitemShut {NoStop}%
\bibitem [{\citenamefont {Tiwary}\ and\ \citenamefont
  {Berne}(2016)}]{tiwary2016spectral}%
  \BibitemOpen
  \bibfield  {author} {\bibinfo {author} {\bibfnamefont {P.}~\bibnamefont
  {Tiwary}}\ and\ \bibinfo {author} {\bibfnamefont {B.~J.}\ \bibnamefont
  {Berne}},\ }\bibfield  {title} {\enquote {\bibinfo {title} {{Spectral Gap
  Optimization of Order Parameters for Sampling Complex Molecular Systems}},}\
  }\href {https://doi.org/https://doi.org/10.1073/pnas.1600917113} {\bibfield
  {journal} {\bibinfo  {journal} {Proc. Natl. Acad. Sci. U.S.A.}\ }\textbf
  {\bibinfo {volume} {113}},\ \bibinfo {pages} {2839} (\bibinfo {year}
  {2016})}\BibitemShut {NoStop}%
\bibitem [{\citenamefont {Tiwary}\ and\ \citenamefont
  {Berne}(2017)}]{tiwary2017predicting}%
  \BibitemOpen
  \bibfield  {author} {\bibinfo {author} {\bibfnamefont {P.}~\bibnamefont
  {Tiwary}}\ and\ \bibinfo {author} {\bibfnamefont {B.}~\bibnamefont {Berne}},\
  }\bibfield  {title} {\enquote {\bibinfo {title} {{Predicting Reaction
  Coordinates in Energy Landscapes with Diffusion Anisotropy}},}\ }\href
  {https://doi.org/https://doi.org/10.1063/1.4983727} {\bibfield  {journal}
  {\bibinfo  {journal} {J. Chem. Phys.}\ }\textbf {\bibinfo {volume} {147}},\
  \bibinfo {pages} {152701} (\bibinfo {year} {2017})}\BibitemShut {NoStop}%
\bibitem [{\citenamefont {Pant}\ \emph {et~al.}(2020)\citenamefont {Pant},
  \citenamefont {Smith}, \citenamefont {Wang}, \citenamefont {Tajkhorshid},\
  and\ \citenamefont {Tiwary}}]{10.1063/5.0030931}%
  \BibitemOpen
  \bibfield  {author} {\bibinfo {author} {\bibfnamefont {S.}~\bibnamefont
  {Pant}}, \bibinfo {author} {\bibfnamefont {Z.}~\bibnamefont {Smith}},
  \bibinfo {author} {\bibfnamefont {Y.}~\bibnamefont {Wang}}, \bibinfo {author}
  {\bibfnamefont {E.}~\bibnamefont {Tajkhorshid}},\ and\ \bibinfo {author}
  {\bibfnamefont {P.}~\bibnamefont {Tiwary}},\ }\bibfield  {title} {\enquote
  {\bibinfo {title} {{Confronting Pitfalls of AI-Augmented Molecular Dynamics
  using Statistical Physics}},}\ }\href {https://doi.org/10.1063/5.0030931}
  {\bibfield  {journal} {\bibinfo  {journal} {J. Chem. Phys.}\ }\textbf
  {\bibinfo {volume} {153}},\ \bibinfo {pages} {234118} (\bibinfo {year}
  {2020})}\BibitemShut {NoStop}%
\bibitem [{\citenamefont {Tsai}, \citenamefont {Smith},\ and\ \citenamefont
  {Tiwary}(2021)}]{tsai2021sgoop}%
  \BibitemOpen
  \bibfield  {author} {\bibinfo {author} {\bibfnamefont {S.-T.}\ \bibnamefont
  {Tsai}}, \bibinfo {author} {\bibfnamefont {Z.}~\bibnamefont {Smith}},\ and\
  \bibinfo {author} {\bibfnamefont {P.}~\bibnamefont {Tiwary}},\ }\bibfield
  {title} {\enquote {\bibinfo {title} {{SGOOP-d: Estimating Kinetic Distances
  and Reaction Coordinate Dimensionality for Rare Event Systems from
  Biased/Unbiased Simulations}},}\ }\href
  {https://doi.org/https://doi.org/10.1021/acs.jctc.1c00431} {\bibfield
  {journal} {\bibinfo  {journal} {J. Chem. Theory Comput.}\ }\textbf {\bibinfo
  {volume} {17}},\ \bibinfo {pages} {6757--6765} (\bibinfo {year}
  {2021})}\BibitemShut {NoStop}%
\bibitem [{\citenamefont {Zou}, \citenamefont {Tsai},\ and\ \citenamefont
  {Tiwary}(2021)}]{zou2021toward}%
  \BibitemOpen
  \bibfield  {author} {\bibinfo {author} {\bibfnamefont {Z.}~\bibnamefont
  {Zou}}, \bibinfo {author} {\bibfnamefont {S.-T.}\ \bibnamefont {Tsai}},\ and\
  \bibinfo {author} {\bibfnamefont {P.}~\bibnamefont {Tiwary}},\ }\bibfield
  {title} {\enquote {\bibinfo {title} {{Toward Automated Sampling of Polymorph
  Nucleation and Free Energies with the SGOOP and Metadynamics}},}\ }\href
  {https://doi.org/https://doi.org/10.1021/acs.jpcb.1c07595} {\bibfield
  {journal} {\bibinfo  {journal} {J. Phys. Chem. B}\ }\textbf {\bibinfo
  {volume} {125}},\ \bibinfo {pages} {13049--13056} (\bibinfo {year}
  {2021})}\BibitemShut {NoStop}%
\bibitem [{\citenamefont {Ghosh}\ \emph {et~al.}(2020)\citenamefont {Ghosh},
  \citenamefont {Dixit}, \citenamefont {Agozzino},\ and\ \citenamefont
  {Dill}}]{ghosh2020maximum}%
  \BibitemOpen
  \bibfield  {author} {\bibinfo {author} {\bibfnamefont {K.}~\bibnamefont
  {Ghosh}}, \bibinfo {author} {\bibfnamefont {P.~D.}\ \bibnamefont {Dixit}},
  \bibinfo {author} {\bibfnamefont {L.}~\bibnamefont {Agozzino}},\ and\
  \bibinfo {author} {\bibfnamefont {K.~A.}\ \bibnamefont {Dill}},\ }\bibfield
  {title} {\enquote {\bibinfo {title} {{The Maximum Caliber Variational
  Principle for Nonequilibria}},}\ }\href
  {https://doi.org/https://doi.org/10.1146/annurev-physchem-071119-040206}
  {\bibfield  {journal} {\bibinfo  {journal} {Annu. Rev. Phys. Chem.}\ }\textbf
  {\bibinfo {volume} {71}},\ \bibinfo {pages} {213--238} (\bibinfo {year}
  {2020})}\BibitemShut {NoStop}%
\bibitem [{\citenamefont
  {Rydzewski}(2023{\natexlab{b}})}]{rydzewski2023spectral}%
  \BibitemOpen
  \bibfield  {author} {\bibinfo {author} {\bibfnamefont {J.}~\bibnamefont
  {Rydzewski}},\ }\bibfield  {title} {\enquote {\bibinfo {title} {{Spectral
  Map: Embedding Slow Kinetics in Collective Variables}},}\ }\href
  {https://doi.org/https://doi.org/10.1021/acs.jpclett.3c01101} {\bibfield
  {journal} {\bibinfo  {journal} {J. Phys. Chem. Lett.}\ }\textbf {\bibinfo
  {volume} {14}},\ \bibinfo {pages} {5216--5220} (\bibinfo {year}
  {2023}{\natexlab{b}})}\BibitemShut {NoStop}%
\bibitem [{\citenamefont {Rydzewski}(2024)}]{rydzewski2024tse}%
  \BibitemOpen
  \bibfield  {author} {\bibinfo {author} {\bibfnamefont {J.}~\bibnamefont
  {Rydzewski}},\ }\bibfield  {title} {\enquote {\bibinfo {title} {{Spectral Map
  for Slow Collective Variables, Markovian Dynamics, and Transition State
  Ensembles}},}\ }\href
  {https://doi.org/https://doi.org/10.1021/acs.jctc.4c00428} {\bibfield
  {journal} {\bibinfo  {journal} {J. Chem. Theory Comput.}\ }\textbf {\bibinfo
  {volume} {20}},\ \bibinfo {pages} {7775--7784} (\bibinfo {year}
  {2024})}\BibitemShut {NoStop}%
\bibitem [{\citenamefont {Rydzewski}\ and\ \citenamefont
  {G{\"o}kdemir}(2024)}]{rydzewski2024learning}%
  \BibitemOpen
  \bibfield  {author} {\bibinfo {author} {\bibfnamefont {J.}~\bibnamefont
  {Rydzewski}}\ and\ \bibinfo {author} {\bibfnamefont {T.}~\bibnamefont
  {G{\"o}kdemir}},\ }\bibfield  {title} {\enquote {\bibinfo {title} {{Learning
  Markovian Dynamics with Spectral Maps}},}\ }\href
  {https://doi.org/https://doi.org/10.1063/5.0189241} {\bibfield  {journal}
  {\bibinfo  {journal} {J. Chem. Phys.}\ }\textbf {\bibinfo {volume} {160}},\
  \bibinfo {pages} {091102} (\bibinfo {year} {2024})}\BibitemShut {NoStop}%
\bibitem [{\citenamefont {Hummer}\ and\ \citenamefont
  {Szabo}(2015)}]{hummer2015optimal}%
  \BibitemOpen
  \bibfield  {author} {\bibinfo {author} {\bibfnamefont {G.}~\bibnamefont
  {Hummer}}\ and\ \bibinfo {author} {\bibfnamefont {A.}~\bibnamefont {Szabo}},\
  }\bibfield  {title} {\enquote {\bibinfo {title} {{Optimal Dimensionality
  Reduction of Multistate Kinetic and Markov-State Models}},}\ }\href
  {https://doi.org/https://doi.org/10.1021/jp508375q} {\bibfield  {journal}
  {\bibinfo  {journal} {J. Phys. Chem. B}\ }\textbf {\bibinfo {volume} {119}},\
  \bibinfo {pages} {9029--9037} (\bibinfo {year} {2015})}\BibitemShut {NoStop}%
\bibitem [{\citenamefont {Martini}\ \emph {et~al.}(2017)\citenamefont
  {Martini}, \citenamefont {Kells}, \citenamefont {Covino}, \citenamefont
  {Hummer}, \citenamefont {Buchete},\ and\ \citenamefont
  {Rosta}}]{martini2017variational}%
  \BibitemOpen
  \bibfield  {author} {\bibinfo {author} {\bibfnamefont {L.}~\bibnamefont
  {Martini}}, \bibinfo {author} {\bibfnamefont {A.}~\bibnamefont {Kells}},
  \bibinfo {author} {\bibfnamefont {R.}~\bibnamefont {Covino}}, \bibinfo
  {author} {\bibfnamefont {G.}~\bibnamefont {Hummer}}, \bibinfo {author}
  {\bibfnamefont {N.-V.}\ \bibnamefont {Buchete}},\ and\ \bibinfo {author}
  {\bibfnamefont {E.}~\bibnamefont {Rosta}},\ }\bibfield  {title} {\enquote
  {\bibinfo {title} {{Variational Identification of Markovian Transition
  States}},}\ }\href
  {https://doi.org/https://doi.org/10.1103/physrevx.7.031060} {\bibfield
  {journal} {\bibinfo  {journal} {Phys. Rev. X}\ }\textbf {\bibinfo {volume}
  {7}},\ \bibinfo {pages} {031060} (\bibinfo {year} {2017})}\BibitemShut
  {NoStop}%
\bibitem [{\citenamefont {Ray}, \citenamefont {Trizio},\ and\ \citenamefont
  {Parrinello}(2023)}]{ray2023deep}%
  \BibitemOpen
  \bibfield  {author} {\bibinfo {author} {\bibfnamefont {D.}~\bibnamefont
  {Ray}}, \bibinfo {author} {\bibfnamefont {E.}~\bibnamefont {Trizio}},\ and\
  \bibinfo {author} {\bibfnamefont {M.}~\bibnamefont {Parrinello}},\ }\bibfield
   {title} {\enquote {\bibinfo {title} {{Deep Learning Collective Variables
  from Transition Path Ensemble}},}\ }\href
  {https://doi.org/https://doi.org/10.1063/5.0148872} {\bibfield  {journal}
  {\bibinfo  {journal} {J. Chem. Phys.}\ }\textbf {\bibinfo {volume} {158}},\
  \bibinfo {pages} {204102} (\bibinfo {year} {2023})}\BibitemShut {NoStop}%
\bibitem [{\citenamefont {Best}\ and\ \citenamefont
  {Hummer}(2005)}]{best2005reaction}%
  \BibitemOpen
  \bibfield  {author} {\bibinfo {author} {\bibfnamefont {R.~B.}\ \bibnamefont
  {Best}}\ and\ \bibinfo {author} {\bibfnamefont {G.}~\bibnamefont {Hummer}},\
  }\bibfield  {title} {\enquote {\bibinfo {title} {{Reaction Coordinates and
  Rates from Transition Paths}},}\ }\href
  {https://doi.org/https://doi.org/10.1073/pnas.0408098102} {\bibfield
  {journal} {\bibinfo  {journal} {Proc. Natl. Acad. Sci. U.S.A.}\ }\textbf
  {\bibinfo {volume} {102}},\ \bibinfo {pages} {6732--6737} (\bibinfo {year}
  {2005})}\BibitemShut {NoStop}%
\bibitem [{\citenamefont {Tribello}\ \emph {et~al.}(2014)\citenamefont
  {Tribello}, \citenamefont {Bonomi}, \citenamefont {Branduardi}, \citenamefont
  {Camilloni},\ and\ \citenamefont {Bussi}}]{plumed}%
  \BibitemOpen
  \bibfield  {author} {\bibinfo {author} {\bibfnamefont {G.~A.}\ \bibnamefont
  {Tribello}}, \bibinfo {author} {\bibfnamefont {M.}~\bibnamefont {Bonomi}},
  \bibinfo {author} {\bibfnamefont {D.}~\bibnamefont {Branduardi}}, \bibinfo
  {author} {\bibfnamefont {C.}~\bibnamefont {Camilloni}},\ and\ \bibinfo
  {author} {\bibfnamefont {G.}~\bibnamefont {Bussi}},\ }\bibfield  {title}
  {\enquote {\bibinfo {title} {{\textsc{plumed} 2: New Feathers for an Old
  Bird}},}\ }\href {https://doi.org/https://doi.org/10.1016/j.cpc.2013.09.018}
  {\bibfield  {journal} {\bibinfo  {journal} {Comp. Phys. Commun.}\ }\textbf
  {\bibinfo {volume} {185}},\ \bibinfo {pages} {604--613} (\bibinfo {year}
  {2014})}\BibitemShut {NoStop}%
\bibitem [{\citenamefont {{\textsc{plumed} Consortium,}}(2019)}]{plumed-nest}%
  \BibitemOpen
  \bibfield  {author} {\bibinfo {author} {\bibnamefont {{\textsc{plumed}
  Consortium,}}},\ }\bibfield  {title} {\enquote {\bibinfo {title} {{Promoting
  Transparency and Reproducibility in Enhanced Molecular Simulations}},}\
  }\href {https://doi.org/https://doi.org/10.1038/s41592-019-0506-8} {\bibfield
   {journal} {\bibinfo  {journal} {Nat. Methods}\ }\textbf {\bibinfo {volume}
  {16}},\ \bibinfo {pages} {670--673} (\bibinfo {year} {2019})}\BibitemShut
  {NoStop}%
\bibitem [{\citenamefont {Tribello}\ \emph {et~al.}(2024)\citenamefont
  {Tribello}, \citenamefont {Bonomi}, \citenamefont {Bussi}, \citenamefont
  {Camilloni} \emph {et~al.}}]{plumed-tutorials}%
  \BibitemOpen
  \bibfield  {author} {\bibinfo {author} {\bibfnamefont {G.~A.}\ \bibnamefont
  {Tribello}}, \bibinfo {author} {\bibfnamefont {M.}~\bibnamefont {Bonomi}},
  \bibinfo {author} {\bibfnamefont {G.}~\bibnamefont {Bussi}}, \bibinfo
  {author} {\bibfnamefont {C.}~\bibnamefont {Camilloni}}, \emph {et~al.},\
  }\bibfield  {title} {\enquote {\bibinfo {title} {{PLUMED Tutorials: A
  Collaborative, Community-Driven Learning Ecosystem}},}\ }\href@noop {}
  {\bibfield  {journal} {\bibinfo  {journal} {arXiv preprint arXiv:2412.03595}\
  } (\bibinfo {year} {2024})}\BibitemShut {NoStop}%
\bibitem [{\citenamefont {Bonati}\ \emph {et~al.}(2023)\citenamefont {Bonati},
  \citenamefont {Trizio}, \citenamefont {Rizzi},\ and\ \citenamefont
  {Parrinello}}]{bonati2023unified}%
  \BibitemOpen
  \bibfield  {author} {\bibinfo {author} {\bibfnamefont {L.}~\bibnamefont
  {Bonati}}, \bibinfo {author} {\bibfnamefont {E.}~\bibnamefont {Trizio}},
  \bibinfo {author} {\bibfnamefont {A.}~\bibnamefont {Rizzi}},\ and\ \bibinfo
  {author} {\bibfnamefont {M.}~\bibnamefont {Parrinello}},\ }\bibfield  {title}
  {\enquote {\bibinfo {title} {{A Unified Framework for Machine Learning
  Collective Variables for Enhanced Sampling Simulations: \tt mlcolvar}},}\
  }\href {https://doi.org/https://doi.org/10.1063/5.0156343} {\bibfield
  {journal} {\bibinfo  {journal} {J. Chem. Phys.}\ }\textbf {\bibinfo {volume}
  {159}},\ \bibinfo {pages} {014801} (\bibinfo {year} {2023})}\BibitemShut
  {NoStop}%
\bibitem [{\citenamefont {Trizio}\ \emph {et~al.}(2024)\citenamefont {Trizio},
  \citenamefont {Rizzi}, \citenamefont {Piaggi}, \citenamefont {Invernizzi},\
  and\ \citenamefont {Bonati}}]{trizio2024advanced}%
  \BibitemOpen
  \bibfield  {author} {\bibinfo {author} {\bibfnamefont {E.}~\bibnamefont
  {Trizio}}, \bibinfo {author} {\bibfnamefont {A.}~\bibnamefont {Rizzi}},
  \bibinfo {author} {\bibfnamefont {P.~M.}\ \bibnamefont {Piaggi}}, \bibinfo
  {author} {\bibfnamefont {M.}~\bibnamefont {Invernizzi}},\ and\ \bibinfo
  {author} {\bibfnamefont {L.}~\bibnamefont {Bonati}},\ }\bibfield  {title}
  {\enquote {\bibinfo {title} {{Advanced Simulations with PLUMED: OPES and
  Machine Learning Collective Variables}},}\ }\href@noop {} {\bibfield
  {journal} {\bibinfo  {journal} {arXiv preprint arXiv:2410.18019}\ } (\bibinfo
  {year} {2024})}\BibitemShut {NoStop}%
\bibitem [{\citenamefont {Maragliano}\ \emph {et~al.}(2006)\citenamefont
  {Maragliano}, \citenamefont {Fischer}, \citenamefont {Vanden-Eijnden},\ and\
  \citenamefont {Ciccotti}}]{maragliano2006string}%
  \BibitemOpen
  \bibfield  {author} {\bibinfo {author} {\bibfnamefont {L.}~\bibnamefont
  {Maragliano}}, \bibinfo {author} {\bibfnamefont {A.}~\bibnamefont {Fischer}},
  \bibinfo {author} {\bibfnamefont {E.}~\bibnamefont {Vanden-Eijnden}},\ and\
  \bibinfo {author} {\bibfnamefont {G.}~\bibnamefont {Ciccotti}},\ }\bibfield
  {title} {\enquote {\bibinfo {title} {{String Method in Collective Variables:
  Minimum Free Energy Paths and Isocommittor Surfaces}},}\ }\href
  {https://doi.org/https://doi.org/10.1063/1.2212942} {\bibfield  {journal}
  {\bibinfo  {journal} {J. Chem. Phys.}\ }\textbf {\bibinfo {volume} {125}},\
  \bibinfo {pages} {024106} (\bibinfo {year} {2006})}\BibitemShut {NoStop}%
\bibitem [{\citenamefont {Lange}\ and\ \citenamefont
  {Grubm{\"u}ller}(2006)}]{lange2006collective}%
  \BibitemOpen
  \bibfield  {author} {\bibinfo {author} {\bibfnamefont {O.~F.}\ \bibnamefont
  {Lange}}\ and\ \bibinfo {author} {\bibfnamefont {H.}~\bibnamefont
  {Grubm{\"u}ller}},\ }\bibfield  {title} {\enquote {\bibinfo {title}
  {{Collective Langevin Dynamics of Conformational Motions in Proteins}},}\
  }\href {https://doi.org/https://doi.org/10.1063/1.2199530} {\bibfield
  {journal} {\bibinfo  {journal} {J. Chem. Phys.}\ }\textbf {\bibinfo {volume}
  {124}},\ \bibinfo {pages} {214903} (\bibinfo {year} {2006})}\BibitemShut
  {NoStop}%
\bibitem [{\citenamefont {Legoll}\ and\ \citenamefont
  {Lelievre}(2010)}]{legoll2010effective}%
  \BibitemOpen
  \bibfield  {author} {\bibinfo {author} {\bibfnamefont {F.}~\bibnamefont
  {Legoll}}\ and\ \bibinfo {author} {\bibfnamefont {T.}~\bibnamefont
  {Lelievre}},\ }\bibfield  {title} {\enquote {\bibinfo {title} {{Effective
  Dynamics using Conditional Expectations}},}\ }\href
  {https://doi.org/https://doi.org/10.1088/0951-7715/23/9/006} {\bibfield
  {journal} {\bibinfo  {journal} {Nonlinearity}\ }\textbf {\bibinfo {volume}
  {23}},\ \bibinfo {pages} {2131} (\bibinfo {year} {2010})}\BibitemShut
  {NoStop}%
\bibitem [{\citenamefont {Zhang}, \citenamefont {Hartmann},\ and\ \citenamefont
  {Sch{\"u}tte}(2016)}]{zhang2016effective}%
  \BibitemOpen
  \bibfield  {author} {\bibinfo {author} {\bibfnamefont {W.}~\bibnamefont
  {Zhang}}, \bibinfo {author} {\bibfnamefont {C.}~\bibnamefont {Hartmann}},\
  and\ \bibinfo {author} {\bibfnamefont {C.}~\bibnamefont {Sch{\"u}tte}},\
  }\bibfield  {title} {\enquote {\bibinfo {title} {{Effective Dynamics along
  Given Reaction Coordinates, and Reaction Rate Theory}},}\ }\href
  {https://doi.org/https://doi.org/10.1039/c6fd00147e} {\bibfield  {journal}
  {\bibinfo  {journal} {Faraday Discuss.}\ }\textbf {\bibinfo {volume} {195}},\
  \bibinfo {pages} {365--394} (\bibinfo {year} {2016})}\BibitemShut {NoStop}%
\bibitem [{\citenamefont {Rhee}\ and\ \citenamefont
  {Pande}(2005)}]{rhee2005one}%
  \BibitemOpen
  \bibfield  {author} {\bibinfo {author} {\bibfnamefont {Y.~M.}\ \bibnamefont
  {Rhee}}\ and\ \bibinfo {author} {\bibfnamefont {V.~S.}\ \bibnamefont
  {Pande}},\ }\bibfield  {title} {\enquote {\bibinfo {title} {{One-Dimensional
  Reaction Coordinate and the Corresponding Potential of Mean Force from
  Commitment Probability Distribution}},}\ }\href
  {https://doi.org/https://doi.org/10.1021/jp045544s} {\bibfield  {journal}
  {\bibinfo  {journal} {J. Phys. Chem. B}\ }\textbf {\bibinfo {volume} {109}},\
  \bibinfo {pages} {6780--6786} (\bibinfo {year} {2005})}\BibitemShut {NoStop}%
\bibitem [{\citenamefont {Krivov}\ and\ \citenamefont
  {Karplus}(2008)}]{krivov2008diffusive}%
  \BibitemOpen
  \bibfield  {author} {\bibinfo {author} {\bibfnamefont {S.~V.}\ \bibnamefont
  {Krivov}}\ and\ \bibinfo {author} {\bibfnamefont {M.}~\bibnamefont
  {Karplus}},\ }\bibfield  {title} {\enquote {\bibinfo {title} {{Diffusive
  Reaction Dynamics on Invariant Free Energy Profiles}},}\ }\href
  {https://doi.org/https://doi.org/10.1073/pnas.0800228105} {\bibfield
  {journal} {\bibinfo  {journal} {Proc. Natl. Acad. Sci. U.S.A.}\ }\textbf
  {\bibinfo {volume} {105}},\ \bibinfo {pages} {13841--13846} (\bibinfo {year}
  {2008})}\BibitemShut {NoStop}%
\bibitem [{\citenamefont {Best}\ and\ \citenamefont
  {Hummer}(2010)}]{best2010coordinate}%
  \BibitemOpen
  \bibfield  {author} {\bibinfo {author} {\bibfnamefont {R.~B.}\ \bibnamefont
  {Best}}\ and\ \bibinfo {author} {\bibfnamefont {G.}~\bibnamefont {Hummer}},\
  }\bibfield  {title} {\enquote {\bibinfo {title} {{Coordinate-Dependent
  Diffusion in Protein Folding}},}\ }\href
  {https://doi.org/https://doi.org/10.1073/pnas.0910390107} {\bibfield
  {journal} {\bibinfo  {journal} {Proc. Natl. Acad. Sci. U.S.A.}\ }\textbf
  {\bibinfo {volume} {107}},\ \bibinfo {pages} {1088--1093} (\bibinfo {year}
  {2010})}\BibitemShut {NoStop}%
\bibitem [{\citenamefont {Dietschreit}\ \emph {et~al.}(2022)\citenamefont
  {Dietschreit}, \citenamefont {Diestler}, \citenamefont {Hulm}, \citenamefont
  {Ochsenfeld},\ and\ \citenamefont
  {G{\'o}mez-Bombarelli}}]{dietschreit2022free}%
  \BibitemOpen
  \bibfield  {author} {\bibinfo {author} {\bibfnamefont {J.~C.}\ \bibnamefont
  {Dietschreit}}, \bibinfo {author} {\bibfnamefont {D.~J.}\ \bibnamefont
  {Diestler}}, \bibinfo {author} {\bibfnamefont {A.}~\bibnamefont {Hulm}},
  \bibinfo {author} {\bibfnamefont {C.}~\bibnamefont {Ochsenfeld}},\ and\
  \bibinfo {author} {\bibfnamefont {R.}~\bibnamefont {G{\'o}mez-Bombarelli}},\
  }\bibfield  {title} {\enquote {\bibinfo {title} {{From Free-Energy Profiles
  to Activation Free Energies}},}\ }\href
  {https://doi.org/https://doi.org/10.1063/5.0102075} {\bibfield  {journal}
  {\bibinfo  {journal} {J. Chem. Phys.}\ }\textbf {\bibinfo {volume} {157}},\
  \bibinfo {pages} {084113} (\bibinfo {year} {2022})}\BibitemShut {NoStop}%
\bibitem [{\citenamefont {Nakamura}(2024)}]{nakamura2024derivation}%
  \BibitemOpen
  \bibfield  {author} {\bibinfo {author} {\bibfnamefont {T.}~\bibnamefont
  {Nakamura}},\ }\bibfield  {title} {\enquote {\bibinfo {title} {{Derivation of
  the Invariant Free-Energy Landscape Based on Langevin Dynamics}},}\ }\href
  {https://doi.org/https://doi.org/10.1103/physrevlett.132.137101} {\bibfield
  {journal} {\bibinfo  {journal} {Phys. Rev. Lett.}\ }\textbf {\bibinfo
  {volume} {132}},\ \bibinfo {pages} {137101} (\bibinfo {year}
  {2024})}\BibitemShut {NoStop}%
\bibitem [{\citenamefont {Boyd}, \citenamefont {Diaconis},\ and\ \citenamefont
  {Xiao}(2004)}]{boyd2004fastest}%
  \BibitemOpen
  \bibfield  {author} {\bibinfo {author} {\bibfnamefont {S.}~\bibnamefont
  {Boyd}}, \bibinfo {author} {\bibfnamefont {P.}~\bibnamefont {Diaconis}},\
  and\ \bibinfo {author} {\bibfnamefont {L.}~\bibnamefont {Xiao}},\ }\bibfield
  {title} {\enquote {\bibinfo {title} {{Fastest Mixing Markov Chain on a
  Graph}},}\ }\href {https://doi.org/https://doi.org/10.1137/s0036144503423264}
  {\bibfield  {journal} {\bibinfo  {journal} {SIAM Rev.}\ }\textbf {\bibinfo
  {volume} {46}},\ \bibinfo {pages} {667--689} (\bibinfo {year}
  {2004})}\BibitemShut {NoStop}%
\bibitem [{\citenamefont {Boyd}(2006)}]{boyd2006convex}%
  \BibitemOpen
  \bibfield  {author} {\bibinfo {author} {\bibfnamefont {S.}~\bibnamefont
  {Boyd}},\ }\bibfield  {title} {\enquote {\bibinfo {title} {{Convex
  Optimization of Graph Laplacian Eigenvalues}},}\ }\href
  {https://doi.org/https://doi.org/10.4171/022-3/63} {\bibfield  {journal}
  {\bibinfo  {journal} {Proc. ICM}\ }\textbf {\bibinfo {volume} {3}},\ \bibinfo
  {pages} {1311--1319} (\bibinfo {year} {2006})}\BibitemShut {NoStop}%
\bibitem [{\citenamefont {Donati}, \citenamefont {Hartmann},\ and\
  \citenamefont {Keller}(2017)}]{donati2017girsanov}%
  \BibitemOpen
  \bibfield  {author} {\bibinfo {author} {\bibfnamefont {L.}~\bibnamefont
  {Donati}}, \bibinfo {author} {\bibfnamefont {C.}~\bibnamefont {Hartmann}},\
  and\ \bibinfo {author} {\bibfnamefont {B.~G.}\ \bibnamefont {Keller}},\
  }\bibfield  {title} {\enquote {\bibinfo {title} {{Girsanov Reweighting for
  Path Ensembles and Markov State Models}},}\ }\href
  {https://doi.org/10.1063/1.4989474} {\bibfield  {journal} {\bibinfo
  {journal} {J. Chem. Phys.}\ }\textbf {\bibinfo {volume} {146}},\ \bibinfo
  {pages} {244112} (\bibinfo {year} {2017})}\BibitemShut {NoStop}%
\bibitem [{\citenamefont {Kieninger}, \citenamefont {Donati},\ and\
  \citenamefont {Keller}(2020)}]{kieninger2020dynamical}%
  \BibitemOpen
  \bibfield  {author} {\bibinfo {author} {\bibfnamefont {S.}~\bibnamefont
  {Kieninger}}, \bibinfo {author} {\bibfnamefont {L.}~\bibnamefont {Donati}},\
  and\ \bibinfo {author} {\bibfnamefont {B.~G.}\ \bibnamefont {Keller}},\
  }\bibfield  {title} {\enquote {\bibinfo {title} {{Dynamical Reweighting
  Methods for Markov Models}},}\ }\href
  {https://doi.org/https://doi.org/10.1016/j.sbi.2019.12.018} {\bibfield
  {journal} {\bibinfo  {journal} {Curr. Opin. Struct. Biol.}\ }\textbf
  {\bibinfo {volume} {61}},\ \bibinfo {pages} {124--131} (\bibinfo {year}
  {2020})}\BibitemShut {NoStop}%
\bibitem [{\citenamefont {Donati}, \citenamefont {Weber},\ and\ \citenamefont
  {Keller}(2022)}]{donati2022review}%
  \BibitemOpen
  \bibfield  {author} {\bibinfo {author} {\bibfnamefont {L.}~\bibnamefont
  {Donati}}, \bibinfo {author} {\bibfnamefont {M.}~\bibnamefont {Weber}},\ and\
  \bibinfo {author} {\bibfnamefont {B.~G.}\ \bibnamefont {Keller}},\ }\bibfield
   {title} {\enquote {\bibinfo {title} {{A Review of Girsanov Reweighting and
  of Square Root Approximation for Building Molecular Markov State Models}},}\
  }\href {https://doi.org/https://doi.org/10.1063/5.0127227} {\bibfield
  {journal} {\bibinfo  {journal} {J. Math. Phys.}\ }\textbf {\bibinfo {volume}
  {63}},\ \bibinfo {pages} {123306} (\bibinfo {year} {2022})}\BibitemShut
  {NoStop}%
\bibitem [{\citenamefont {Chen}\ and\ \citenamefont
  {Ferguson}(2018)}]{chen2018molecular}%
  \BibitemOpen
  \bibfield  {author} {\bibinfo {author} {\bibfnamefont {W.}~\bibnamefont
  {Chen}}\ and\ \bibinfo {author} {\bibfnamefont {A.~L.}\ \bibnamefont
  {Ferguson}},\ }\bibfield  {title} {\enquote {\bibinfo {title} {{Molecular
  Enhanced Sampling with Autoencoders: On-the-fly Collective Variable Discovery
  and Accelerated Free Energy Landscape Exploration}},}\ }\href
  {https://doi.org/https://doi.org/10.1002/jcc.25520} {\bibfield  {journal}
  {\bibinfo  {journal} {J. Comput. Chem.}\ }\textbf {\bibinfo {volume} {39}},\
  \bibinfo {pages} {2079--2102} (\bibinfo {year} {2018})}\BibitemShut {NoStop}%
\bibitem [{\citenamefont {Ribeiro}\ \emph {et~al.}(2018)\citenamefont
  {Ribeiro}, \citenamefont {Bravo}, \citenamefont {Wang},\ and\ \citenamefont
  {Tiwary}}]{ribeiro2018reweighted}%
  \BibitemOpen
  \bibfield  {author} {\bibinfo {author} {\bibfnamefont {J.~M.~L.}\
  \bibnamefont {Ribeiro}}, \bibinfo {author} {\bibfnamefont {P.}~\bibnamefont
  {Bravo}}, \bibinfo {author} {\bibfnamefont {Y.}~\bibnamefont {Wang}},\ and\
  \bibinfo {author} {\bibfnamefont {P.}~\bibnamefont {Tiwary}},\ }\bibfield
  {title} {\enquote {\bibinfo {title} {{Reweighted Autoencoded Variational
  Bayes for Enhanced Sampling (RAVE)}},}\ }\href
  {https://doi.org/https://doi.org/10.1063/1.5025487} {\bibfield  {journal}
  {\bibinfo  {journal} {J. Chem. Phys.}\ }\textbf {\bibinfo {volume} {149}},\
  \bibinfo {pages} {072301} (\bibinfo {year} {2018})}\BibitemShut {NoStop}%
\bibitem [{\citenamefont {Brotzakis}\ and\ \citenamefont
  {Parrinello}(2018)}]{brotzakis2018enhanced}%
  \BibitemOpen
  \bibfield  {author} {\bibinfo {author} {\bibfnamefont {Z.~F.}\ \bibnamefont
  {Brotzakis}}\ and\ \bibinfo {author} {\bibfnamefont {M.}~\bibnamefont
  {Parrinello}},\ }\bibfield  {title} {\enquote {\bibinfo {title} {{Enhanced
  Sampling of Protein Conformational Transitions via Dynamically Optimized
  Collective Variables}},}\ }\href
  {https://doi.org/https://doi.org/10.1021/acs.jctc.8b00827} {\bibfield
  {journal} {\bibinfo  {journal} {J. Chem. Theory Comput.}\ }\textbf {\bibinfo
  {volume} {15}},\ \bibinfo {pages} {1393--1398} (\bibinfo {year}
  {2018})}\BibitemShut {NoStop}%
\bibitem [{\citenamefont {Mehdi}\ \emph {et~al.}(2022)\citenamefont {Mehdi},
  \citenamefont {Wang}, \citenamefont {Pant},\ and\ \citenamefont
  {Tiwary}}]{mehdi2022accelerating}%
  \BibitemOpen
  \bibfield  {author} {\bibinfo {author} {\bibfnamefont {S.}~\bibnamefont
  {Mehdi}}, \bibinfo {author} {\bibfnamefont {D.}~\bibnamefont {Wang}},
  \bibinfo {author} {\bibfnamefont {S.}~\bibnamefont {Pant}},\ and\ \bibinfo
  {author} {\bibfnamefont {P.}~\bibnamefont {Tiwary}},\ }\bibfield  {title}
  {\enquote {\bibinfo {title} {{Accelerating All-Atom Simulations and Gaining
  Mechanistic Understanding of Biophysical Systems through State Predictive
  Information Bottleneck}},}\ }\href
  {https://doi.org/https://doi.org/10.1021/acs.jctc.2c00058} {\bibfield
  {journal} {\bibinfo  {journal} {J. Chem. Theory Comput.}\ }\textbf {\bibinfo
  {volume} {18}},\ \bibinfo {pages} {3231--3238} (\bibinfo {year}
  {2022})}\BibitemShut {NoStop}%
\bibitem [{\citenamefont {Shmilovich}\ and\ \citenamefont
  {Ferguson}(2023)}]{shmilovich2023girsanov}%
  \BibitemOpen
  \bibfield  {author} {\bibinfo {author} {\bibfnamefont {K.}~\bibnamefont
  {Shmilovich}}\ and\ \bibinfo {author} {\bibfnamefont {A.~L.}\ \bibnamefont
  {Ferguson}},\ }\bibfield  {title} {\enquote {\bibinfo {title} {{Girsanov
  Reweighting Enhanced Sampling Technique (GREST): On-the-Fly Data-Driven
  Discovery of and Enhanced Sampling in Slow Collective Variables}},}\ }\href
  {https://doi.org/https://doi.org/10.1021/acs.jpca.3c00505} {\bibfield
  {journal} {\bibinfo  {journal} {J. Phys. Chem. A}\ }\textbf {\bibinfo
  {volume} {127}},\ \bibinfo {pages} {3497--3517} (\bibinfo {year}
  {2023})}\BibitemShut {NoStop}%
\bibitem [{\citenamefont {Molgedey}\ and\ \citenamefont
  {Schuster}(1994)}]{molgedey1994separation}%
  \BibitemOpen
  \bibfield  {author} {\bibinfo {author} {\bibfnamefont {L.}~\bibnamefont
  {Molgedey}}\ and\ \bibinfo {author} {\bibfnamefont {H.~G.}\ \bibnamefont
  {Schuster}},\ }\bibfield  {title} {\enquote {\bibinfo {title} {{Separation of
  a Mixture of Independent Signals using Time Delayed Correlations}},}\ }\href
  {https://doi.org/10.1103/PhysRevLett.72.3634} {\bibfield  {journal} {\bibinfo
   {journal} {Phys. Rev. Lett.}\ }\textbf {\bibinfo {volume} {72}},\ \bibinfo
  {pages} {3634--3637} (\bibinfo {year} {1994})}\BibitemShut {NoStop}%
\bibitem [{\citenamefont {Wiskott}\ and\ \citenamefont
  {Sejnowski}(2002)}]{wiskott2002slow}%
  \BibitemOpen
  \bibfield  {author} {\bibinfo {author} {\bibfnamefont {L.}~\bibnamefont
  {Wiskott}}\ and\ \bibinfo {author} {\bibfnamefont {T.~J.}\ \bibnamefont
  {Sejnowski}},\ }\bibfield  {title} {\enquote {\bibinfo {title} {{Slow Feature
  Analysis: Unsupervised Learning of Invariances}},}\ }\href
  {https://doi.org/https://doi.org/10.1162/089976602317318938} {\bibfield
  {journal} {\bibinfo  {journal} {Neural Comput.}\ }\textbf {\bibinfo {volume}
  {14}},\ \bibinfo {pages} {715--770} (\bibinfo {year} {2002})}\BibitemShut
  {NoStop}%
\bibitem [{\citenamefont {Ceriotti}, \citenamefont {Tribello},\ and\
  \citenamefont {Parrinello}(2011)}]{ceriotti2011simplifying}%
  \BibitemOpen
  \bibfield  {author} {\bibinfo {author} {\bibfnamefont {M.}~\bibnamefont
  {Ceriotti}}, \bibinfo {author} {\bibfnamefont {G.~A.}\ \bibnamefont
  {Tribello}},\ and\ \bibinfo {author} {\bibfnamefont {M.}~\bibnamefont
  {Parrinello}},\ }\bibfield  {title} {\enquote {\bibinfo {title} {{Simplifying
  the Representation of Complex Free-Energy Landscapes using Sketch-Map}},}\
  }\href {https://doi.org/https://doi.org/10.1073/pnas.1108486108} {\bibfield
  {journal} {\bibinfo  {journal} {Proc. Natl. Acad. Sci. U.S.A.}\ }\textbf
  {\bibinfo {volume} {108}},\ \bibinfo {pages} {13023--13028} (\bibinfo {year}
  {2011})}\BibitemShut {NoStop}%
\bibitem [{\citenamefont {Naritomi}\ and\ \citenamefont
  {Fuchigami}(2011)}]{naritomi2011slow}%
  \BibitemOpen
  \bibfield  {author} {\bibinfo {author} {\bibfnamefont {Y.}~\bibnamefont
  {Naritomi}}\ and\ \bibinfo {author} {\bibfnamefont {S.}~\bibnamefont
  {Fuchigami}},\ }\bibfield  {title} {\enquote {\bibinfo {title} {{Slow
  Dynamics in Protein Fluctuations Revealed by Time-Structure based Independent
  Component Analysis: the Case of Domain Motions}},}\ }\href
  {https://doi.org/https://doi.org/10.1063/1.3554380} {\bibfield  {journal}
  {\bibinfo  {journal} {J. Chem. Phys.}\ }\textbf {\bibinfo {volume} {134}},\
  \bibinfo {pages} {065101} (\bibinfo {year} {2011})}\BibitemShut {NoStop}%
\bibitem [{\citenamefont {Ceriotti}, \citenamefont {Tribello},\ and\
  \citenamefont {Parrinello}(2013)}]{ceriotti2013demonstrating}%
  \BibitemOpen
  \bibfield  {author} {\bibinfo {author} {\bibfnamefont {M.}~\bibnamefont
  {Ceriotti}}, \bibinfo {author} {\bibfnamefont {G.~A.}\ \bibnamefont
  {Tribello}},\ and\ \bibinfo {author} {\bibfnamefont {M.}~\bibnamefont
  {Parrinello}},\ }\bibfield  {title} {\enquote {\bibinfo {title}
  {{Demonstrating the Transferability and the Descriptive Power of
  Sketch-Map}},}\ }\href {https://doi.org/https://doi.org/10.1021/ct3010563}
  {\bibfield  {journal} {\bibinfo  {journal} {J. Chem. Theory Comput.}\
  }\textbf {\bibinfo {volume} {9}},\ \bibinfo {pages} {1521--1532} (\bibinfo
  {year} {2013})}\BibitemShut {NoStop}%
\bibitem [{\citenamefont {Pérez-Hernández}\ and\ \citenamefont
  {Noé}(2016)}]{perez2016hierarchical}%
  \BibitemOpen
  \bibfield  {author} {\bibinfo {author} {\bibfnamefont {G.}~\bibnamefont
  {Pérez-Hernández}}\ and\ \bibinfo {author} {\bibfnamefont {F.}~\bibnamefont
  {Noé}},\ }\bibfield  {title} {\enquote {\bibinfo {title} {{Hierarchical
  Time-Lagged Independent Component Analysis: Computing Slow Modes and Reaction
  Coordinates for Large Molecular Systems}},}\ }\href
  {https://doi.org/10.1021/acs.jctc.6b00738} {\bibfield  {journal} {\bibinfo
  {journal} {J. Chem. Theory Comput.}\ }\textbf {\bibinfo {volume} {12}},\
  \bibinfo {pages} {6118--6129} (\bibinfo {year} {2016})}\BibitemShut {NoStop}%
\bibitem [{\citenamefont {McGibbon}, \citenamefont {Husic},\ and\ \citenamefont
  {Pande}(2017)}]{mcgibbon2017identification}%
  \BibitemOpen
  \bibfield  {author} {\bibinfo {author} {\bibfnamefont {R.~T.}\ \bibnamefont
  {McGibbon}}, \bibinfo {author} {\bibfnamefont {B.~E.}\ \bibnamefont
  {Husic}},\ and\ \bibinfo {author} {\bibfnamefont {V.~S.}\ \bibnamefont
  {Pande}},\ }\bibfield  {title} {\enquote {\bibinfo {title} {{Identification
  of Simple Reaction Coordinates from Complex Dynamics}},}\ }\href
  {https://doi.org/https://doi.org/10.1063/1.4974306} {\bibfield  {journal}
  {\bibinfo  {journal} {J. Chem. Phys.}\ }\textbf {\bibinfo {volume} {146}},\
  \bibinfo {pages} {044109} (\bibinfo {year} {2017})}\BibitemShut {NoStop}%
\bibitem [{\citenamefont {Sidky}, \citenamefont {Chen},\ and\ \citenamefont
  {Ferguson}(2019)}]{sidky2019high}%
  \BibitemOpen
  \bibfield  {author} {\bibinfo {author} {\bibfnamefont {H.}~\bibnamefont
  {Sidky}}, \bibinfo {author} {\bibfnamefont {W.}~\bibnamefont {Chen}},\ and\
  \bibinfo {author} {\bibfnamefont {A.~L.}\ \bibnamefont {Ferguson}},\
  }\bibfield  {title} {\enquote {\bibinfo {title} {{High-Resolution Markov
  State Models for the Dynamics of Trp-Cage Miniprotein Constructed Over Slow
  Folding Modes Identified by State-Free Reversible VAMPnets}},}\ }\href
  {https://doi.org/10.1021/acs.jpcb.9b05578} {\bibfield  {journal} {\bibinfo
  {journal} {J. Phys. Chem. B}\ }\textbf {\bibinfo {volume} {123}},\ \bibinfo
  {pages} {7999--8009} (\bibinfo {year} {2019})}\BibitemShut {NoStop}%
\bibitem [{\citenamefont {Li}, \citenamefont {Lin},\ and\ \citenamefont
  {Ren}(2019)}]{li2019computing}%
  \BibitemOpen
  \bibfield  {author} {\bibinfo {author} {\bibfnamefont {Q.}~\bibnamefont
  {Li}}, \bibinfo {author} {\bibfnamefont {B.}~\bibnamefont {Lin}},\ and\
  \bibinfo {author} {\bibfnamefont {W.}~\bibnamefont {Ren}},\ }\bibfield
  {title} {\enquote {\bibinfo {title} {{Computing Committor Functions for the
  Study of Rare Events using Deep Learning}},}\ }\href
  {https://doi.org/https://doi.org/10.1063/1.5110439} {\bibfield  {journal}
  {\bibinfo  {journal} {J. Chem. Phys.}\ }\textbf {\bibinfo {volume} {151}},\
  \bibinfo {pages} {054112} (\bibinfo {year} {2019})}\BibitemShut {NoStop}%
\bibitem [{\citenamefont {Chen}, \citenamefont {Sidky},\ and\ \citenamefont
  {Ferguson}(2019{\natexlab{b}})}]{chen2019capabilities}%
  \BibitemOpen
  \bibfield  {author} {\bibinfo {author} {\bibfnamefont {W.}~\bibnamefont
  {Chen}}, \bibinfo {author} {\bibfnamefont {H.}~\bibnamefont {Sidky}},\ and\
  \bibinfo {author} {\bibfnamefont {A.~L.}\ \bibnamefont {Ferguson}},\
  }\bibfield  {title} {\enquote {\bibinfo {title} {{Capabilities and
  Limitations of Time-Lagged Autoencoders for Slow Mode Discovery in Dynamical
  Systems}},}\ }\href {https://doi.org/https://doi.org/10.1063/1.5112048}
  {\bibfield  {journal} {\bibinfo  {journal} {J. Chem. Phys.}\ }\textbf
  {\bibinfo {volume} {151}},\ \bibinfo {pages} {064123} (\bibinfo {year}
  {2019}{\natexlab{b}})}\BibitemShut {NoStop}%
\bibitem [{\citenamefont {Tribello}\ and\ \citenamefont
  {Gasparotto}(2019)}]{tribello2019using}%
  \BibitemOpen
  \bibfield  {author} {\bibinfo {author} {\bibfnamefont {G.~A.}\ \bibnamefont
  {Tribello}}\ and\ \bibinfo {author} {\bibfnamefont {P.}~\bibnamefont
  {Gasparotto}},\ }\bibfield  {title} {\enquote {\bibinfo {title} {{Using
  Dimensionality Reduction to Analyze Protein Trajectories}},}\ }\href
  {https://doi.org/https://doi.org/10.3389/fmolb.2019.00046} {\bibfield
  {journal} {\bibinfo  {journal} {Front. Mol. Biosci.}\ }\textbf {\bibinfo
  {volume} {6}},\ \bibinfo {pages} {46} (\bibinfo {year} {2019})}\BibitemShut
  {NoStop}%
\bibitem [{\citenamefont {Wang}, \citenamefont {Ribeiro},\ and\ \citenamefont
  {Tiwary}(2019)}]{wang2019past}%
  \BibitemOpen
  \bibfield  {author} {\bibinfo {author} {\bibfnamefont {Y.}~\bibnamefont
  {Wang}}, \bibinfo {author} {\bibfnamefont {J.~M.~L.}\ \bibnamefont
  {Ribeiro}},\ and\ \bibinfo {author} {\bibfnamefont {P.}~\bibnamefont
  {Tiwary}},\ }\bibfield  {title} {\enquote {\bibinfo {title} {{Past--Future
  Information Bottleneck for Sampling Molecular Reaction Coordinate
  Simultaneously with Thermodynamics and Kinetics }},}\ }\href
  {https://doi.org/https://doi.org/10.1038/s41467-019-11405-4} {\bibfield
  {journal} {\bibinfo  {journal} {Nat. Commun.}\ }\textbf {\bibinfo {volume}
  {10}},\ \bibinfo {pages} {3573} (\bibinfo {year} {2019})}\BibitemShut
  {NoStop}%
\bibitem [{\citenamefont {Zhang}, \citenamefont {Yang},\ and\ \citenamefont
  {No{\'e}}(2019)}]{zhang2019targeted}%
  \BibitemOpen
  \bibfield  {author} {\bibinfo {author} {\bibfnamefont {J.}~\bibnamefont
  {Zhang}}, \bibinfo {author} {\bibfnamefont {Y.~I.}\ \bibnamefont {Yang}},\
  and\ \bibinfo {author} {\bibfnamefont {F.}~\bibnamefont {No{\'e}}},\
  }\bibfield  {title} {\enquote {\bibinfo {title} {{Targeted Adversarial
  Learning Optimized Sampling}},}\ }\href
  {https://doi.org/https://doi.org/10.1021/acs.jpclett.9b02173} {\bibfield
  {journal} {\bibinfo  {journal} {J. Phys. Chem. Lett.}\ }\textbf {\bibinfo
  {volume} {10}},\ \bibinfo {pages} {5791--5797} (\bibinfo {year}
  {2019})}\BibitemShut {NoStop}%
\bibitem [{\citenamefont {Bonati}, \citenamefont {Rizzi},\ and\ \citenamefont
  {Parrinello}(2020)}]{bonati2020data}%
  \BibitemOpen
  \bibfield  {author} {\bibinfo {author} {\bibfnamefont {L.}~\bibnamefont
  {Bonati}}, \bibinfo {author} {\bibfnamefont {V.}~\bibnamefont {Rizzi}},\ and\
  \bibinfo {author} {\bibfnamefont {M.}~\bibnamefont {Parrinello}},\ }\bibfield
   {title} {\enquote {\bibinfo {title} {{Data-Driven Collective Variables for
  Enhanced Sampling}},}\ }\href
  {https://doi.org/https://doi.org/10.1021/acs.jpclett.0c00535} {\bibfield
  {journal} {\bibinfo  {journal} {J. Phys. Chem. Lett.}\ }\textbf {\bibinfo
  {volume} {11}},\ \bibinfo {pages} {2998--3004} (\bibinfo {year}
  {2020})}\BibitemShut {NoStop}%
\bibitem [{\citenamefont {Morishita}(2021)}]{morishita2021time}%
  \BibitemOpen
  \bibfield  {author} {\bibinfo {author} {\bibfnamefont {T.}~\bibnamefont
  {Morishita}},\ }\bibfield  {title} {\enquote {\bibinfo {title}
  {{Time-Dependent Principal Component Analysis: A Unified Approach to
  High-Dimensional Data Reduction using Adiabatic Dynamics}},}\ }\href
  {https://doi.org/10.1063/5.0061874} {\bibfield  {journal} {\bibinfo
  {journal} {J. Chem. Phys.}\ }\textbf {\bibinfo {volume} {155}},\ \bibinfo
  {pages} {134114} (\bibinfo {year} {2021})}\BibitemShut {NoStop}%
\bibitem [{\citenamefont {Wang}\ and\ \citenamefont
  {Tiwary}(2021)}]{wang2021state}%
  \BibitemOpen
  \bibfield  {author} {\bibinfo {author} {\bibfnamefont {D.}~\bibnamefont
  {Wang}}\ and\ \bibinfo {author} {\bibfnamefont {P.}~\bibnamefont {Tiwary}},\
  }\bibfield  {title} {\enquote {\bibinfo {title} {{State Predictive
  Information Bottleneck}},}\ }\href
  {https://doi.org/https://doi.org/10.1063/5.0038198} {\bibfield  {journal}
  {\bibinfo  {journal} {J. Chem. Phys.}\ }\textbf {\bibinfo {volume} {154}},\
  \bibinfo {pages} {134111} (\bibinfo {year} {2021})}\BibitemShut {NoStop}%
\bibitem [{\citenamefont {Belkacemi}\ \emph {et~al.}(2021)\citenamefont
  {Belkacemi}, \citenamefont {Gkeka}, \citenamefont {Leli{\`e}vre},\ and\
  \citenamefont {Stoltz}}]{belkacemi2021chasing}%
  \BibitemOpen
  \bibfield  {author} {\bibinfo {author} {\bibfnamefont {Z.}~\bibnamefont
  {Belkacemi}}, \bibinfo {author} {\bibfnamefont {P.}~\bibnamefont {Gkeka}},
  \bibinfo {author} {\bibfnamefont {T.}~\bibnamefont {Leli{\`e}vre}},\ and\
  \bibinfo {author} {\bibfnamefont {G.}~\bibnamefont {Stoltz}},\ }\bibfield
  {title} {\enquote {\bibinfo {title} {{Chasing Collective Variables using
  Autoencoders and Biased Trajectories}},}\ }\href
  {https://doi.org/https://doi.org/10.1021/acs.jctc.1c00415} {\bibfield
  {journal} {\bibinfo  {journal} {J. Chem. Theory Comput.}\ }\textbf {\bibinfo
  {volume} {18}},\ \bibinfo {pages} {59--78} (\bibinfo {year}
  {2021})}\BibitemShut {NoStop}%
\bibitem [{\citenamefont {Novelli}\ \emph {et~al.}(2022)\citenamefont
  {Novelli}, \citenamefont {Bonati}, \citenamefont {Pontil},\ and\
  \citenamefont {Parrinello}}]{novelli2022characterizing}%
  \BibitemOpen
  \bibfield  {author} {\bibinfo {author} {\bibfnamefont {P.}~\bibnamefont
  {Novelli}}, \bibinfo {author} {\bibfnamefont {L.}~\bibnamefont {Bonati}},
  \bibinfo {author} {\bibfnamefont {M.}~\bibnamefont {Pontil}},\ and\ \bibinfo
  {author} {\bibfnamefont {M.}~\bibnamefont {Parrinello}},\ }\bibfield  {title}
  {\enquote {\bibinfo {title} {{Characterizing Metastable States with the Help
  of Machine Learning}},}\ }\href
  {https://doi.org/https://doi.org/10.1021/acs.jctc.2c00393} {\bibfield
  {journal} {\bibinfo  {journal} {J. Chem. Theory Comput.}\ }\textbf {\bibinfo
  {volume} {18}},\ \bibinfo {pages} {5195--5202} (\bibinfo {year}
  {2022})}\BibitemShut {NoStop}%
\bibitem [{\citenamefont {Ketkaew}\ and\ \citenamefont
  {Luber}(2022)}]{ketkaew2022deepcv}%
  \BibitemOpen
  \bibfield  {author} {\bibinfo {author} {\bibfnamefont {R.}~\bibnamefont
  {Ketkaew}}\ and\ \bibinfo {author} {\bibfnamefont {S.}~\bibnamefont
  {Luber}},\ }\bibfield  {title} {\enquote {\bibinfo {title} {{DeepCV: A Deep
  Learning Framework for Blind Search of Collective Variables in Expanded
  Configurational Space}},}\ }\href
  {https://doi.org/https://doi.org/10.1021/acs.jcim.2c00883} {\bibfield
  {journal} {\bibinfo  {journal} {J. Chem. Inf. Model.}\ }\textbf {\bibinfo
  {volume} {62}},\ \bibinfo {pages} {6352--6364} (\bibinfo {year}
  {2022})}\BibitemShut {NoStop}%
\bibitem [{\citenamefont {Sun}\ \emph {et~al.}(2022)\citenamefont {Sun},
  \citenamefont {Vandermause}, \citenamefont {Batzner}, \citenamefont {Xie},
  \citenamefont {Clark}, \citenamefont {Chen},\ and\ \citenamefont
  {Kozinsky}}]{sun2022multitask}%
  \BibitemOpen
  \bibfield  {author} {\bibinfo {author} {\bibfnamefont {L.}~\bibnamefont
  {Sun}}, \bibinfo {author} {\bibfnamefont {J.}~\bibnamefont {Vandermause}},
  \bibinfo {author} {\bibfnamefont {S.}~\bibnamefont {Batzner}}, \bibinfo
  {author} {\bibfnamefont {Y.}~\bibnamefont {Xie}}, \bibinfo {author}
  {\bibfnamefont {D.}~\bibnamefont {Clark}}, \bibinfo {author} {\bibfnamefont
  {W.}~\bibnamefont {Chen}},\ and\ \bibinfo {author} {\bibfnamefont
  {B.}~\bibnamefont {Kozinsky}},\ }\bibfield  {title} {\enquote {\bibinfo
  {title} {{Multitask Machine Learning of Collective Variables for Enhanced
  Sampling of Rare Events}},}\ }\href
  {https://doi.org/https://doi.org/10.1021/acs.jctc.1c00143} {\bibfield
  {journal} {\bibinfo  {journal} {J. Chem. Theory Comput.}\ }\textbf {\bibinfo
  {volume} {18}},\ \bibinfo {pages} {2341--2353} (\bibinfo {year}
  {2022})}\BibitemShut {NoStop}%
\bibitem [{\citenamefont {Song}\ and\ \citenamefont
  {Zhao}(2022)}]{song2022slow}%
  \BibitemOpen
  \bibfield  {author} {\bibinfo {author} {\bibfnamefont {P.}~\bibnamefont
  {Song}}\ and\ \bibinfo {author} {\bibfnamefont {C.}~\bibnamefont {Zhao}},\
  }\bibfield  {title} {\enquote {\bibinfo {title} {{Slow Down to go Better: A
  Survey on Slow Feature Analysis}},}\ }\href
  {https://doi.org/https://doi.org/10.1109/TNNLS.2022.3201621} {\bibfield
  {journal} {\bibinfo  {journal} {IEEE Trans. Neural Netw. Learn. Syst.}\
  }\textbf {\bibinfo {volume} {35}},\ \bibinfo {pages} {3416--3436} (\bibinfo
  {year} {2022})}\BibitemShut {NoStop}%
\bibitem [{\citenamefont {Chen}, \citenamefont {Roux},\ and\ \citenamefont
  {Chipot}(2023)}]{chen2023discovering}%
  \BibitemOpen
  \bibfield  {author} {\bibinfo {author} {\bibfnamefont {H.}~\bibnamefont
  {Chen}}, \bibinfo {author} {\bibfnamefont {B.}~\bibnamefont {Roux}},\ and\
  \bibinfo {author} {\bibfnamefont {C.}~\bibnamefont {Chipot}},\ }\bibfield
  {title} {\enquote {\bibinfo {title} {{Discovering Reaction Pathways, Slow
  Variables, and Committor Probabilities with Machine Learning}},}\ }\href
  {https://doi.org/https://doi.org/10.1021/acs.jctc.3c00028} {\bibfield
  {journal} {\bibinfo  {journal} {J. Chem. Theory Comput.}\ }\textbf {\bibinfo
  {volume} {19}},\ \bibinfo {pages} {4414--4426} (\bibinfo {year}
  {2023})}\BibitemShut {NoStop}%
\bibitem [{\citenamefont {Jung}\ \emph {et~al.}(2023)\citenamefont {Jung},
  \citenamefont {Covino}, \citenamefont {Arjun}, \citenamefont {Leitold},
  \citenamefont {Dellago}, \citenamefont {Bolhuis},\ and\ \citenamefont
  {Hummer}}]{jung2023machine}%
  \BibitemOpen
  \bibfield  {author} {\bibinfo {author} {\bibfnamefont {H.}~\bibnamefont
  {Jung}}, \bibinfo {author} {\bibfnamefont {R.}~\bibnamefont {Covino}},
  \bibinfo {author} {\bibfnamefont {A.}~\bibnamefont {Arjun}}, \bibinfo
  {author} {\bibfnamefont {C.}~\bibnamefont {Leitold}}, \bibinfo {author}
  {\bibfnamefont {C.}~\bibnamefont {Dellago}}, \bibinfo {author} {\bibfnamefont
  {P.~G.}\ \bibnamefont {Bolhuis}},\ and\ \bibinfo {author} {\bibfnamefont
  {G.}~\bibnamefont {Hummer}},\ }\bibfield  {title} {\enquote {\bibinfo {title}
  {{Machine-Guided Path Sampling to Discover Mechanisms of Molecular
  Self-Organization}},}\ }\href
  {https://doi.org/https://doi.org/10.1038/s43588-023-00428-z} {\bibfield
  {journal} {\bibinfo  {journal} {Nat. Comput. Sci.}\ }\textbf {\bibinfo
  {volume} {3}},\ \bibinfo {pages} {334--345} (\bibinfo {year}
  {2023})}\BibitemShut {NoStop}%
\end{thebibliography}%

\end{document}